\documentclass[11pt]{article}
\pdfoutput=1
\usepackage{amssymb,dsfont,amsmath,amsfonts,hyperref,mdframed,mathrsfs,stmaryrd}
\usepackage[active]{srcltx}
\usepackage{slashed}
\usepackage{color}
\usepackage{amsthm}
\usepackage{scrtime}
\usepackage{cite}
\usepackage[normalem]{ulem}
\usepackage{framed}
\usepackage{cancel}
\usepackage{comment}
\usepackage{empheq}
\usepackage{rotating}
\usepackage{array}
\usepackage{datetime}
\usepackage[utf8]{inputenc} 
\usepackage{enumitem}

\usepackage[letterpaper]{geometry}
\numberwithin{equation}{section}

\let\oldsqrt\sqrt

\def\sqrt{\mathpalette\DHLhksqrt}
\def\DHLhksqrt#1#2{%
	\setbox0=\hbox{$#1\oldsqrt{#2\,}$}\dimen0=\ht0
	\advance\dimen0-0.2\ht0
	\setbox2=\hbox{\vrule height\ht0 depth -\dimen0}%
	{\box0\lower0.4pt\box2}}

\def\be{\begin{equation}}
\def\ee{\end{equation}}
\def\bea{\begin{eqnarray}}
\def\eea{\end{eqnarray}}

\def\a{\alpha}
\def\b{\beta}
\def\g{\gamma}
\def\d{\delta}
\def\e{\epsilon}
\def\k{\kappa}
\def\l{\lambda}
\def\m{\mu}
\def\n{\nu}
\def\s{\sigma}
\def\Comp{\mathbb{C}}
\def\Real{\mathbb{R}}

\def\o{\omega}
\def\O{\Omega}
\def\ad{\dot\alpha}
\def\bd{\dot\beta}
\def\cd{\dot\gamma}
\def\dd{\dot\delta}
\def\sb{\bar\sigma}
\def\ub{\bar u}

\def\ty{\widetilde y}
\def\vark{\varkappa}
\def\yb{{\bar y}}

\def\varkb{{\bar \varkappa}}
\def\mso{{\mathfrak{so}}}
\def\msp{{\mathfrak{sp}}}
\def\msl{{\mathfrak{sl}}}
\def\ft#1#2{{\textstyle{{\scriptstyle #1}
\over {\scriptstyle #2}}}} 

\def\lb{\boldsymbol{\lambda}}

\def\t{\tau}
\def\D{\Delta}
\def\G{\Gamma}

\input{glyphtounicode}
\newcommand{\eq}[1]{(\ref{#1})}

\newcommand{\nn}{\nonumber}

\begin{document}
\begin{center} 
{\huge \textbf{Higher spin fluctuations }\\[10pt]\textbf{on spinless 4D BTZ black hole}}
\\[15pt]
{\large Rodrigo Aros $^{a,\dag}$, \ \  
	Carlo Iazeolla $^{b,\ddag}$, \ \   
	Per Sundell $^{a,\S}$, \ \  
	Yihao Yin  $^{c,a,\sharp}$}

{\let\thefootnote\relax\footnote{\tt 
		\linebreak  
		$^\dag$ raros@unab.cl \linebreak 
		$^\ddag$ c.iazeolla@gmail.com \linebreak  
		$^\S$ per.anders.sundell@gmail.com \linebreak 
		$^\sharp$ yinyihao@nuaa.edu.cn , \ yinyihao@gmail.com}}
\\
$^{a}\ $\textit{Departamento de Ciencias F\'isicas,
	Universidad Andres Bello\\ Republica 220, Santiago de Chile}\\[0pt]
\bigskip
$^{b}\ $\textit{ NSR Physics Department, G. Marconi University \\
	via Plinio 44, Rome, Italy \quad \& \\
	INFN, Sezione di Napoli\\ Complesso Universitario di Monte S. Angelo, Napoli, Italy}\\[0pt]
\bigskip
$^{c}\ $\textit{ College of Science, Nanjing University of Aeronautics and Astronautics  \\
	Jiangjun Avenue 29, Nanjing, China}
\end{center}

\paragraph{Abstract.}

We construct linearized solutions to Vasiliev's four-dimensional higher spin gravity on warped $AdS_3 \times_\xi S^1$ which is an $Sp(2)\times U(1)$ invariant non-rotating BTZ-like black hole with $\mathbb{R}^2\times T^2$ topology.
The background can be obtained from $AdS_4$ by means of identifications along a Killing boost $K$ in the region where $\xi^2\equiv K^2\geqslant 0$, or, equivalently, by gluing together \emph{two} Ba\~nados--Gomberoff--Martinez eternal black holes along their past and future space-like singularities (where $\xi$ vanishes) as to create a periodic (non-Killing) time.
The fluctuations are constructed from gauge functions and initial data obtained by quantizing \emph{inverted} harmonic oscillators providing an oscillator realization of $K$ and of a commuting Killing boost $\widetilde K$.
The resulting solution space has two main branches in which $K$ star commutes and anti-commutes, respectively, to Vasiliev's twisted-central closed two-form $J$.
Each branch decomposes further into two subsectors generated from ground states with zero momentum on $S^1$.
We examine the subsector in which $K$ anti-commutes to $J$ and the ground state is $U(1)_K\times U(1)_{\widetilde K}$-invariant of which $U(1)_K$ is broken by momenta on $S^1$ and $U(1)_{\widetilde K}$ by quasi-normal modes.
We show that a set of $U(1)_{\widetilde K}$-invariant modes (with $n$ units of $S^1$ momenta) are singularity-free as master fields living on a total bundle space, although the individual Fronsdal fields have membrane-like singularities at $\widetilde K^2=1$.
We interpret our findings as an example where Vasiliev's theory completes singular classical Lorentzian geometries into smooth higher spin geometries.
\bigskip

\setcounter{footnote}{0}
\newpage

\tableofcontents

\section{Introduction}  \label{SecVasilieveqs}

\subsection{Higher spin resolution of gravitational singularities}

An interesting problem in gravity is whether classical spacetime singularities can be resolved by switching on higher spin gauge fields.
Indeed, the resulting non-abelian interactions are spacetime non-local already at the classical level, akin to those of a full quantum effective field theory.
Moreover, higher spin gravities contain infinite towers of massless fields at weak coupling that one may argue become massive due to quantum effects, hence associated to screened charges in weakly coupled asymptotic regions, while supporting moduli spaces of classical solutions interpolating between asymptotic regions and strongly coupled core regions with nontrivial topology. 
This motivates examining whether classical spacetime singularities can be completed into smooth higher spin geometries given by classical solutions to unbroken higher spin gravities with \emph{bounded} field configurations and \emph{finite} observables accessible to asymptotic observers, providing semi-classical realizations of geometrically \emph{entangled} quantum states.

To concord with basic properties of the holographic correspondence between generally covariant theories with anti-de Sitter vacua and conformal field theories in the context of higher spin theory \cite{Sundborg:1999ue,Sezgin:2002rt,Klebanov:2002ja,Girardello:2002pp}, we shall
\begin{itemize}
	\item[a)] presume a higher spin symmetry breaking mechanism whereby weakly coupled gauge fields with spins greater than two (and possibly also some fields with spins less than or equal to two) acquire masses so as to leave a spectrum with massless subsector corresponding to matter-coupled gravity; and
	\item[b)] construct exact solutions to unbroken higher spin gravities that describe smooth higher spin geometries containing asymptotically locally anti-de Sitter (ALAdS) (or de Sitter) regions where the full theory can be approximated by (free) Fronsdal fields\footnote{For the literature on solutions of this type in four spacetime dimensions, see 
\cite{Sezgin:2005pv,Iazeolla:2007wt,Didenko:2009td,Iazeolla:2017vng,Iazeolla:2017dxc,Iazeolla:2011cb,Iazeolla:2012nf,Bourdier:2014lya,Sundell:2016mxc,Aros:2017ror};
see also \cite{Prokushkin:1998bq,Didenko:2006zd,Iazeolla:2015tca}
for exact solutions in three spacetime dimensions, and \cite{Gubser:2014ysa} 
for solutions obtained in axial gauge in oscillator space.}. 
\end{itemize}
Thus, in the broken phase, the asymptotic fall-off of the fields that have acquired mass is enhanced, ensuring that they do not affect the leading orders of the Fefferman--Graham expansion of an effective lower-spin theory containing gravity (even though the spectrum of the broken phase is not fully gapped). 
By this screening mechanism, we envisage weakly coupled asymptotic regions described by an effective gravity theory \emph{glued} to strongly coupled core regions described by an unbroken higher spin gravity; that is, we trust the latter when its curvatures are large, and the former when its curvatures are small.

Moreover, drawing on recent progress in assigning entanglement entropy to topologically nontrivial spacetimes \cite{Arias:2019pzy}, our working hypothesis is that the emerging higher spin geometries are not only smooth but also entangled in the sense that 
\begin{itemize}
	\item[c)] the higher spin resolution of a gravitational singularity yields a set of topologies including manifolds with numerous boundaries; and
	\item[d)] manifolds with multiple boundaries are represented quantum mechanically by geometrically entangled states.
\end{itemize} 
Combining the dynamical higher spin symmetry breaking mechanisms (a) and (b) with the geometric entangling mechanisms (c) and (d), we envisage asymptotic observers represented by operators acting in the Hilbert space of the broken phase (with asymptotically enhanced mass-gap), sandwiched between geometrically entangled states with one ``external'' leg in the broken state space and multiple ``internal'' legs in unbroken state spaces, represented semi-classically by ALAdS higher spin geometries with non-trivial core topology.
In other words, we propose that gravitational singularities are resolved into moduli spaces of smooth higher spin geometries, whereby physical observables are given by sums over unbroken core states: the latter are organized into ensembles by geometrically entangled states \cite{Arias:2019pzy} represented semi-classically by classical solutions to higher spin gravities.

In this paper, we shall focus on (b) by exploring classical solutions to Vasiliev's equations in four spacetime dimensions \cite{Vasiliev:1990en}.
Vasiliev's theory has been conjectured \cite{Sezgin:2002rt,Klebanov:2002ja,Leigh:2003ez,Girardello:2002pp} to undergo dynamical symmetry breaking due to mixing between (massless) one-particle states and multi-particle Goldstone modes in the presence of special boundary conditions in anti-de Sitter spacetime.
As this mechanism does not require any coupling to additional fields, the theory, possibly including Yang-Mills-like gauge fields and fermions \cite{Konstein:1989ij,Sezgin:2012ag}, provides a relatively minimalistic framework for studying singularity resolutions already at the classical level in accordance with (a) and (b)\footnote{Stringy extensions \cite{Sezgin:2002rt,Engquist:2005yt,Chang:2012kt,Vasiliev:2018zer} by extra massive fields are likely required in order to admit flat space limits with significant mass-gaps.}.

The following classical singularities of matter-coupled gravity will be of interest for this work:
\begin{itemize}
\item[i)] Degenerate metrics;
\item[ii)] Analytic singularities\footnote{We refer to a singularity in an otherwise real-analytic function as an analytic singularity.} in generalized Weyl curvatures\footnote{In a matter-coupled gravity theory without fermions, the generalized Weyl curvatures consist of the spin-two Weyl curvature, the spin-one Faraday tensors and the scalar fields.};  
\item[iii)] Delta function sources in the equations of motion.
\end{itemize} 
At the center of the Schwarzschild black hole, all three types of singularities arise: the metric degenerates on trapped spheres (leading to geodesic incompleteness); the Weyl tensor blows up; and the linearized equations of motion have a delta function source.
In order to disentangle these types singularities, it is useful to instead consider fluctuations over constantly curved black holes, as the background only exhibits degenerate metrics related to trapped submanifolds, while the fluctuations can be made to exhibit curvature singularities.

Constantly curved black holes were first constructed in three dimensions by Ba\~nados, Teitelboim and Zanelli (BTZ) \cite{Banados:1992wn}, and further studied by Ba\~nados, Henneaux, Teitelboim and Zanelli (BHTZ) \cite{Banados:1992gq} within the context of a more general moduli space of three-dimensional constant curvature geometries coordinatized by conjugacy classes of $\mathfrak{so}(2,2)$, including extremal black holes, conical singularities and proper three-dimensional anti-de Sitter spacetime itself.
BHTZ-like geometries in four spacetime dimensions associated to conjugacy classes of $\mathfrak{so}(2,3)$, were first studied by \AA minneborg, Bengtsson, Holst and Peldan in \cite{Aminneborg:1996iz,Aminneborg:1997pz}, who observed that the uplift of the spinless BTZ black hole only has quasi-horizons that fail to trap any  two-dimensional subspaces.
The latter geometry was later revisited by Ba\~nados, Gomberoff and Martinez (BGM) \cite{Banados:1998dc}, who properly interpreted it (by representing it using a three-dimensional Penrose diagram) as a black hole that traps (one-dimensional) circles rather than any two-manifold.

In this paper, we shall examine fluctuations around the eternal spinless BGM black hole thought of as a classical solution of Vasiliev's bosonic higher spin theory in four dimensions.
More precisely, we shall construct linearized massless Weyl tensors of arbitrary integer spin obeying Bargmann--Wigner equations of motion on the aforementioned background and subject to various boundary conditions corresponding to different representations of the background symmetry group, including modes with momenta around the trapped sphere and quasi-normal modes.
Pending a fully non-linear construction, we shall verify our main hypothesis, namely that the linearized Vasiliev master fields admit analytic continuations across singularities as well as horizons, so as to create field configurations on extended manifolds with topologies that differ from that of the original BGM geometry.  
We shall focus on resolved geometries with a single asymptotic region, though the formalism readily produces resolutions with multiple asymptotic regions as well.

We remark that Vasiliev's four-dimensional theory contains a higher spin connection in spacetime  
\begin{itemize}
\item[1)] that remains flat at the fully non-linear level;
\item[2)] the holonomies of which can be combined with open Wilson lines in twistor space \cite{Sezgin:2011hq,Colombo:2012jx,Bonezzi:2017vha,DeFilippi:2019jqq} so as to provide a set of classical observables;
\item[3)] reduces to the background connection solutions to Vasiliev's equations in which the Weyl zero-form vanishes.
\end{itemize}
Thus, in rather sharp contrast to the extraction of classical observables for four-dimensional BHTZ-like geometries in gravity, which has so far turned out to be problematic \cite{Guilleminot:2017hrp,Guilleminot:2017hfy}, the (topologically extended) BHTZ-like higher spin geometries can be labelled faithfully by the aforementioned holonomies around the circle resulting from the  identification.
Moreover, from the higher spin point-of-view, there is nothing preventing switching on general charges in $\mathfrak{so}(2,3)$ including parameters associated to frame-fields containing rotation and various conical singularities. 
In this paper, we shall focus, however, on the issue of topological extensions of the background and higher spin fluctuation fields in the spinless case, leaving the construction and analysis of more complicated vacuum solutions to future work\footnote{It is worth mentioning that, from the point of view of the standard spin-$2$ geometry, there is no four-dimensional uplift of the three-dimensional rotating BTZ black hole, since, differently from the spinless case, the presence of an extra spatial dimension erases the horizon \cite{Holst:1997tm}. Since one of the issues to be studied in this paper is the resolution of singularities of fluctuation fields at the horizon within Vasiliev's higher spin gravity, we shall choose to investigate the linearized dynamics around a vacuum solution corresponding to the four-dimensional (topologically extended) non-rotating BTZ-like black hole.}.

\subsection{Resolution mechanisms}

More broadly speaking, Vasiliev's higher spin gravity tempers (i) using differential algebras and (ii) and (iii) using non-commutative algebras. 
In particular, 
\begin{enumerate}[label=\roman*)]
\item[I)] Degenerate metrics are handled by abandoning the Fronsdal formulation in favour of the \emph{unfolded} formulation \cite{Vasiliev:1988sa,Chernbook,Vasiliev:1999ba,Bekaert:2005vh}, in which the fundamental fields are differential forms obeying covariant constancy conditions referring to backgrounds with differential Poisson structures rather than Lorentzian structures. 
As we shall see, this formalism permits the construction of fluctuation fields from objects defined in coordinate-free bases that remain well-defined as the frame field degenerates, and that hence admit continuation across singularities of type (i).
To our best understanding, this mechanism for sending fluctuations through singularities associated to degenerate metrics has so far not been exhibited in the higher spin literature\footnote{Rather, in constructing unfolded systems of equations it is usually assumed that if the frame field is invertible then the system must admit a dual interpretation as a complex for an algebraic differential whose cohomology in different degrees consists of the dynamical Fronsdal fields, their gauge parameters, and equations of motion and Bianchi identities \cite{Vasiliev:1999ba}; for analogous treatment of mixed symmetry fields, see \cite{Skvortsov:2008vs,Boulanger:2008up}.}, though degenerate background metrics have been considered within the context of ``wormholes'' of three-dimensional Chern--Simons higher spin gravity \cite{Ammon:2011nk}, and the first-order formulation of gravity in the context of topology change \cite{Horowitz:1990qb}.

\item[II)] As for resolving analytic Weyl curvature singularities in higher spin gravity, the basic mechanism involves assembling infinite-dimensional towers of fields into \emph{horizontal} forms on fibered spaces with \emph{non-commutative} fibers, that we shall refer to as \emph{correspondence spaces}.
Locally, the horizontal forms, that we shall often refer to as \emph{master} fields, are forms on the base manifold valued in algebras of quantum mechanical operators realized as various distributions on the fiber including real-analytic functions and non-real analytic objects such as delta functions and their derivatives.
Above generic points of the base manifold, the master fields are real-analytic with Lorentz-covariant Taylor expansions in the fiber coordinates, the coefficients of which are bounded component fields.
Closing in on special points, however, the master fields approach non-real analytic distributions in the fiber that nonetheless remain well-defined as symbols of quantum mechanical operators belonging to a star product algebra with a trace, though their naive interpretation in terms of Lorentz-covariant component fields clearly breaks down.
So far, this mechanism has been shown to resolve Coulomb-like singularities (of codimension three) in the Weyl curvatures of four-dimensional higher spin black hole-like solutions of Vasiliev's theory \cite{Iazeolla:2011cb}. 
In this paper, we shall extend this result to membrane-like singularities (of codimension one) in linearized fluctuation fields over BGM black holes as well as $AdS_4$.
More precisely, the mechanism at work trades the analytic spacetime singularities in the Weyl curvatures for delta function singularities in the fiber supported on fiber submanifolds of codimension two, which can be shown to be well-defined operators in the above sense upon using a certain regular presentations \cite{Iazeolla:2011cb,Iazeolla:2017vng} to be outlined below.

\item[III)] Delta function sources in equations of motion typically accompany the singularities in (II), at least at the linearized level.
As for the analytic curvature singularities of odd codimension referred to above, we expect that the linearized Vasiliev system \cite{Vasiliev:1990vu,Vasiliev:1999ba} provides a map that transfers corresponding delta function sources in the Fronsdal field equations to delta function sources of codimension two for the noncommutative twistor space connection.
In fact, independently of whether the spacetime fields are singular or not, \emph{any} Vasiliev higher spin geometry exhibits a twistor space delta function source of codimensions two; for a recent treatment of these singularity structures, see also \cite{DeFilippi:2019jqq}.
Moreover, the latter arises via a vacuum expectation value of a dynamical two-form of an extension of Vasiliev's theory off-shell based on an internal 3-graded Frobenius algebra \cite{Boulanger:2015kfa,Bonezzi:2016ttk}, referred to as Frobenius--Chern--Simons (FCS) gauge theory.
We expect that the FCS two-form can develop various expectation values in spacetime as well including codimension-two delta function sources, as these can be regularized by embedding spacetime as a Lagrangian submanifold into its non-commutative cotangent bundle (or phase-spacetime) \cite{Iazeolla:2011cb}.
This suggests the existence of fully nonlinear higher spin geometries serving as resolutions of conical singularities arising in BHTZ-like geometries interpretable as entanglement surfaces extended into the bulk \cite{Ryu:2006bv,Ammon:2013hba,Arias:2019pzy}.
\end{enumerate}

In this paper, mechanism (I) is ubiquitous, as the unfolded formalism serves as part of the definition of the theory.
Some of its consequences for topological black holes are spelled out in Section \ref{Sec:BTZ}, where we contrast the metric-like and unfolded formulations.
We do so by constructing a number of gauge functions that describe various extensions of spinless BTZ geometries in three and four dimensions that provide solutions to the unfolded equations of motion but that do not admit an interpretation in terms of the metric-like formulation simply due to the fact that the extended geometries contain codimension-one surfaces where the BTZ warp factor vanishes; in particular, the extended BGM geometry, which is most direct relevance for the rest of the paper, is described in Section \ref{HPcoord}.

As for (II), the non-commutative geometry framework for higher spin resolution of curvature singularities is outlined in Section \ref{Sec:reso}; the specific mechanism found in \cite{Iazeolla:2011cb} for resolving Coulomb-like singularities of spherically symmetric genealized Petrov Type D solutions found in \cite{Didenko:2009td,Iazeolla:2011cb} is spelled out in Section \ref{Sec:resolution}.
We tend to the extension of this mechanism to membrane-like singularities in Section \ref{Sec zeroform-L} and Appendix \ref{Sec:singularity}.

As for (III), we have less to report on in this paper.
We would nonetheless like to remark that the higher spin singularity resolution mechanisms introduced so far can be implemented off- as well as on-shell, using an adaptation of the  Alexandrov--Kontsevich--Schwarz--Zaboronsky (AKSZ) formalism \cite{Alexandrov:1995kv} to Cartan integrable systems on non-commutative manifolds \cite{Boulanger:2012bj}. 
In fact, as far as one is concerned with resolving degenerate metrics in ordinary gravity, the AKSZ formalism permits the inclusion of degenerate frame fields into the classical theory, though quantum corrections are more delicate as they require a balance between even and odd forms in order for topological anomalies \cite{Wu:1990ci} to combine into a finite one-loop partition function.
On the other hand, the FCS model is manifestly topologically supersymmetric in the sense that its spectrum of even and odd forms is identical thus ensuring a finite one-loop normalization of the partition function on a given manifold \cite{Boulanger:2015kfa,Bonezzi:2016ttk,Arias:2016agc}.
As for resolving curvature singularities, the attendant non-commutative geometries are not visible in ordinary gravity nor in the perturbatively defined Fronsdal formulation of higher spin gravity.

In summary, to our best understanding, the higher spin resolution of classical singularities in gravity relies crucially not only on the higher spin extension as such, but also on its implementation using Vasiliev's unfolded formulation in terms of master fields.
Thus, before going further into the details of higher spin singularity resolutions, we would like to briefly point to a few key geometric features of Vasiliev's higher spin gravity that distinguishes it from the perturbative metric-like Fronsdal formulation, and how to think of these two different frameworks as being dual to each other in asymptotically anti-de Sitter geometries.

\subsection{Vasiliev versus Fronsdal formulations}

While the deformed Fronsdal formulation of higher spin gravity refers to a Lorentzian spacetime background, Vasiliev's formalism \cite{Vasiliev:1990vu,Vasiliev:1999ba,Bekaert:2005vh} introduces a non-commutative background for a differential graded (homotopy) associative algebra (DGA) of differential forms.
This algebra is a deformation of the classical algebra of differential forms (with its compatible wedge product and de Rham differential) along a differential Poisson structure so as to produce a space of symbols equipped with an associative star product and a mutually compatible differential.

The DGA operations can be realized together with compatible trace and hermitian conjugation operations by attaching differential forms as boundary vertex operators to an induced first-quantized differential Poisson sigma model \cite{Arias:2015wha}, which is a two-dimensional topological field theory with an ${\cal N}=1$ supersymmetry (of degree one) generated by the de Rham differential\footnote{Within the context of higher spin gravity, one may think of the differential Poisson sigma model as describing first-quantized conformal particles making up the partons of a tensionless string \cite{Engquist:2005yt}.}.
The fibration of the correspondence space (giving rise to the horizontal forms) arises from additional supersymmetries (of degree minus one) generated by inner derivatives along vector fields that preserve the differential Poisson structure, which are hence special fundamental vector fields \cite{Arias:2016agc}.

The DGA operations induce a class of \emph{star product local} functionals given by traces of star products of horizontal forms and their exterior derivatives.
This class remains closed under the Batalin--Vilkovisky (BV) bracket modulo a set of boundary conditions (usually solved by choosing a polarization and setting all momenta to zero at the boundary).
Thus, the BV master equation poses a well-defined deformation problem for a gauge invariant BV path integral measure based on a star product local master action, leading to a notion of star product local (quasi-)topological non-commutative field theories of AKSZ-type \cite{Boulanger:2012bj}.

Assuming the existence of a topological open string on $T^\ast Sp(4)\times \Comp^2$ (with holomorphic symplectic structure on $\Comp^2$) obtained from deformation quantization of a single conformal particle \cite{Engquist:2005yt}, we think of the FCS theory as a truncation that retains the zero- and winding modes, which thus coordinatize the correspondence space with non-commutative fibers arising from fermionic zero-modes on $\Comp^2$ induced via the aforementioned special fundamental vector fields \cite{Arias:2016agc}.
We then embed Vasiliev's theory into the FCS theory as an on-shell branch with ``order parameter'' given by the aforementioned two-forn vacuum expectation value\footnote{We expect that reductions of the FCS model in the presence of various vacuum expectation values create a moduli space of unfolded systems on the reduced correspondence spaces (with four-dimensional commuting base manifold and non-commutative $\Comp^2$ fiber), containing the plethora of ``formal'' higher spin gravities \cite{Sharapov:2019vyd} obtained by deformations of the fiber star product.}.

The Vasiliev branch contains ALAdS solutions, which are master field configurations (on the total non-commutative fibered space) subject to boundary conditions giving rise to asymptotically free Fronsdal fields \cite{Iazeolla:2017vng,Iazeolla:2017dxc,DeFilippi:2019jqq}\footnote{The ALAdS boundary conditions add non-trivial perturbative corrections to the gauge function already at the linearized level which steer the perturbative expansion away from the singular gauge found in \cite{Boulanger:2015ova}.}.
Our basic hypothesis is that the free energy, \emph{i.e.} the on-shell action, of the FCS model is finite on these ALAdS configurations.
The FCS free energy given by the on-shell value of a topological vertex operator (TVO) \cite{Sezgin:2011hq,Boulanger:2015kfa,Bonezzi:2016ttk}, \emph{i.e.} a higher spin invariant star product local boundary functional whose total variation vanish on-shell (such that it can be added to the AKSZ bulk action without affecting the smoothness and nilpotency of the BRST operator).
The FCS theory only admits a finite number of TVO's, given by Chern classes and Chern--Simons forms, that is, the FCS free energy functional contains only a finite number of free parameters.
In a stark contrast, Vasiliev's theory admits an infinite number of TVO's, suggesting that these can be used as building blocks for the FCS free energy functional.

As for classical observables in the Vasiliev branch of the FCS theory, the simplest ones are zero-form charges \cite{Sezgin:2005pv,Colombo:2010fu,Iazeolla:2011cb,Sezgin:2011hq,Colombo:2012jx}, which are integrals over the non-commutative twistor space of constructs formed out of spacetime curvatures and their derivatives evaluated at a single point in spacetime.
These observables have cluster decomposition properties characteristic of extensive variables \cite{Colombo:2010fu,Iazeolla:2011cb}, and hence serve as natural building blocks for higher spin amplitudes, referred to in \cite{Colombo:2010fu} as quasi-amplitudes.
Indeed, their classical perturbative expansion around $AdS_4$ backgrounds reveal a direct correspondence between the first-quantized topological open string amplitudes and the correlation functions of holographically dual conformal field theories \cite{Colombo:2012jx,Didenko:2012tv,Didenko:2013bj,Gelfond:2013xt,Bonezzi:2017vha}.

We expect the deformed Fronsdal theory to be perturbatively equivalent to the Vasiliev branch of FCS model at the level of amplitudes rather than at the level of spacetime vertices \cite{Boulanger:2015ova,Bekaert:2015tva,Vasiliev:2016xui,Didenko:2018fgx} (or microscopic field configurations).
In other words, we propose that the Fronsdal program set up on a Lorentzian spacetime manifold and the Vasiliev program set up on a non-commutative manifold are dual (at the level of free energy functionals) provided the two sides are supplemented with corresponding ALAdS boundary conditions.

The Fronsdal and Vasiliev formulations exhibit an important conceptual difference. 
The simplest TVO of the FCS model does not receive any quantum corrections, as it is built from forms in degree one; for details, see \cite{Boulanger:2011dd,Boulanger:2015kfa}.
This simple result is qualitatively in agreement with the holographically dual conformal field theory\footnote{Non-trivial quantum corrections to the FCS free energy can be generated by adding TVO's that depend on forms in higher form degrees \cite{Boulanger:2011dd}.}.
As for the perturbatively defined Fronsdal action, on the other hand, its natural interpretation is as a quantum effective action fixed essentially by uplifting the conformal bootstrap approach into the bulk \cite{Sleight:2017fpc,Sleight:2017pcz}.
Moreover, from its $1/N$-expansion it follows that it has \emph{no} (non-trivial) classical limit.
In other words, it appears that the deformed Fronsdal theory does not provide any path integral measure based on a classical (possibly quasi-local) action formulated directly on spacetime\footnote{
Nonetheless, it has been demonstrated \cite{Giombi:2013fka} that perturbative ``re-quantization'' of the deformed Fronsdal theory dual to the free theory can be interpreted sensibly at least at one-loop, suggesting that its realm of validity can be extended so as to include boundary conditions corresponding to non-trivial conformal field theories.}.

In summary, in order to embed the results of this paper (which hold on their own) into the above broader physical context, we will assume that 
\begin{itemize}
\item[--] Vasiliev's equations describe a quantum effective field theory including quantum effects from second as well as first quantization analogously to string field theory \cite{Witten:1985cc}; 
\item[--] there exists a free energy functional that makes Vasiliev's equations dual to perturbatively defined Fronsdal formulation on ALAdS backgrounds.
\end{itemize}
Thus, \emph{to the extent that one expects that quantum corrections are important in order to smoothen out classical spacetime singularities, Vasiliev's equations provide a background-independent formulation for studying such effects within the context of higher spin gravity}.

\subsection{Outline of the paper}

The scope of the paper is to show that probing the spinless BGM black hole in four dimensions using linearized higher spin master fields leads to smooth linearized higher spin geometries.

In Section \ref{Sec:reso}, we first recall the key geometric structures arising in Vasiliev's formalism of relevance for resolving curvature singularities.
We then demonstrate the resolution of the analytic part of the Coulomb-like singularity (in the Weyl curvature) at the level of linearized master fields; the generalization of this mechanism to membrane-like singularities is analyzed in Section \ref{Sec:fluct} and Appendix \ref{Sec:singularity}. 

In Section \ref{Sec:BTZ}, we first recall the basics of the spinless BTZ black hole in three dimensions and its uplift to the spinless BGM black hole in four dimensions.
We then show how the base manifolds of these black holes can be extended using the gauge function approach so as to cross over singularities as well as include additional boundaries; as we shall see, the extended topologies are simpler than those in gravity in the sense that there is no longer any need to attach boundaries to the future and past singularities.

In Section \ref{Sec:HSBTZ}, we discuss some generalities of switching on higher spin fluctuations around the topologically extended spinless BGM black hole as a vacuum solution to Vasiliev's four-dimensional higher spin gravity; as for the details of the Vasiliev system, we refer to the literature \cite{Vasiliev:1990en,Vasiliev:1990vu,Vasiliev:1992av}.

In Section \ref{Sec:indata}, we construct a space of building blocks for the integration constant of the Weyl zero-form that gives rise to fluctuation modes on the spinless BGM black hole background.
These building blocks are stargenfunctions of two number operators with complex eigenvalues that obey kinematical conditions as well as the quantization condition induced by the BHTZ-like identification on the BGM background.
In particular, we find that quantizing the identification Killing vector implies that the spectrum of the dual Killing vector has imaginary parts, which we interpret as quasi-normal modes. 

In Section \ref{Sec:fluct}, we unfold the initial datum and extract fluctuation fields in a simple case (when there are no quasi-normal modes) which allows us to examine the fiber real-analyticity properties of the Weyl zero-form in detail and exhibit the resolution of its membrane-like singularity.

In Section \ref{Sec:concl}, we conclude, stressing the limitations in our approach visavi non-linear perturbative corrections, that we hope to present elsewhere. 

In the Appendices, we spell out our conventions; collect various formulae that are used in the body of the paper; and analyze in detail the fiber distribution arising at the membrane-like singularity; and discard an apparent singularity of no physical importance.

\section{Resolving of curvature singularities in ALAdS backgrounds}\label{Sec:reso}

In this Section, we outline key features of the unfolded formulation of higher spin gravity of relevance for resolving singularities and generating vacua with nontrivial topology associated to geometrically entangled vacuum states.
We exemplify the resolution of analytic Weyl curvature singularities in the context of codimension-three Coulomb-like singularities using an extension of the Weyl algebra by delta function distributions, referred to as the extended Weyl algebra \cite{Iazeolla:2011cb,Boulanger:2015kfa,Iazeolla:2017vng}.

\subsection{Horizontal forms and quasi-topological noncommutative field theories}

The fundamental field of the FCS model is a flat horizontal odd multi-form, or Quillen superconnection, on a fibered non-commutative manifold, or correspondence space, valued in an internal 3-graded Frobenius algebra.
This master field decomposes under the internal algebra into a set of differential forms of different degrees, including zero, on the total space, all of which are horizontal, that is, given locally by differential forms on the base space valued in a space of zero-forms on the fiber space forming an associative operator algebra.
Finally, the flatness condition on the Quillen superconnection implies that all its horizontal components obey Cartan integrable covariant constancy conditions on the correspondence space.

The appearance of horizontal forms has two immediate consequences for resolving singularities:
\begin{itemize}
	\item[a)] A finite set of covariantly constant master fields contains an infinite set of covariantly constant differential forms on the base manifold, capable of capturing ordinary local degrees of freedom propagating on commutative spacetime leafs of the base manifold;
	\item[b)] The associative fiber algebra contains various higher spin representations including delta functions as well as real analytic functions, capable of capturing spacetime singularities as well as regular, possibly ALAdS, configurations.
\end{itemize}
Further below, we shall exemplify how (a) and (b) play a crucial role in resolving classical singularities associated with degenerate frame fields and analytic Weyl curvature singularities, respectively.

The FCS model provides an example of a quasi-topological field theory, \emph{i.e.} a functorial map \cite{Atiyah:1989vu,Segal:2001dc} into a category of infinite-dimensional tensors, which one may think of as a set of generalized representation spaces, from a category of topological manifolds with geometrical decorations, which one may think of as a generalized group; in the case of the FCS model, differential Poisson manifolds (with conformal infinities and other defects) are encoded into differential graded star product algebras (with vacuum gauge functions and other cohomologically nontrivial elements). 
These maps provide a natural generalization of the representations used in ordinary quantum mechanics, whereby manifolds with boundaries and other defects are mapped to geometrically entangled ``vacuum'' states on which locally defined quantum fields act modulo overlap conditions encoding transition functions and other boundary conditions.

A natural quasi-topological field theory is the two-dimensional gauged Poisson sigma model of a quantum mechanical system with symplectic manifold $S$ on which acts a group $G$.
Putting this AKSZ model on a disc with boundaries with marked points for insertions of boundary vertex operators gives rise to a boundary functor that maps a boundary point to a space of functions on $S$ realized as operators in a Hilbert space ${\cal H}$, and an oriented open boundary interval to a representation of $G$ in terms of quantum mechanical evolution operators in ${\cal H}\otimes {\cal H}^{\ast}$ with group parameters given by vacuum expectation values for the embedding of the interval into $G$.
Thus, the quasi-topological treatment of ordinary quantum mechanics gives rise to an interplay between layers of functors acting on points, intervals and discs, providing an example of a two-category topological field theory \cite{Baez:2004pa}. 

To reach a quasi-topological re-formulation of an ordinary quantum field theory (on a Lorentzian manifold), one first switches to its unfolded formulation on-shell as a Cartan integrable system which can be taken off-shell as an two-category AKSZ model in one higher dimension.
One may then ask whether the original $S$-matrix (or holographic correlation functions) admits a dual realization as a TVO activated on-shell by combinations of gauge functions for the frame field and Weyl zero-form integration constants \cite{Boulanger:2008up,Boulanger:2012bj}.
A closely related topic is the re-formulation of two-dimensional (matter-coupled) gravities in terms of topological open membranes of AKSZ type \cite{Hofman:2001zt} as part of a background independent formulation of string field theory.
In this context, the boundaries of the topological bulk theory are two-dimensional surfaces with multiple defects mapped functorially to geometrically entangled multi-string states obeying overlap conditions induced by transition functions as synthesized within the group theoretic operator formalism for string scattering amplitudes \cite{Neveu:1987mn,AlvarezGaume:1988bg,Engberg:1991yp}.

As for the quasi-topological re-formulation of quantum field theories containing gravity in higher dimensions, higher spin gravity provides an interesting testing ground\footnote{Several interesting questions can be addressed directly within an AKSZ-inspired semi-classical treatment of pure gravity; for example, the Gibbons--Hawking entropy can be interpreted as being due to geometrical entanglement involving de Sitter vacua with boundaries and defects representing static observers \cite{Arias:2019pzy}.}.
To this end, we treat AdS spacetime as a manifold with the topology of a circle and a sphere, and with a circle defect representing the conformal infinity where the frame fields blow up \cite{Boulanger:2015kfa}.
This closed manifold can be taken to be the boundary of an AKSZ bulk manifold in one higher dimension, to which suitable TVO's can be attached.
The resulting quantum field theory provides a functorial map from the topological bulk manifold, viewed as a morphisms of boundary manifolds with defects to, to a space of geometrically entangled states built from boundary states that in their turn have substructures representing spacetimes with conformal infinities and other defects inducing local degrees of freedom in infinite-dimensional representations of higher spin algebras (or other non-compact gauge algebras).
Thus, in the above sense, the quasi-topological formulation of higher spin gravity (as well as ordinary gravity) is not much different from that if ordinary quantum mechanics in that both are multi-category gauge theories of AKSZ type.

The resulting partition functions are thus given in the semi-classical approximation by sums over on-shell boundary states weighted by TVO's.
The latter thus play a role that is analogous to that of Boltzmann factors in the standard approach to quantum mechanics and field theory, in that they act as convergence factors in sums over infinite-dimensional spaces of boundary states, though they arise in quite a different fashion essentially as homogeneous solutions to the BV master equation triggered by the AKSZ boundary condition.

In what follows, we shall detach ourselves from the above larger picture and limit ourselves to the construction of semi-classical boundary states, that is, classical solutions to higher spin gravity with multiple conformal infinities and curvature singularities giving rise to finite free energies.

\subsection{Extended Weyl algebra}

To exhibit this resolution mechanism, we take the fiber to be the non-commutative holomorphic symplectic $\Comp^2$ with canonical coordinates $Y_{\underline\alpha}=(y_\alpha,\bar y_{\dot\alpha})$ subject to the canonical star product commutation rules
\be [y_\alpha,y_\beta]_\star = 2i \epsilon_{\alpha\beta}\ ,\qquad
[y_\alpha,\bar y_{\dot\beta}]_\star = 0\ ,\qquad
[\bar y_{\dot\alpha},\bar y_{\dot\beta}]_\star = 2i \epsilon_{\dot\alpha\dot\beta}\ .\ee
The chiral star product 
\be f(y,\bar y)\star g(y,\bar y)=
\int \frac{d^{2}\xi d^{2}\bar{\xi}d^{2}\eta d^{2}\bar{\eta}}{(2\pi )^{4}}
e^{i\left( \eta^{\alpha }\xi_{\alpha }+\bar{\eta}^{\dot{\alpha}}\bar{\xi}_{\dot{\alpha}%
	}\right) }f_{1}\left( y+\xi,\bar{y}+\bar{\xi}\right) 
f_{2}\left( y+\eta,\bar{y}+\bar{\eta}\right) \ ,\ee
where each auxiliary doublet is integrated over $\Real^2$, is equivalent to the Moyal product for the space ${\cal P}$ of Weyl ordered polynomials.
It admits the following compatible hermitian conjugation operation:
\be (y_\alpha)^\dagger=\bar y_{\dot \alpha}\ ,\qquad \dagger\circ \dagger={\rm Id}\ .\ee
Realizing the Lie algebras $\msp(4)$ and $\msl(2;\Comp)$ as Weyl-ordered bilinears, $Y_{\underline\alpha}$ form a real $\msp(4)$-quartet, and $y_\alpha$ a complex $\msl(2;\Comp)$-doublet.

The \emph{extended Weyl algebra} is defined by 
\be {\cal W}={\cal P}\oplus ({\cal P}\star \kappa_y)\oplus ({\cal P}\star\bar\kappa_{\bar y})\oplus({\cal P}\star \kappa_y\star \bar\kappa_{\bar y})\ ,\ee
where
\begin{equation}
\kappa_{y}=2\pi \delta^{2}(y)\ ,\qquad \bar\kappa_{\bar y}=2\pi \delta^{2}(\bar y)\ ,
\end{equation}
are Klein operators obeying 
\begin{equation}
\kappa_{y}\star \kappa_{y}=1\ ,\qquad \pi (f)=\kappa_y \star f\star \kappa_y \text{ \ ,}\label{ky21}
\end{equation}
\begin{equation}
\bar\kappa_{\bar y}\star \bar \kappa_{\bar y}=1\ ,\qquad \bar\pi (f)=\bar \kappa_{\bar y} \star f\star \bar \kappa_{\bar y}\text{ \ ,}\label{kby21}
\end{equation}
where the inner automorphisms 
\be \pi(y_\alpha,\bar y_{\dot\alpha})=(-y_\alpha,\bar y_{\dot\alpha})\ ,\qquad
\bar\pi(y_\alpha,\bar y_{\dot\alpha})=(y_\alpha,-\bar y_{\dot\alpha})\ .\ee
In other words, ${\cal W}$ is spanned by polynomials and derivatives of holomorphic, anti-holomorphic and full fiber delta functions.
It follows that ${\cal W}$ is an associative algebra that is left invariant under chiral Fourier transformation \cite{Didenko:2009td} of the Weyl ordered symbols, \emph{viz.}
\be {\cal W}\cong {\cal W}\star \kappa_y\ .\ee

\subsection{Vacuum gauge functions and topology change via degenerate metrics}

Vasiliev's four-dimensional higher spin gravity contains ALAdS vacuum solutions $({\cal M}^{(n)}_4,\Omega)$, where $\Omega$ is an $\msp(4)$-valued connection obeying 
\begin{itemize}
\item[i)] the flatness condition
\be d\Omega+\Omega\star \Omega=0\ ,\ee
on
\be {\cal M}^{(n)}_4\stackrel{\rm top}{\cong} S_1 \times (S^3\setminus \{P_1,\dots,P_n\})\ ,\ee
where $P_\xi$, $\xi=1,\dots,n$, are points in $S^3$;
\item[ii)] ALAdS boundary conditions at $S^1\times\{P_\xi\}$, that is, the $\pi$-odd component of $\Omega$ is an invertible frame field in a tubular neighbourhood of $S^1\times\{P_\xi\}$, for $\xi=1,\dots,n$.
\end{itemize}
Thus, we may view the theory as a field theory on 
\be {\cal M}_4 \stackrel{\rm top}{\cong} S^1\times S^3\ ,\ee
with a set of marked submanifolds where the $\pi$-odd  projection of $\Omega$ is allowed to blow up.
This boundary value formulation provides an alternative to the standard conformal compactification of anti-de Sitter spacetime that is convenient in order to describe vacua of higher spin gravity with multiple boundaries, but that becomes crucial, however, in order to re-formulate higher spin gravity (or unfolded gravity for that matter) as a quasi-topoligal AKSZ quantum field theory in one higher dimension \cite{Boulanger:2015kfa} (for which $D_2\times S^3$ appears to be the most natural choice). 

Thus, in any simple region $U_4\subset {\cal M}^{(n)}_4$ we have 
\be \Omega=L^{-1}\star dL\ ,\ee
where $L:U_4\to Sp(4)$ is a gauge function.
Two gauge functions are considered equivalent as long as they are homotopic in the interior of ${\cal M}^{(n)}_4$ and obey the boundary conditions; keeping the boundary conditions at $S^1\times \{P_\xi\}$ fixed there is nothing preventing the gauge functions from collapsing in the interior of ${\cal M}^{(n)}_4$ so as to create a degenerate frame field as long as no new singularities arise in $\Omega$.

A one-parameter family of AdS vacua with one boundary arises on 
\be {\cal M}^{(1)}_4:={\cal M}_4\setminus (S^1\times\{N\})\stackrel{\rm top}{\cong} S^1\times \Real^3\ ,\ee
where $N\in S^3$, by coordinatizing the $S^1$ using 
\be T\in [0,2\pi\beta)\ ,\qquad \beta>0\ ,\ee
and $\Real^3$ using $(\rho,n^r)$, $r=1,2,3$, obeying
\be \rho\geqslant 0\ ,\qquad n^r n_r=1\ ,\ee
and taking 
\be L=\exp_\star (iET)\star \exp_\star (iP_r n^r {\rm arcsinh} f^{(1)}(\rho))\ ,\label{Lg}\ee
where $E$ is the energy operator; $P_r$ are the spatial transvections in $\mathfrak{so}(2,3)$ (for conventions, see Appendix \ref{App:conv}; and the radial function 
\be f^{(1)}(\rho)=\rho\ ,\ee
up to homotopic deformations (as these do not affect any physical observables).
For any value of $\beta$, the vacuum connection $\Omega$ is periodic on $S^1$ with holonomy 
\be H_{S^1}(\Omega)=\exp_\star (2\pi i \beta E)=(\kappa_y\star \bar\kappa_{\bar y})^{\star[\beta]}\star \exp_\star (2\pi i (\beta-[\beta]) E)\ ,\ee
where $[\beta]$ denotes the integer part of $\beta$, as can be seen using 
\be \exp_\star (2\pi i  E)=\kappa_y\star \bar\kappa_{\bar y}\ .\label{exp2piiE}\ee
Thus, standard global $AdS_4$, given by the hyperbola in embedding space covered once, corresponds to 
\be \mbox{Global $AdS_4$:}\quad \beta=1\ ,\qquad H_{S^1}(\Omega)=\kappa_y\star \bar\kappa_{\bar y}\ ,\ee
which thus has a non-trivial holonomy\footnote{In constructing higher spin fluctuation fields in Section \ref{Sec:fluct}, we shall use the stereographic gauge function, which is not globally defined but makes explicit an $\mso(1,3)$ subalgebra of the isometry $\msp(4)$.}.

For $n=2$, one has
\be {\cal M}^{(2)}_4:={\cal M}_4\setminus S^1\times\{N,S\}\stackrel{\rm top}{\cong} S^1\times S^2\times \Real\ ,\ee
where $N,S\in S^3$.
One may now take 
\be L=\exp_\star (iET)\star \exp_\star (iP_r n^r {\rm arcsinh} f^{(2)}(\rho))\ ,\label{Lg(2)}\ee
with radial function $f^{(2)}(\rho)$ homotopic in the bulk to
\be f^{(2)}(\rho)=\sqrt{a^2+\rho^2}\ ,\qquad \rho\in \Real\ ,\qquad a\geqslant 0\ ,\label{frho}\ee
where $a$ is a constant.
The resulting constantly curved manifold has two ALAdS regions in the tubular neighbourhoods of $S^1\times\{N,S\}$ with conformal infinities given by Lorentzian $S^1\times S^2$.
These asymptotic regions are connected by a cylinder with a degenerate metric
\be ds^2=-(1+a^2+\rho^2)dT^2+\frac{\rho^2 d\rho^2}{(a^2+\rho^2)(1+a^2+\rho^2)}+
(a^2+\rho^2)d\Omega_2^2\ ,\ee
which is Lorentzian except at $\rho=0$; if $a>0$, then the non-metricity is due to the degeneration of $\vec\partial_\rho$, and if $a=0$ then the non-metricity us due to the degeneration of the metric on the $S^2$ at $\rho=0$.
One may view the above quasi-Lorentzian geometry as a semi-classical description of an entangled vacuum state \cite{Arias:2019pzy} arising upon taking the massless limit of the eternal Kruskal--Szekeres black hole in $AdS_4$.

The above construction can be generalized so as to introduce further asymptotic regions; we leave the study of the resulting moduli spaces for future work.

\subsection{Linearized Weyl zero-form}

The higher spin fluctuations around $({\cal M}^{(n)}_4,\Omega)$ are contained in a zero-form $\Phi$, referred to as the Weyl zero-form, valued in the \emph{extended} twisted-adjoint representation
\be \mathfrak{T}:=\left\{ T\in {\cal W}: T^\dagger=\pi(T)\ ,\quad \pi\bar\pi(T)=T\right\}\ ,\label{bosandreal}\ee
of the \emph{extended} higher spin Lie algebra
\be \mathfrak{hs}(4):=\left\{ X\in {\cal W}: X^\dagger=-X\ ,\quad \pi\bar\pi(X)=X\right\}\ ;\ee
the twisted adjoint representation map $\rho:\mathfrak{hs}(4)\to {\rm End}(\mathfrak{T})$ is defined by
\be \rho(X)T:= X\star T-T\star \pi(X)\ .\ee
The corresponding unextended representations are given by
\be \check{\mathfrak{hs}}(4):= \left.\mathfrak{hs}(4)\right|_{\cal P}\ ,\qquad
\check{\mathfrak{T}}:= \left.\mathfrak{T}\right|_{\cal P}\ .\ee
The linearized Weyl zero-form obeys
\be D^{(0)}\Phi:=d\Phi+\Omega\star \Phi-\Phi\star\pi(\Omega)=0\ ,\ee
whose general solution is given by
\be \Phi^{(L)}_{\Psi}:=L^{-1}\star\Phi' \star\pi(L )\ ,\qquad \Phi':=\Psi\star \kappa_y\ ,\label{LrotPhi}\ee
where $\Psi$ is a constant in ${\cal W}$ and $L$ is a gauge function.
The constant $\Phi'$ contains all spacetime derivatives of the physical fields evaluated at a spacetime point, which we shall refer to as the \emph{unfolding point}, that are invariant under linearized (or abelian) higher spin gauge transformations.
The gauge function $L$ ``spreads’’, or ``unfolds’’ this local datum, which we shall also refer to as \emph{initial datum}, on the spacetime chart $U_4$ where $L$ is defined \cite{Vasiliev:1999ba,Bekaert:2005vh}. 

To construct globally defined configurations for $n=1,2$, we may use the gauge functions in \eq{Lg} and \eq{Lg(2)}, respectively, with $\beta\in \mathbb{N}$; in particular, for $\beta=1$, the periodicity of the linearized Weyl zero-form under $T\to T+2\pi$ follows from \eq{exp2piiE} and the fact that $\pi\bar\pi(\Phi')=\Phi'$.

\subsection{Particle and black hole states in $AdS_4$}

Families of (exact) biaxially symmetric, generalized Petrov-type D solutions to Vasiliev's equations have been constructed in \cite{Iazeolla:2011cb,Iazeolla:2012nf,Sundell:2016mxc} using gauge functions and Weyl zero-form integration constants.
These integration constants are expanded in special bases such that each distinct (micro)state consists of an infinite tower of Fronsdal fields.
The corresponding master fields are valued in a fiber algebras spanned by delta functions as well as real-analytic functions \cite{Iazeolla:2011cb,Sundell:2016mxc,Iazeolla:2017vng,Iazeolla:2017dxc,Aros:2017ror}.

In particular, there are two branches with two compact Killing symmetries, of which one consists of black-hole states with ALAdS regions, including the linearized fields of the charged Kerr--AdS black hole of the Einstein--Maxwell theory (which we think of as a broken phase of the higher spin gravity theory).

At the linearized level, the black hole states arise naturally together with particles states by taking 
\be \Psi\in {\rm End}({\cal F})\ ,\qquad {\cal F}={\cal F}_+ \oplus {\cal F}_-\ ,\ee
where ${\cal F}_+$ and ${\cal F}_-$ consist of the direct product of two Fock and anti-Fock spaces, respectively, as $E$ is the Hamiltonian of the two-dimensional harmonic oscillator.
The resulting linearized solution spaces consist of superpositions of generalized Type-D modes $\Phi^{(L)}_{\Psi_D}$ with initial data  
\be \Psi_D\in {\cal A}_{D}:= {\rm Hom}({\cal F}_+,{\cal F}_+)\oplus {\rm Hom}({\cal F}_-,{\cal F}_-)\ ,\ee
and particle modes $\Phi^{(L)}_{\Psi_P}$ with initial data 
\be \Psi_P\in {\cal A}_{P}:= {\rm Hom}({\cal F}_+,{\cal F}_-)\oplus {\rm Hom}({\cal F}_-,{\cal F}_+)\ .\ee
From $\kappa_y\star {\cal F}_\pm={\cal F}_\pm$, it follows that these two types of modes are exchanged by the duality transformation 
\be {\cal A}_{D}= \kappa_y\star {\cal A}_{P}\ .\label{calad}\ee
Presenting the initial data using regular presentations \cite{Iazeolla:2011cb,Iazeolla:2017vng,Aros:2017ror}, yields the orthogonality relations
\be {\rm Hom}({\cal F}_\sigma,{\cal F}_\sigma')\star {\rm Hom}({\cal F}_{\sigma''},{\cal F}_{\sigma'''})=\delta_{\sigma',\sigma''} {\rm Hom}({\cal F}_\sigma,{\cal F}_\sigma')\star {\rm Hom}({\cal F}_{\sigma'},{\cal F}_{\sigma'''})\ ,\ee
where $\sigma,\sigma',\sigma'',\sigma'''=\pm$, that is
\be {\cal A}_{D}\star {\cal A}_{D}={\cal A}_{D}\ ,\qquad {\cal A}_{D}\star {\cal A}_{P}={\cal A}_{P}\ ,\ee
\be {\cal A}_{P}\star {\cal A}_{D}={\cal A}_{P}\ ,\qquad {\cal A}_{P}\star {\cal A}_{P}={\cal A}_{D}\ ,\ee
which turn out to dictate the self-interactions among particle and black hole states governed by the quadratic terms in Vasiliev's equations \cite{Iazeolla:2011cb}.

\subsection{Resolving Coulomb-like singularities}\label{Sec:resolution}

The linearized black hole geometries contain Coulomb-like singularities, which consist of analytic singularities in the Weyl curvatures and delta function sources for the Fronsdal curvatures.
They arise by first expanding the horizontal forms into Lorentz tensors in the ALAdS region, and then following these fields towards the origin.
On the other hand, the horizontal forms remain well-defined as symbols of operator algebra elements defined on the entire correspondence space. 

To exhibit this resolution mechanism, we start by observing that if $\Phi$ is real-analytic on all fibers above a region $U_4\subset {\cal M}_4$, then it follows from the master field equations of motion that
\be \check\Phi_{\a(m),\dot\a(n)}:= \left.\left.\frac{\partial^m}{\partial y^{\alpha(m)}} \frac{\partial^n}{\partial \bar y^{\dot\alpha(n)}} \Phi\right|_{(y_\alpha,\bar y_{\dot\alpha})=(0,0)}\right|_{U_4}\ ,\ee
are higher spin generalized Weyl tensors and background covariant derivatives thereof obeying \emph{source-free} Bargmann--Wigner equations in $U_4$.
Assembling $\check \Phi_{\a(m),\dot\a(n)}$ into an \emph{unextended} twisted adjoint master field 
\be \check \Phi:=\sum_{m,n} \left.\frac{1}{m!n!}y^{\alpha(m)}\bar y^{\dot\alpha(n)} \check\Phi_{\a(m),\dot\a(n)}\right|_{U_4}\in \check{\mathfrak{T}}\ ,\ee
it follows that $(\check\Phi-\Phi)|_{U_4}=0$.
The extension of $\check \Phi$ to all of ${\cal M}_4$ is a (singular) distribution on spacetime valued in $\check{\mathfrak{T}}$ obeying 
\be D^{(0)} \check\Phi=\check T_{\Phi}\ ,\qquad D^{(0)} \check T_{\Phi}=0\ ,\qquad
\mbox{on ${\cal M}_4$}\ ,\ee
where the spacetime one-form $\check T_{\Phi}\in \check{\mathfrak{T}}$ is given by a distribution on ${\cal M}_4$ with support on ${\cal M}_4\setminus U_4$, which we refer to as the Bargmann--Wigner source of $\check\Phi$.

Taking $L$ to be the global gauge function on ${\cal M}^{(1)}_4$ (with a single conformal boundary), the particle modes $\Phi^{(L)}_{P}\equiv \Phi^{(L)}_{\Psi_P}$ consists of real-analytic Gaussian functions on the fiber for all points on ${\cal M}^{(1)}_4$.
Thus, the corresponding Bargmann--Wigner source on ${\cal M}_4$ vanishes, \emph{viz.} $\check T_{\Phi^{(L)}_P}=0$.

On the other hand, the black hole modes $\Phi^{(L)}_{D}\equiv \Phi^{(L)}_{\Psi_D}$ are real-analytic on the fiber in the asymptotic regions, while they become fiber delta functions over the codimension-three submanifold of ${\cal M}_4$ where $\rho=0$.
The corresponding Bargmann--Wigner sources $\check T_{\Phi^{(L)}_D}$ are given by Hodge duals of codimension-three delta functions on ${\cal M}_4$ with support at $\rho=0$.
This source is singular in the sense that $\check T_{\Phi^{(L)}_D}\star \pi(\check T_{\Phi^{(L)}_D})$ is ill-defined.

The higher spin resolution of these Coulomb-like singularities amounts to the fact that from the integrability condition it follows that $\check T=D^{(0)} \chi$ locally, such that the extended Weyl zero-form
\be \Phi=\check \Phi-\chi\ ,\ee
obeys a source free equation with initial data $\Psi\in {\cal W}$ that is regular in the sense that $\Psi\in {\cal A}$ and hence $\Psi\star \Psi$ is well-defined (by the assumption that ${\cal A}$ has a well-defined star product). 

We expect that the Fronsdal fields carrying the black hole modes have delta function sources on ${\cal M}_4$ \cite{Segal:2001di}.
The Fronsdal fields are assembled together with distributions in $Y$-space into a spacetime one-form master field valued in the extended higher spin algebra $\mathfrak{hs}(4)$.
Vasiliev's equations maps this spacetime one-form to a horizontal twistor space one-form field with a source in non-commutative twistor space of codimension two.
Remarkably, in the FCS model, the latter source has a finite free energy, given by the on-shell value of a TVO given by the second Chern class on twistor space.
Thus, Vasiliev's formalism replaces the ill-defined free energy for a Coulomb-like configurations, computed from singular sources in spacetime using the Fronsdal on-shell action, by a well-defined finite free energy, computed from the regular source in non-commutative twistor space.

An interesting problem is to extend the black hole solutions of \cite{Iazeolla:2011cb}, which were constructed in trivial topology, to higher spin eternal black holes (with topology $\Real\times S^2\times S^1$) by using gauge functions of the form \eq{Lg} with $f(\rho)$ given by \eq{frho}.
As $f(\rho)$ is bounded from below, it follows that the resulting solutions will consist of infinite towers of Lorentz tensors that are bounded.
We leave this for future work.


\section{Topologically extended BTZ-like geometries}\label{Sec:BTZ}


The BTZ black hole \cite{Banados:1992gq} has contributed in many respects to our understanding of gravity.
It provides a remarkable toy model comprehending many crucial aspects of black holes in higher dimensions: mass and angular momentum; area law for entropy; and a causal structure making it a proper background geometry for the study of properties of quantum fields in curved spacetime.

In this section, we outline the following two dual descriptions of the spinless BTZ black hole in three dimensions and its direct BGM uplift to four dimensions:
\begin{itemize}
\item[---] \emph{Metric-like formulation}: Viewing the black hole geometry as a non-compact Lorentzian generalization \cite{Banados:1992gq} of a compact Riemannian Clifford-Klein form $\Gamma\backslash G/H$ where $G/H$ is a maximally symmetric space and $\Gamma\subset G$ is a discrete subgroup acting without fixed points  \cite{Borel:1962,KobayashiYoshino:2005}, leads to Lorentzian coset spacetimes with geometry as well as topology induced from extrinsic covering spaces.
\item[---] \emph{Unfolded formulation}: Viewing black hole geometry as a flat one-form section of an $H$-bundle with fiber given by the Lie algebra of $G$, leads to locally defined $G$-valued functions glued together with transition functions from $H$ into global configurations subject to (asymptotic) boundary conditions. 
\end{itemize}
The switch from the metric to the unfolded formulation replaces the metricity condition inside the bulk with the requirement of well-defined holonomies in $G$, which leads to \emph{topological extensions} of the base manifolds, as will be analysed in three and four dimensions, respectively, in Sections \ref{3DBTZ} and \ref{HPcoord}.
We would like to stress that the extended geometries are vacua of Vasiliev's higher spin gravity in four as well as three dimensions.   
Indeed, holonomies, which are classical observables in three-dimensional gravity as well as higher spin gravity, remain classical observables in four-dimensional higher spin gravity. 
Moreover, the higher spin fluctuations around the four-dimensional topologically extended BTZ-like geometries, which an be constructed using group algebra methods as we shall spell out in Section \ref{Sec:HSBTZ}, give rise to master fields that are bounded on the entire topologically extended base manifold thought of as horizontal forms, which is the topic of Sections \ref{Sec:indata} and \ref{Sec:fluct}.

\subsection{Generalities}

\paragraph{Three-dimensional gravity.}

Einstein gravity in three dimensions can be thought of as a topological theory with structure group $SO(1,2)$ and dynamical field given by a one-form $\Omega$ valued in the Lie algebra $\mathfrak{g}$ of $G=SO(3,1)$, $SO(2,2)$ or $ISO(2,1)$ depending on whether the cosmological constant is positive, negative or null.
On-shell, the connection obeys 
\begin{equation}\label{TheBasicF=0}
 d\Omega + \Omega \star \Omega = 0\ , 
\end{equation}
in charts $U_3\subseteq {\cal M}'_3$, a non-compact manifold obtained from a closed three-manifold ${\cal M}_3$ by removing conformal infinities and other defects (such as conical singularities).
Locally, the flat one-form is given by
\be \Omega=L^{-1} \star dL\ ,\qquad L:U_3\to G\ ,\ee
where the gauge function $L$ is defined modulo 
\be L\sim g_0 \star L\star H\ ,\qquad H:U_3\to SO(1,2)\ ,\qquad g_0\in G\ ,\ee
with chartwise defined constants $g_0$.
The gauge functions obey overlap conditions with transition functions from $SO(1,2)$ and boundary conditions at conformal boundaries.
The resulting classical moduli spaces consist of boundary states \cite{Brown:1986nw,Coussaert:1995zp,Cacciatori:2001un,Klemm:2002ir}, holonomies, and defects \cite{Ammon:2013hba,Hulik:2016ifr}, coordinatized by 
\begin{itemize}
\item[i)] Asymptotic charges given by generators of large gauge transformations evaluated at conformal infinites; and
\item[ii)] Holonomies $H_C(\Omega)$ attached to closed curves $C\in {\cal M}'_3$.
\end{itemize}
These quantities serve as classical observables in terms of which one may express the free energy given by the on-shell Chern--Simons action.

\paragraph{Metric-like approach.}

Three-dimensional Einstein manifolds with non-trivial topology can be obtained as quotients
\be \Gamma\backslash({\cal M}^{(K)}_3,ds^2)\ ,\qquad \Gamma\cong \{\gamma^n\}_{n\in \mathbb{Z}}\ ,\qquad \gamma\cong e^{2\pi \overrightarrow{K}}\ee
of Lorentzian covering spaces $({\cal M}^{(K)}_3,ds^2)$ given by restrictions of $G/SO(1,2)$ adapted to identification  Killing vector fields $\overrightarrow K$ in conjugacy classed of $\mathfrak{g}/G$.
The identification procedure presents three problems:
\begin{itemize}
\item[a)] Closed time-like curves arise upon identifying points in $G/SO(1,2)$ connected by time-like curves;
\item[b)] Conical singularities in the Riemann curvature arise at fixed surfaces of $\Gamma$ of co-dimension two;
\item[c)] Causal singularities may arise at fixed surfaces of $\Gamma$ of co-dimension one and two;
\item[d)] The induced topology of ${\cal M}'_3$ may turn out to be non-Hausdorff at fixed points of $\Gamma$.
\end{itemize}
The closed time-like curves in (a) can be excised by taking 
\be {\cal M}^{(K)}_3=\left\{p\in G/SO(1,2)|\xi(p)>0\right\}\ ,\qquad \xi^2:=\overrightarrow K^2\ ,\ee
where thus ${\cal M}^{(K)}_3$ is obtained by first restricting $G/SO(1,2)$ to the submanifold where $\overrightarrow K^2$ is space-like, that is, $\xi$ is real, and then restricting further to the subspace where $\xi$ is in addition positive.
As a result, singularities of type (b)--(c) may arise at $\xi=0$, depending on the nature of $\overrightarrow K$.

In the case of the spinless BTZ black hole, its Riemann tensor is bounded while it exhibits singularities of type (c) and (d) at $\xi=0$.
However, as we shall see below, the latter are artifacts of the metric-like formulation that are absent in the unfolded formulation. 

\paragraph{Gauge function approach.} 

The unfolded description of BHTZ-like geometries associated to the identification Killing vector $\overrightarrow{K}$ is obtained by taking
\be {\cal M}'_3=S^1_K\times {\cal M}'_2\ ,\ee
and consider a classical moduli space of gauge functions 
\be L=\exp_\star(i K\phi)\star \check{L}\ ,\qquad \check{L}:{\cal M}'_3\to G/SO(1,3)\ ,\ee
where $K\in \mathfrak{g}/G$ corresponds to $\overrightarrow{K}$;
\be \phi\in [0,2\pi)\ ,\ee
coordinatize $S^1_K$; and $\check{L}$ is strictly periodic on $S^1_K$ and subject to conditions at boundaries or other defects of ${\cal M}'_3$.
The resulting holonomy
\be H_{S^1_K}(\Omega)=\exp_\star(2\pi i K)\ ,\ee
is thus given by $\gamma$, the generator of $\Gamma$.

Whether the gauge function contains a conical singularity or a BTZ-like black hole depends on the topology of ${\cal M}'_3$.
The conical singularity arises on ${\cal M}'_3\stackrel{\rm top}{\cong} \Real^3\setminus C\stackrel{\rm top}{\cong}\Real\times (\Real^2\setminus(0,0))$, where $C\cong \Real$, which yields $dd\phi=\delta_{[2]}(C)$ hence $d\Omega+\Omega\star \Omega=iK\delta_{[2]}(C)$ on $\Real^3$, that is, $\Omega$ is source free on ${\cal M}'_3$.
The BTZ-like black hole instead arises by taking 
\be {\cal M}'_2\stackrel{\rm top}{\cong} \Real^2\ ,\ee
which yields $dd\phi\equiv 0$ hence $d\Omega+\Omega\star \Omega=0$ on ${\cal M}'_3$ as well.

\paragraph{Four-dimensional uplift.} 

The classical moduli spaces of unfolded BHTZ-like geometries can be lifted relatively uneventfully from three to four dimensions, that is, to locally flat one-forms valued in the Lie algebra of the isometry group $G$ of four-dimensional spacetime with a non-trivial cosmological constant and structure group $SO(1,3)$.
The lifting of the corresponding classical observables is problematic, however, as four-dimensional gravitational gauge fields are deformed on-shell by Weyl tensors which appear to obstruct any intrinsically defined functional that reduces on-shell to a holonomy.

Vasiliev's higher spin gravity, on the other hand, contains a flat one-form valued in a higher spin algebra (even in the presence of a non-trivial Weyl zero-form).
Thus, the theory maps closed curves in spacetime to holonomies valued in the higher spin group; for a recent review, see Section 6 of \cite{DeFilippi:2019jqq}.
Indeed, these holonomies reduce to those of the unfolded BHTZ-like geometries upon embedding the latter into higher spin gravity as vacua.

\subsection{3D spinless BTZ black hole}\label{3DBTZ}

In what follows, we embed the Lorentzian eternal spinless BTZ black hole, obtained by means of identifications, into a topologically extended unfolded geometry, described by a gauge function.

\paragraph{Ambient metric-like approach.}
The eternal BTZ black hole geometry with negative cosmological constant arises as $\Gamma\backslash AdS^{(K)}_3$, where 
\begin{itemize}
\item[---] $\Gamma$ is the discrete subgroup of $SO(2,2)$ generated by $\gamma= \exp{2\pi \overrightarrow K}$; 
\item[---] the identification Killing vector $\overrightarrow K$ is a boost in $\mathfrak{so}(2,2)$; and
\item[---] $AdS^{(K)}_3\subset AdS_3$ consists of all the points in $AdS_3$ with $\xi>0$ where $\xi^2:=\overrightarrow K^2$.
\end{itemize}
The identification Killing vector belongs to a specific conjugacy class of $\mathfrak{so}(2,2)$, referred to in the literature \cite{Banados:1992gq} as $I_b$, spanned by $\overrightarrow{P}:=\overrightarrow{P}_1=\overrightarrow{M}_{0'1}$ and $\overrightarrow{B}:=\overrightarrow{B}_2=\overrightarrow{M}_{02}$ modulo the large $SO(2,2)$ transformation that exchanges $\overrightarrow{P}$ and $\overrightarrow{B}$.
Thus, 
\be \overrightarrow K=\alpha_1\overrightarrow{P}+\alpha_2\overrightarrow{B}\ ,\ee
where $\alpha_i\in \Real$ are defined modulo $\alpha_1\leftrightarrow \alpha_2$. 
The group of Killing symmetries of the black hole is given by ${\rm Stab}_G(\overrightarrow K)$, that is, $U(1)_{\overrightarrow P}\times U(1)_{\overrightarrow B}$.
The parameters $\alpha_i$ are related to the mass $M$ and spin $J$ of the black hole.
Taking $\alpha_2=0$ yields a non-rotating BTZ black hole with mass $M=(\alpha_1)^2$ and identification Killing vector
\be \overrightarrow{K}=\sqrt{M}\, \overrightarrow{P}\ .\ee

To exhibit the orbispacetime geometry, one may use the embedding $\imath:AdS_3\to \Real^{2,2}$ of (proper) $AdS_3$ into flat four-dimensional ambient space $\Real^{2,2}$ with signature $(-1,-1,1,1)$ as the quadratic form
\begin{equation}\label{QuadraticForm}
-\left( X^{0^{\prime }}\right) ^{2}-\left( X^{0}\right) ^{2}+\left(
X^{1}\right) ^{2}+\left( X^{2}\right) ^{2}=-1\ ,
\end{equation}
whose isometries are generated by the Killing vectors $\overrightarrow{M}_{AB} = X_A \overrightarrow{\partial}_B - X_B \overrightarrow\partial_A$.
From $\xi^2 \equiv \overrightarrow K^2= M\left( (X^{0'})^2 - (X^1)^2\right)$, it follows that 
\be (X^{0'},X^1)=\frac{\xi}{\sqrt{M}}(\cosh \sqrt{M} \phi,\sinh \sqrt{M} \phi)\ ,\qquad \xi>0\ ,\ee
on $\imath(AdS^{(K)}_3)$, such that 
\be (X^{0'},X^1)=\frac{\xi}{\sqrt{M}}(\cosh \sqrt{M} \phi,\sinh \sqrt{M} \phi)\ ,\qquad \xi>0\ ,\qquad \phi\in [0,2\pi)\ ,\ee
on $\imath(\Gamma\backslash AdS^{(K)}_3)$.

The induced geometry is thus given by the warped product\footnote{We use a notation in which $ds^2_{M\times_f N}=ds^2_M+f^2 ds^2_N$ where $f:M\to \Real$.}
\be \Gamma\backslash AdS^{(K)}_3={\rm CMink}_2\times_\xi S^1_K\ ,\ee
where 
\be ds_{\rm CMink_2}^2:= \left(-d\xi^2/M-(dX^0)^2+(dX^2)^2\right)|_{-\xi^2/M-(X^0)^2+(X^2)^2=-1}\ ,\ee
is the metric on one of the two stereographic coordinate charts of $AdS_2$.
Kruskal--Szekeres-like coordinates can be introduced via the embedding ($m=0,2$)
\be X^m= \frac{2x^m}{1-x^2}\ ,\qquad 1>x^2>-1\ ,\ee
for which the two-dimensional line element and warp factor, respectively, take the form
\be ds_{\rm CMink_2}^2= \frac{4dx^2}{(1-x^2)^2}\ ,\qquad \xi =\sqrt{M}\frac{1+x^2}{1-x^2}\ .\ee
In other words, the orbispacetime $\Gamma\backslash AdS^{(K)}_3$ is a \emph{eternal spinless BTZ black hole} with metric
\be ds^2_{\rm EBTZ}=\frac{4dx^2}{(1-x^2)^2}+\xi^2 d\phi^2\ ,\ee
topology
\be \Gamma\backslash AdS^{(K)}_3\stackrel{\rm top}{\cong}\Real^2\times S^1\ ,\ee
two conformal infinities, no closed time-like curves, and past and future singularities of $\Real\times S^1$ topology at $\xi=0$ hidden behind future and past horizons at $\xi=\sqrt{M}$.
Its Killing vectors are given by the identification Killing vector $\overrightarrow{K}=\sqrt{M}\overrightarrow{P}= -\overrightarrow \partial_{\phi }$, and $\overrightarrow{B}$, that is, the Killing vector of ${\rm CMink}_2$ that annihilates $\xi$.

The eternal black hole can be restricted further to a \emph{Schwarzschild BTZ black hole}, with line element 
\begin{equation}\label{BTZnonRotating}
ds^{2}_{\rm SBTZ}=-\left( r^{2}-M\right) dt^{2}+\left( r^{2}%
-M\right) ^{-1}dr^{2}+r^{2}d\phi ^{2}\ ,\qquad \xi=r\geqslant 0\ ,\qquad t\in \Real\ ,\ee
corresponding to the embedding
\bea
\mbox{Outer region ($r\geqslant \sqrt{M}$):}\quad 
X^{0}& =&\sqrt{\frac{r^{2}}{M}-1}\ \mathrm{sinh}\left( \sqrt{M}t\right)\ ,\\ X^{2} &=&\sqrt{\frac{r^{2}}{M}-1}\ \mathrm{cosh}\left(\sqrt{M}t\right),\\[5pt]
\mbox{Inner region ($\sqrt{M} \geqslant r >0$):}\quad  
X^{0}& =&\sqrt{1-\frac{r^{2}}{M}}\ \mathrm{cosh}\left( \sqrt{M}t\right)\ ,\\ X^{2} &=&\sqrt{1-\frac{r^{2}}{M}}\ \mathrm{sinh}\left(\sqrt{M}t\right)\ .
\end{eqnarray}
The Killing vectors are now given by $\overrightarrow{K}\equiv \sqrt{M} \overrightarrow{P}= -\vec \partial_{\phi }$, and $\overrightarrow{B}=\frac{1}{\sqrt{M}}\vec\partial_{t}$.

\paragraph{Intrinsic unfolded approach.}

We first observe that the two eternal spinless BTZ black holes with $\xi>0$ and $\xi<0$, respectively, can be glued smoothly together across their causal singularities into a single \emph{topologically extended eternal spinless BTZ black hole}
\be ({\cal M}'_3,ds^2)_{\rm ExtEBTZ}=AdS_2\times_\xi S^1_K\ ,\ee
with topology\footnote{The closed time-like curve can be removed by going to the covering space of proper $AdS_2$ leading to a three-dimensional geometry with topology $\Real^2\times S^1$.}
\be {\cal M}'_3\stackrel{\rm top}{\cong}\Real\times T^2\ ,\ee
and singularities of $\Real\times S^1$ topology at $\xi=0$ hidden behind future and past horizons at $\xi=\pm \sqrt{M}$.

The corresponding globally defined gauge function\footnote{A gauge function adapted to the stereographic coordinate system on ${\rm CMink}_2$ is given by 
\be L= \exp_\star (iK\phi)\star \exp_\star (iP_m \xi^m)\ ,\ee
where $\xi^m=4\Upsilon(x^2) x^m$ with $\Upsilon$ given in Appendix C.1 of \cite{Aros:2017ror}.}
\be L=\exp_\star (iK\phi)\star \exp_\star (iET)\star \exp_\star(iP_2 {\rm arcsinh} \rho)\ ,\ee
where
\be K=\sqrt{M} P\ ,\qquad P=P_1=M_{0'1}\ ,\qquad P_2=M_{0'2}\ ,\ee
are the $\mso(2,2)$ generators corresponding to the identification Killing vector and its dual, and
\be \phi\in [0,2\pi)\ ,\qquad T\in [0,2\pi)\ ,\qquad \rho\in\Real\ .\ee
The gauge function is $2\pi$-periodic in $T$, as $\exp_\star (2\pi i E)$ is a central element in $SO(2,2)$.

The corresponding $\mathfrak{so}(2,2)$-valued one-form $\Omega=L^{-1}\star dL$ consists of a quasi-frame field $e^a$ and Lorentz connection $\omega^{ab}$ that are is bounded and constantly curved in the interior of ${\cal M}'_3$, though $e^a$ fails to be non-degenerate at $\xi=0$.
The resulting quasi-Lorentzian metric
\be ds_{\rm ExtEBTZ}^2=ds_{AdS_2}^2+ \xi^2 d\phi^2\ ,\ee
where 
\be ds_{AdS_2}^2=-(1+\rho^2)dT^2+ \frac{d\rho^2}{1+\rho^2}\ ,\quad \xi=\sqrt{1+\rho^2}\cos T\ ,\ee
has two conformal infinities at $\rho=\pm \infty$ with conformal quasi-Lorentzian metric
\be \left[ds_2^2\right]_{\pm\infty} =\left[ -dT^2 + \cos^2 T d\phi^2 \right]\ .\ee
This geometry is an extension of the eternal spinless BTZ black hole obtained by gluing together two ${\rm CMink}_2$ into a (proper) $AdS_2$ across the two surfaces where $\xi$ vanishes, that is, at $T=\pi/2$ and $T=3\pi/2$, where thus the trapped warped circle shrinks to zero size.
Indeed, restricting $T$ to $(\pi/2,3\pi/2)$ yields the eternal spinless BTZ black hole; as this restriction respects the flow lines of the globally defined Killing vectors, the restricted vector fields remain globally defined.

\subsection{4D spinless BGM black hole\label{HPcoord}}

In what follows, we first describe the eternal spinless BTZ black hole obtained from the ambient space metric using identifications.
We then construct two topologically extended versions using intrinsic gauge functions.

\paragraph{Ambient metric-like approach.}

Higher-dimensional orbispacetimes $AdS_n/\Gamma$ with $n>3$ are more complex than their three dimensional counterparts, as the identification Killig vector leaves more than one ambient plane invariant, which may lead to non-abelian (residual) Killing symmetries. 
An identification Killing vector in a conjugacy class of Type I preserves the foliation defined by its norm, whose leaves are constantly curved manifolds whose signature as well as radius may vary along the foliation.

Four-dimensional constantly curved orbispacetimes were studied in \cite{Aminneborg:1996iz,Aminneborg:1997pz,Banados:1998dc}.
The direct uplift of the three-dimensional eternal spinless BTZ black hole to four dimensions is the \emph{eternal spinless BGM black hole} \cite{Banados:1998dc}
\be \Gamma\backslash AdS^{(K)}_4={\rm CMink}_3\times_\xi S^1_K\stackrel{\rm top}{\cong}\Real^3\times S^1\ ,\label{EBGM}\ee
where 
\be AdS_4^{(K)}=\{p\in AdS_4| \xi(p)>0\}\ ,\qquad \xi^2:= \overrightarrow K^2\ ,\ee
and the identification Killing vector
\be \overrightarrow{K}=\sqrt{M}\overrightarrow P\ ,\qquad \overrightarrow P=\overrightarrow P_1=\overrightarrow{M}_{0'1}\ .\ee
More precisely, expressing $AdS_4$ as the quadratic form
\begin{equation}
-\left( X^{0^{\prime }}\right) ^{2}-\left( X^{0}\right) ^{2}+\left(
X^{1}\right) ^{2}+\left( X^{2}\right) ^{2}+\left( X^{3}\right) ^{2}=-1\ ,
\label{embcoord4D}
\end{equation}
one has $\xi^2= M\left((X^{0'})^2 - (X^1)^2\right)$, hence 
\be (X^{0^{\prime }},X^1)=\frac{\xi}{\sqrt{M}} (\mathrm{cosh}\left( \sqrt{M}\phi \right),\mathrm{sinh}\left( \sqrt{M}\phi \right))\ ,\qquad \xi>0\ ,\qquad \phi\in[0,2\pi)\ ,\ee
on $\Gamma\backslash AdS^{(K)}_4$.
Kruskal--Szekeres-like coordinates can be introduced via the embedding ($m=0,2,3$)
\be X^m= \frac{2x^m}{1-x^2}\ ,\qquad 1>x^2>-1\ ,\ee
manifesting \eq{EBGM} with 
\be ds_{\rm CMink_3}^2= \frac{4dx^2}{(1-x^2)^2}\ ,\qquad \xi =\sqrt{M}\frac{1+x^2}{1-x^2}\ .\ee
Thus, the geometry has no closed time-like curves, past and future singularities of $\Real^2\times S^1$ topology at $\xi=0$ hidden behind future and past horizons at $\xi=\sqrt{M}$.
Its Killing vectors are given by $\overrightarrow{K}\equiv \sqrt{M} \overrightarrow{P}= -\overrightarrow \partial_{\phi }$, and the Killing vectors of ${\rm CMink}_3$ that annihilate $\xi$, which form an $\msl(2)$ containing $\overrightarrow{B}$.

A Schwarzschild-like patch with coordinates $\{r,t,\phi,\theta \}$ can be obtained by taking
\be \xi=r\ ,\ee
and
\begin{eqnarray}
\mbox{Outer coordinates ($r>\sqrt{M}$):}\qquad
X^{0} &=&\sqrt{\frac{r^{2}}{M}-1}\ \mathrm{sinh}\left( \sqrt{M}
t\right) \text{ ,}  \notag\\
X^{2} &=&\sqrt{\frac{r^{2}}{M}-1}\ \mathrm{cosh}\left( \sqrt{M}%
t\right) \ \mathrm{sin}\ \theta \text{ ,}  \notag \\
X^{3} &=&\sqrt{\frac{r^{2}}{M}-1}\ \mathrm{cosh}\left( \sqrt{M}%
t\right) \ \mathrm{cos}\ \theta \text{ ,}  \label{embed-HP}\\[5pt]
\mbox{Inner coordinates ($\sqrt{M}>r>0$):}\qquad
X^{0} &=&\sqrt{1 - \frac{r^{2}}{M}} \cosh \left( \sqrt{M} t \right) \nonumber ,\\
X^{2} &=&\sqrt{1 - \frac{r^{2}}{M}}\sinh \left( \sqrt{M} t \right) \sin \theta\nonumber\ , \\
X^{3} &=&\sqrt{1 - \frac{r^{2}}{M}} \sinh\left( \sqrt{M} t \right)  \cos \theta\ ,  \label{embed-HPSecond}
\end{eqnarray}
which yield the Holst--Peldan (HP) line elements
\begin{equation}\label{PeldanMetric}
\mbox{Outer:}\quad ds^{2}_{\rm HP}= \left( \frac{r^{2}}{M}-1\right)\left(- Mdt^{2} + \cosh^{2}\left( \sqrt{M}t\right) d\theta^{2}\right) + \left( r^{2} 
-M\right)^{-1} dr^{2} + r^{2} d\phi ^{2}\ ,
\end{equation}
\begin{equation}\label{PeldanMetricInterior}
\mbox{Inner:}\quad ds^{2}_{\rm HP}= \left( 1 - \frac{r^{2}}{M} \right)\left( M dt^{2} + \sinh^{2}\left( \sqrt{M} t \right) d\theta ^{2}\right) + \left( r^{2}%
-M\right) ^{-1} dr^2 + r^2 d\phi ^{2}
\end{equation}
Its Killing vectors are $\overrightarrow{K}=-\vec\partial _{\phi}$, and  $\overrightarrow{M}_{02},\overrightarrow{M}_{03}$ and $\overrightarrow{M}_{23}$, of which $\overrightarrow{M}_{23}\propto \vec{\partial}_{\theta}$ is manifest in the Schwarzschild--like coordinates.
As stressed by BGM, the spinless HP black hole 
\begin{itemize}
\item[---] traps circles (instead of spheres as in the case of the Schwarzschild black hole), resulting in a three-dimensional Penrose diagram; 
\item[---] does not admit any globally defined time-like Killing vector fields (the globally time-like vector field $\overrightarrow{\partial}_{t}$ is not a symmetry of the geometry).
\end{itemize}
From Eqs.(\ref{QuadraticForm}) and (\ref{PeldanMetric}), it follows that at constant $r>\sqrt{M}$ the submanifold $\Sigma_2$ left invariant by the identification has a line element
\be ds^{2}_{\Sigma_2}=\left(\frac{r^{2}}{M}-1\right)\left(-Mdt^{2}+  
\mathrm{cosh}^{2}\left( \sqrt{M}t\right) d\theta ^{2}\right)\ ,\ee
which is that of two-dimensional de Sitter space of radius $\sqrt{\frac{r^{2}}{M}-1}$. 
At $r^2 = M$, the subspace left invariant by the identification is the two-dimensional light-like cone
\begin{equation}\label{TheHorizon}
    -(X^0)^2 + (X^2)^2+(X^3)^2 = 0.
\end{equation}
On the other hand, for $0 < r< \sqrt{M} $ the subspace left invariant is an Euclidean manifold of negative curvature, i.e., $H^2$ or $H^2/\Gamma_2$ with $\Gamma_2 \in SO(2,1)$ a smooth identification without fixed points \cite{Wolf1984}. 

Returning to the Killing vectors, one observes that only $\overrightarrow P$ and $\overrightarrow B$ are globally defined at conformal infinity, as the Killing vectors in $\msl(2;\Real)/\mathfrak{u}(1)$ contain $\cos T$ and $\sin T$. 
Thus, rather than assigning the spinless BGM black hole asymptotic charges, we assign it a holonomy $H_{S^1_K}(\Omega)$.

\paragraph{Intrinsic unfolded approach.}

Two eternal BGM black holes with $\xi>0$ and $\xi<0$, respectively, can be glued together across their singularities into a single \emph{topologically extended eternal spinless BGM black hole}\footnote{The closed time-like curves can be removed by going to the covering space of $AdS_3$ leading to four-dimensional geometry with topology $\Real^3\times S^1$.}
\be ({\cal M}^{(1)}_4,ds_4^2)_{\rm ExtEBGM}=AdS_3\times_\xi S^1\stackrel{\rm top}{\cong}\Real^2\times T^2\ ,\ee
with singularities of $\Real^2\times S^1$ topology at $\xi=0$ hidden behind future and past horizons at $\xi=\pm \sqrt{M}$, and a single conformal infinity.
The corresponding globally defined gauge function is given by ($r=2,3$)
\be L = \exp_\star iK\phi \star \exp_\star iE T \star \exp_\star i P_r n^r {\rm arcsinh} f^{(1)}(\rho)\ ,\label{adaptedL}\ee
where 
\be \phi\in[0,2\pi)\ ,\qquad T\in [0,2\pi)\ ,\qquad f^{(1)}(\rho)=\rho \geqslant 0\ ,\qquad n^r n_r=1\ ,\ee
and $L$ is $2\pi$-periodic in $T$ as $\exp_\star (2\pi i E)$ is a central element in $SO(2,3)$.
Indeed, this gauge function yields the line-element for $AdS_3\times_\xi S^1$ with
\be \xi= \cos T \sqrt{1+\rho^2}\ .\ee

Taking instead 
\be L = \exp_\star iK\phi \star \exp_\star iE T \star \exp_\star i P_r n^r {\rm arcsinh} f^{(2)}(\rho)\ ,\label{adaptedL(2)}\ee
where 
\be f^{(2)}(\rho)=\sqrt{a^2+\rho^2}\ ,\qquad \rho\in\Real\ ,\qquad a\geqslant 0\ ,\ee
and $a$ is a constant, yields a geometry with topology $\Real\times T^3$ and two conformal infinities given by quasi-Lorentzian $T^3$.
This quasi-Lorentzian geometry provides a semi-classical description of an entangled vacuum state \cite{Arias:2019pzy} with a topology distinct from that of the massless limit of the eternal Kruskal--Szekeres black hole in $AdS_4$ in \eqref{Lg(2)}.


\section{Higher spin fluctuations around 4D spinless BGM black hole}\label{Sec:HSBTZ}


%
In three dimensions, linearized higher spin fluctuations around the spinning BTZ black hole have been constructed in \cite{Didenko:2006zd} from gauge functions defined on Schwarzschild patches and zero-form initial data for conformally coupled scalar and spinor fields.

In what follows, we shall consider various aspects of the construction of a linear space of higher spin fluctuations around the four-dimensional topologically extended eternal spinless BGM black hole that consists of states that
\begin{itemize}
\item[i)] diagonalize the adjoint action of $K$;
\item[ii)] can be composed using the star product so as to form an associative operator algebra.
\end{itemize}
We would like to remark that 
\begin{itemize}
\item[a)] as the topologically extended $AdS_3\times_\xi S^1$ geometry has adapted gauge function, condition (i) is required in order for the linearized master fields to be periodic on the warped $S^1$, whereas the periodicity on the time-like $S^1$ inside $AdS_3$ follows from the $\pi\bar\pi$-projection of the master fields as explained below Eq \eqref{adaptedL};
\item[b)] condition (ii), which is imposed in order for the linearized fluctuations to give rise to a well-defined non-linear extension, can be disregarded as far as a strictly linearized analysis is concerned.
\end{itemize}
In the remainder of this Section, we shall propose a scheme obeying (i) and (ii).
However, from Section \ref{Sec:indata} and onwards, we shall forego condition (ii) and zoom in on particularly simple building blocks for the algebra in a strictly linearized analysis in concordance with remark (b).

\paragraph{Periodic boundary conditions on warped $S^1$ and group algebra.}
The boundary conditions on the warped circle require the Weyl zero-form integration constant $\Psi$ to be expanded over a basis of fiber functions $\Psi_n[\nu]$, $n\in \mathbb{Z}$, where $\nu$ is a set of amplitudes, that diagonalizes the adjoint star product action of the boost $K\in \mso(2,3)$ and forms a basis of an amplitude dependent generalization of the group algebra $\Comp[\mathbb{Z}]$, \emph{viz.}\footnote{As a simpler example, the free particle Hamiltonian $H=p^2$, for which 
\[ 
\Psi_n[\nu]=\int dp \nu(p) \left| \sqrt{p^2+n} \right\rangle \left\langle p\right| ,
\]
obeys ${\rm ad}^\star_H \Psi_n[\nu]=n \Psi_n[\nu]$ and $(\nu \circ_{n,n'} \nu')(p)=\nu(\sqrt{p^2+n})\nu'(p)$, where $n \in \mathbb{Z}^+$.}
\be {\rm ad}^\star_K\Psi_n=n\Psi_n\ ,\qquad \Psi_n[\nu]\star \Psi_{n'}[\nu']=\Psi_{n+n'}[\nu \circ_{n,n'} \nu']\ ,\ee%
where $\circ_{n,n'}$ a composition rule obeying the co-cycle condition 
\be (\nu \circ_{n,n'} \nu')\circ_{n+n',n''} \nu''=\nu \circ_{n,n'+n''}( \nu'\circ_{n',n''} \nu'')\ .\ee

\paragraph{Real and chiral group algebras.}
The boost $K$, which is a non-compact operator, is realized in the fiber as the direct product of two inverted harmonic oscillators, also known as Hubble Hamiltonians, \emph{viz.}
\be K=\sqrt{M} (H_1-H_2)\ ,\qquad H_i:=\frac12(p_i^2-x_i^2)\ .\ee
The spectrum of a single (normalized) Hubble Hamiltonian $H:=p^2-x^2$ has been determined in \cite{math_people}.
The point spectrum of $H$, \emph{i.e.} the eigenstates $|\lambda\rangle$ with complex eigenvalue $\lambda$, belong to the Banach space $L^p$ iff $p>2$ and $|{\rm Im}\lambda|< 1/2-1/p$; one can show that the H\"older dual $(L^p)^\ast\cong L^{\tilde p}$, where $\frac1p+\frac1{\tilde p}=1$, is also in the spectrum, \emph{i.e.} it is not possible to invert $H-\lambda$ in $L^{\tilde p}$ for $\lambda$ in the strip $|{\rm Im}\lambda|< 1/2-1/p$.
Taking the limit $p=2$, it follows that $H-\lambda$ remains non-invertible in $L^2$ for real $\lambda$, \emph{i.e.} the Hubble Hamiltonian has a continuous real spectrum, for which we can use the normalization
\be \langle \mu|\lambda\rangle=\delta(\mu-\lambda)\ ,\qquad \mu,\ \lambda\in\Real\ .\ee
Thus, one has
\bea \Psi^{\rm \Real}_n[\nu]&:=&\sum_m \int_{\Real^4}  d\lambda_1 d\mu_1 d\lambda_2 d\mu_2 \,\delta\left(\lambda_1-\lambda_2-\mu_1+\mu_2-\frac{n}{\sqrt{M}}\right) \,\delta\left(\lambda_1+\lambda_2+\mu_1+\mu_2-2m\right)\nn\\ 
&&\ \ \times\ \ \nu(\lambda_1,\lambda_2;\mu_1,\mu_2;m)\,f^{\rm \Real}_{\lambda_1,\lambda_2\mid \mu_1,\mu_2}\ ,\eea
where $f^{\rm \Real}_{\lambda_1,\lambda_2\mid \mu_1,\mu_2}\cong \mid\lambda_1,\lambda_2\rangle\langle  \mu_1,\mu_2\mid$ via the Wigner--Ville map; the first delta function quantizes ${\rm ad}^\star_K$; and the second delta function imposes the $\pi\bar\pi$-projection.

However, the reality condition on $\Psi$ requires $\kappa_y\star \bar\kappa_{\bar y}$ to have a well-defined one-sided action on $\Psi_n$.
To this end, we shall assume the existence of complexified stargenfunctions $f^{\rm \Comp}_{\lambda_1,\lambda_2\mid \mu_1,\mu_2}$ obeying
\be (H_i-\lambda_i)\star f^{\rm \Comp}_{\lambda_1,\lambda_2\mid \mu_1,\mu_2}=0=f^{\rm \Comp}_{\lambda_1,\lambda_2\mid \mu_1,\mu_2}\star (H_i-\mu_i)\ ,\ee
where $\lambda_i,\mu_i\in\Comp$, and 
\be \kappa_y\star \bar\kappa_{\bar y}\star f^{\rm \Comp}_{\lambda,l-\lambda\mid \mu,m-\mu}=(-1)^l f^{\rm \Comp}_{\lambda,l-\lambda\mid \mu,m-\mu}\ ,\qquad
 f^{\rm \Comp}_{\lambda,l-\lambda\mid \mu,m-\mu}\star \kappa_y\star \bar\kappa_{\bar y}=(-1)^m f^{\rm \Comp}_{\lambda,l-\lambda\mid \mu,m-\mu}\ ,\ee
for $\lambda,\mu\in \Comp$ and $l,m\in\mathbb{Z}$, and
\be f^{\rm \Comp}_{\lambda,l-\lambda\mid \mu,m-\mu}\star f^{\rm \Comp}_{\lambda',l'-\lambda'\mid \mu',m'-\mu'}=\delta_{m,l'}\delta^2(\mu-\lambda')f^{\rm \Comp}_{\lambda,l-\lambda\mid \mu',m'-\mu}\ .\ee
Thus, using such a chiral direct product of two Hubble Hamiltoniams, one has
\be \Psi^{\rm \Comp}_n[\nu]=\sum_{l,m} \int_{\Comp^2} d^2\lambda \,d^2\mu \,\delta^2\left(2\lambda-l-2\mu+m-\frac{n}{\sqrt M}\right) \nu(\lambda,l;\mu,m)f^{\rm \Comp}_{\lambda,l-\lambda\mid \mu,m-\mu}\ ,\ee
and $\nu_{n}(\lambda,l;\mu,m)\equiv \frac12(1+(-1)^{l+m})\nu_{n}(\lambda,l;\mu,m)$ in order to impose the $\pi\bar\pi$-projection.

\paragraph{Regular prescription.} In what follows, we shall
\begin{itemize}
\item[a)] write the group algebra elements, which are special functions in $Y$, as integral transforms
\be \Psi_n(Y)={\cal R}\left[\Psi_n(Y)\right]:=\int dS dT e^{Y^{\underline\alpha}S_{\underline{\alpha\beta}} Y^{\underline\beta}+T_{\underline\alpha} Y^{\underline\alpha}} \widetilde \Psi_n(S,T)\ ,\ee
with contours in the $S$ and $T$ planes corresponding to Mellin and Laplace transforms;
\item[b)] take the unfolding point, where the zero-form initial data is defined, to be an intersection point between a future and a past horizon of the topological black hole background.
\end{itemize}
As we shall see, this yields a Weyl zero-form that is real-analytic in the fiber over generic spacetime points (hence liftable to a master field configuration solving the linearized Vasiliev system), provided that all star products (between the initial data and gauge functions) are performed \emph{prior} to reading off Lorentz tensorial component fields.
In particular, this procedure yields a Weyl zero-form that is real-analytic in the fiber above the original unfolding point, whereas $\Psi\star\kappa_y$ is a non-real-analytic function involving complex powers of oscillators\footnote{A similar computational method, based on displacing the unfolding point away from the horizon, was employed in \cite{Didenko:2006zd}.}.

The above prescription is part of a broader scheme \cite{Iazeolla:2011cb,Iazeolla:2017vng,Aros:2017ror} for perturbative computations in Vasiliev's higher spin gravity according to which\footnote{Assumption (a) follows (i), and the order of operations that we apply to evaluate the Weyl zero-form at the original unfolding point is in accordance with (ii).}
\begin{itemize}
	\item[i)] the perturbatively defined Vasiliev master fields are assigned \emph{regular presentations} in terms of Gaussian functions on the full noncommutative twistor space (including Vasiliev's $Z$-space); and
    \item[ii)] star products (including twistor space derivatives) and traces are performed \emph{prior} to the parametric integrals used for regular presentations and representing twistor space homotopy contractors;
    \item[iii)] the parametric integrals arising at every intermediate stage of classical perturbation theory must provide an unambiguous regular presentation of a function or distribution in twistor space.
\end{itemize} 
The scheme facilitates perturbative computations in Vasiliev's theory using the gauge function method \cite{Iazeolla:2011cb,Iazeolla:2017vng,Aros:2017ror,DeFilippi:2019jqq}, since the initial data (and other twistor space constructs arising in Vasiliev's $Z$-space) indeed admit regular presentations and Gaussian kernels can be star multiplied and traced straightforwardly.
Thus, at every order of perturbation theory condition (iii) serves as an arbitrator among otherwise potentially ambiguous choices of (complex) contours for parametric integrals, thereby removing potential ambiguities from the scheme, though the scheme may clearly break down (provided that there exists either no or multiple consistent nestings of parametric integrals).

In the case of particle and black hole states in $AdS_4$, the scheme has been implemented to all orders in \cite{Iazeolla:2011cb,Iazeolla:2017vng} albeit in a (holomorphic) gauge which does not respect ALAdS boundary conditions, as stipulated by the central on mass-shell theorem, though ALAdS configurations can be reached by perturbative modifications of the gauge function at least at the linearized level \cite{DeFilippi:2019jqq}.

In what follows, the scheme will only be used to extract the linearized Weyl zero-form (which is the first object of the full set of Vasiliev fields to be encountered at every order of perturbations), though this nonetheless constitutes a nontrivial application of the formalism as it removes the aforementioned unphysical singularity at the unfolding point, and, moreover, lifts an apparent ambiguity in the choice of regular presentation of the initial data of the Weyl zero-form; for details, see Appendix \ref{App:trouble}.  

\paragraph{Singularity structure.}
As we shall see, the fiber real-analyticity of the Weyl zero-form only breaks down on two codimension-one submanifolds:
\begin{itemize}
\item[--] at $\xi=0$, \emph{i.e.} at the singularity of the BGM background, where the Weyl zero-form approaches a fiber distribution with a regular presentation given in Section \ref{Sec zeroform-L};
\item[--] at membrane-like singularities, where the Weyl zero-form approaches fiber delta functions with regular presentations, as will be shown in Appendix \ref{Sec:singularity} in a special case, namely when $\Psi$ is an operator in a complexified Fock space, which will be the topic of the next Section.
\end{itemize}
In particular, this means the Weyl zero-form remains real-analytic in the fiber above the entire horizons at $\xi=\pm \sqrt{M}$ \emph{including} the unfolding point (except at possible intersections between the horizons and the membrane-like singularity).

\section{Construction of zero-form initial data using Fock spaces}\label{Sec:indata}

In this Section, we shall provide simple building blocks for the Weyl zero-form integration constant $\Phi'=\Psi\star\kappa_y$ that diagonalize the twisted adjoint action of the oscillator realization $K$ of the identification Killing vector field $\overrightarrow K$ used to construct the four-dimensional BGM black hole background, as discussed in Section \ref{Sec:HSBTZ}.

To this end, we shall start in Section \ref{Section:recallproj} by recalling the construction in \cite{Iazeolla:2011cb,Iazeolla:2017vng} of linearized Weyl zero-forms on $AdS_4$ by expanding $\Psi$ over stargenfunctions obtained by dressing Fock space projectors and twisted projectors by polynomials in corresponding (complexified) creation and annihilation operators introduced so as to create \emph{integer} left and right eigenvalues for Cartan generators in $\mso(2,3)$ whose $Sp(4)$ matrices square to $-1$, namely, $E$, $J$, $iP$ and $iB$.

To obtain linearized Weyl zero-forms on $AdS_4$ on BGM backgrounds with identification Killing vector $\vec K=\alpha_1 \vec P+\alpha_2 \vec B$, we shall 
\begin{itemize}
\item[i)] create stargenfunctions with \emph{integer} left and right eigenvalues for $K$ by modifying the dressings of the generalized projectors built from $iP$ and $iB$ by including complex powers with \emph{quantized imaginary parts} of creation and annihilation operators, as spelled out in Section \ref{Sec differenteigenvalues};
\item[ii)] provide $\Psi$ with a regular presentation by Mellin transforming the complex powers of oscillators and Laplace transforming the generalized projectors, as done in Section \ref{Sec:regpres};
\item[iii)] constrain the eigenvalues so as to implement the bosonic projection; the reality conditions; and the BTZ-like identification, as discussed in Sections \ref{Sec conjf} and \ref{Sec ident}.
\item[iv)] rewrite the stargenfunction on $Sp(4,\Real)$-covariant form, which will be particularly useful in analysing the singularity structure of the Weyl zero-form, which is the topic of Section \ref{Sec gravitational notation}.
\end{itemize}
We stress that the above construction provides a particular type of building blocks for $\Psi$ that diagonalize ${\rm ad}^\star_K$.
The fact that these elements do not span any associative algebra on their own does not pose any problem as long as we limit ourselves to a strictly linearized analysis.

\subsection{Fock spaces associated to different Cartan subalgebras}\label{Section:recallproj}

The basic idea is thus to expand the initial datum $\Psi$ (or, equivalently, $\Phi'$ of \eq{LrotPhi}) in operators that span specific representations of the complexified $AdS_4$ isometry algebra $\msp(4,\Comp)$, and then subject them to the identification condition that characterizes the four-dimensional BGM background.
To this end, we shall modify and extend the method developed in \cite{Iazeolla:2011cb,Iazeolla:2012nf,Iazeolla:2017vng}, which we shall briefly review in what follows for the reader's convenience.

Given a pair $(K_{(+)},K_{(-)})$ of mutually commuting and normalized generators of (the complexified) $\msp(4,\Comp)$ with oscillator realization 
\be K_{(\pm)}=\frac{1}{8} K^{(\pm)}_{\underline{\a\b}}Y^{\underline\a}\star Y^{\underline\b}\ ,\ee
where
\bea [K^{(q)},K^{(q')}]_{\underline{\a\b}}~=~0\ ,\qquad  K^{(q)}_{\underline\a}{}^{\underline\g}\,K^{(q)}_{\underline\g}{}^{\underline\b}~=~-\delta_{\underline\a}{}^{\underline\b}\ ,\label{K2=1}\eea
they can be written in terms of two number operators
\begin{equation}
w_{i} \ := \ a_{i}^+ a_{i}^- \ = \ a_{i}^+ \star a_{i}^- +\frac{1}{2}\text{ ,} \qquad \text{(no sum over \emph{i})}
\end{equation}
as
\be K_{(\pm)}=\frac{1}{2}(w_2\pm w_1)\ ,\ee
where the creation and annihilation operators $a^{\pm}_i=(A^\pm_i)_{\underline\alpha}Y^{\underline\alpha}$, $i=1,2$, using projectors built from $K^{(q)}_{\underline{\a\b}}$.
An extension of the Weyl algebra by delta functions contains operators $P_{{\mathbf n}_L,{\mathbf n}_R}(Y)$ obeying 
\be P_{{\mathbf n}_L,{\mathbf n}_R}=\pi\bar\pi(P_{{\mathbf n}_L,{\mathbf n}_R})\ ,\ee
and 
\be P_{{\mathbf n}_L,{\mathbf n}_R} \star P_{{\mathbf m}_L,{\mathbf m}_R} \ = \ \delta_{{\mathbf n}_R,{\mathbf m}_L} P_{{\mathbf n}_L,{\mathbf m}_R} \ , \label{projalg}\ee
with ${\bf n}_{L,R}=(n_1,n_2)_{L,R}\in ({\mathbb Z}+1/2)\times ({\mathbb Z}+1/2)$, {\it idem} ${\bf m}_{L,R}$, being half-integer eigenvalues under the left or right star-product action of number operators $w_i$,
\be (w_i-n_{iL}) \star  P_{{\mathbf n}_L,{\mathbf n}_R} \ = \ 0 \ = \ P_{{\mathbf n}_L,{\mathbf n}_R} \star (w_i-n_{iR}) \ . \ee
Clearly, the $P_{{\mathbf n}_L,{\mathbf n}_R} $ also diagonalize the adjoint as well as twisted-adjoint actions of $K_{(\pm)}$, \emph{viz.} 
\be K_{(\pm)}\star P_{{\mathbf n}_L,{\mathbf n}_R} -P_{{\mathbf n}_L,{\mathbf n}_R}  \star K_{(\pm)} \ = \ \frac12\left(n_{2L}\pm n_{1L} -(n_{2R}\pm n_{1R}) \right)P_{{\mathbf n}_L,{\mathbf n}_R} \ ,\label{adjeig} \ee
\be K_{(\pm)}\star P_{{\mathbf n}_L,{\mathbf n}_R} -P_{{\mathbf n}_L,{\mathbf n}_R}  \star \pi(K_{(\pm)}) \ = \ \frac12 \left(n_{2L}\pm n_{1L} -(-1)^{\sigma_\pi(K_{(\pm)})} (n_{2R}\pm n_{1R}) \right)P_{{\mathbf n}_L,{\mathbf n}_R} \ ,\label{twadjeig} \ee
where $\pi(K_{(\pm)}) = \sigma_\pi(K_{(\pm)}) K_{(\pm)}$.

The diagonal elements $P_{\bf n,\bf n}\equiv P_{\bf n}=P_{n_1,n_2}$ are projectors and belong to the enveloping algebra of the number operators, and hence factorize as $P_{n_1,n_2}(w_1,w_2)=P_{n_1}(w_1)\star P_{n_2}(w_2)$. 
In particular, the projectors onto the lowest-weight state of the Fock space ($+$) and the highest-weight of the anti-Fock space ($-$) correspond to 
\begin{equation}
P_{\frac{\e_1}{2},\frac{\e_2}{2}} \ = \ 4e^{-2(\e_1 w_1+ \e_2 w_2)}\text{ ,}\qquad \e_1,\e_2=\pm\ .
\label{Bgrndproj}
\end{equation}
In star-product form, the generic projector reads
\begin{equation}
P_{n_{1},n_{2}}=\frac{\left( a_{2}^{\e_2}\right) ^{\star
(|n_{2}|-1/2)}}{\sqrt{(|n_{2}|-1/2) !}}\star \frac{\left( a_{1}^{\e_1}\right)
^{\star (|n_1|-1/2)}}{\sqrt{(|n_1|-1/2) !}}\star P_{\frac{\e_1}{2},\frac{\e_2}{2}
}\star \frac{\left( a_{1}^{-\e_1}\right) ^{\star (|n_1|-1/2)}}{\sqrt{(|n_1|-1/2)!}
}\star \frac{\left( a_{2}^{-\e_2}\right) ^{\star (|n_2|-1/2)}}{\sqrt{(|n_2|-1/2) !}}\text{ ,}\label{projF}
\end{equation}
where $\e_i:={\rm sign} (n_i)$. 

There are \emph{three} distinct pairs of $(K_{(+)},K_{(-)})$ modulo $Sp(4,\Real)$ rotations, given by \cite{Iazeolla:2011cb,Iazeolla:2012nf}\footnote{We refer the reader to the Appendix \ref{App:conv} for our $AdS_4$ and spinor conventions.}
\be (E,J) \ , \qquad (J,iB) \ , \qquad (iB,iP) \ ,\ee
where $E:=P_0=M_{0'0}$ is the AdS energy, $J:=M_{12}$ is a spin, $B:= M_{03}$ is a boost and $P:=P_1=M_{0'1}$ is a transvection; as $E$ and $J$ are compact it follows that $\exp(\pm 4E)$ are projectors, while as $B$ and $P$ are non-compact, the corresponding projectors are given by $\exp(\pm 4iB)$ and $\exp(\pm4i P)$.
Thus, starting from a pair of Cartan generators, one may form four lowest-weight ($\e=-$) or highest-weight ($\e=+$) projectors, namely $\exp(4\e K_{(\e')})$, where $\e,\e'=\pm$, and their twisted counterparts $\exp(4\e K_{(\e')})\star \kappa_y$, which are distinct elements iff $K_{(\e')}=E$ or $iP$ since $\exp(\pm 4J)\star \kappa_y=\exp(\pm 4J)$ and \emph{idem} $iB$.
The orbit of $\exp(4\e K_{(\e')})$ under the left and right actions of the extended Weyl algebra, form an associative algebra ${\cal M}_\e(K_{(\e')};K_{(-\e')})$ with \emph{principal} Cartan generator $K_{(\e')}$; letting ${\cal M}(K_{(\e')};K_{(-\e')})={\cal M}_+(K_{(\e')};K_{(-\e')})\oplus {\cal M}_-(K_{(\e')};K_{(-\e')})$, we thus have\footnote{The orbits ${\cal M}(K_{(\e')};K_{(-\e')})\oplus {\cal M}_\e(K_{(-\e')};K_{(\e')})$ do not account for all stargenfunctions $P_{{\mathbf n}_L,{\mathbf n}_R}$ introduced above.}
\be {\cal M}(E;J)\ ,\quad {\cal M}(J;E)\ ;\qquad {\cal M}(J;iB)\ ,\quad {\cal M}(iB;J)\ ;\qquad {\cal M}(iB;iP)\ ,\quad {\cal M}(iP;iB)\ ,\label{families}\ee
Expanding $\Phi'$ over ${\cal M}(K_{(\e')};K_{(-\e')})$, we refer to the contributions from the Weyl algebra orbits of $\exp(\pm4K_{(\e')})$ and $\exp(\pm4K_{(\e')})\star \kappa_y$, respectively, as the \emph{regular} and \emph{twisted} sectors, since the former gives rise to a Weyl zero-form that is real-analytic in $Y$ at the unfolding point\footnote{The one-sided star multiplication by $\kappa_y$ exchanges a symbol by a dual symbol obtained by chiral Fourier transformation in $y$-space (but not $\yb$-space) followed by replacing the Fourier dual variable by $y$. 
This duality transformation, that need not be a symmetry of the symbols of a generic quantum mechanical system, leaves the solution spaces found in \cite{Iazeolla:2017vng} invariant; whether it is a symmetry of higher spin gravity, possibly related to a GSO-like projection of an underlying topological open string, is an interesting open problem.}.
Thus, the twisted sector is nontrivial iff the principal Killing vector is taken to be $E$ or $iP$, in which case we expand\footnote{The (diagonal) projectors in the regular sector of ${\cal M}(E;J)$ gives rise to massless scalar particle modes in $AdS_4$, while the twisted counterpart yields spherically symmetric higher spin black holes \cite{Iazeolla:2017vng}.}
\bea {\cal M}(E;J)\ ,\quad {\cal M}(iP;iB)\ :\qquad \Psi(Y) \ = \ \sum_{{\mathbf n}_L,{\mathbf n}_R } \left( \n_{{\mathbf n}_L,{\mathbf n}_R}P_{{\mathbf n}_L,{\mathbf n}_R}(Y) +\m_{{\mathbf n}_L,{\mathbf n}_R}P_{{\mathbf n}_L,{\mathbf n}_R}(Y)\star\kappa_y \right)\ , 
\eea
where $\n_{{\mathbf n}_L,{\mathbf n}_R}$ and $\m_{{\mathbf n}_L,{\mathbf n}_R}$ are independent deformation parameters, while in the remaining families we set the $\mu$-parameters to zero, \emph{viz.}
\bea {\cal M}(J;E)\ ,\quad {\cal M}(J;iB)\ ,\quad {\cal M}(iB;J)\ ,\quad {\cal M}(iB;iP)\ :\qquad \Psi(Y) \ = \ \sum_{{\mathbf n}_L,{\mathbf n}_R } \n_{{\mathbf n}_L,{\mathbf n}_R}P_{{\mathbf n}_L,{\mathbf n}_R}(Y) \ . \eea
In the latter case, once a regular presentation ${\cal R}\left[P_{{\mathbf n}_L,{\mathbf n}_R}(Y)\right]$ has been chosen, there remains an apparent ambiguity whether to expand $\Psi$ in terms of ${\cal R}(P_{{\mathbf n}_L,{\mathbf n}_R}(Y))$ or ${\cal R}(P_{{\mathbf n}_L,{\mathbf n}_R}(Y))\star\kappa_y$, as both choices lead to Weyl zero-forms on-shell whose component fields obey the same boundary conditions in spacetime.
However, a closer inspection of how their regular presentations vary over spacetime (see Appendix \ref{App:trouble}) reveals that only former choice is compatible with condition (iii) in Section  \ref{Sec:HSBTZ}.

\paragraph{Four different families of fluctuations around spinless BGM black holes.} In what follows, we shall consider linearized Weyl zero-forms on the $U(1)\times Sp(2)$ invariant spinless BGM black hole.
This black hole has two isomorphic realizations, depending on whether one takes the identification Killing vector to be $\overrightarrow{P}$ or $\overrightarrow{B}$.
In each case, one may consider initial data $\Psi$ for the Weyl zero-form contained in extensions of ${\cal M}(iB;iP)$ and ${\cal M}(iP;iB)$ obtained by acting on their ground states by not only the Weyl algebra but also suitable complex powers of creation and annihilation operators\footnote{As discussed in Section  \ref{Sec:HSBTZ}, we expect that additional states must be added to $\Psi$ in order for the linearized solutions to admit completions into perturbatively defined nonlinear solutions, as this requires $\Psi$ to belong to an associative algebra.}; see Table \ref{Table}.
This results in four linearized moduli spaces with distinct characteristics, given by the unbroken symmetry $H$ and singularity structure of the physical scalar field $C$ of the corresponding ground states.

\begin{table}
\centering
\begin{tabular}{| l || c | c | c |}
\hline
$(K;\widetilde K)$ & $\Psi_0$ & $H$ & $C$ \\
\hline
$(P;B)$ & $e^{\pm 4iP}$ & $U(1)_P \times Sp(2)_B$ & $\frac{1}{\sqrt{1-\xi^2}}$\\
 & $e^{\pm 4iP}\star\kappa_y$ &  $Sp(2)_B$ & $\frac{X^{0'}+X^1}{\xi^2}$ \\
 & $e^{\pm 4iB}$ & $U(1)_P \times U(1)_B$ & $\frac{1}{\sqrt{1-\widetilde\xi^2}}$\\
 \hline
$(B;P)$ & $e^{\pm 4iP}$ & $U(1)_B \times U(1)_P$ &  $\frac{1}{\sqrt{1-\widetilde\xi^2}}$\\
&$e^{\pm 4iP}\star\kappa_y$ & $U(1)_B$ & $\frac{X^{0'}+X^1}{\widetilde \xi^2}$\\
&$e^{\pm 4iB}$& $U(1)_B \times Sp(2)_P$ & $\frac{1}{\sqrt{1-\xi^2}}$\\
    \hline
    \end{tabular}
\caption{\textbf{Ground states for fluctuations spaces on spinless BGM black holes.} $\protect\overrightarrow K$ and $\protect\overrightarrow{\widetilde K}$, respectively, denote the identification Killing vector and its dual of a BGM black hole with mass $M=1$ and spin $J=0$, \emph{i.e.} $AdS_3\times_\xi \times S^1$.
The black hole symmetry group is given by $Stab_{\mso(2,3)}(K)$, \emph{i.e.} $Stab_{\mso(2,3)}(P)=U(1)_P \times Sp(2)_B$ and $Stab_{\mso(2,3)}(B)=U(1)_B \times Sp(2)_P$, which is also the stabilizer of the warp factor $\xi:= \sqrt{\protect\overrightarrow K^2}$.
$H$ and $C$, respectively, denote the symmetry group and scalar field of the ground state $\Psi_0$ of a sector of fluctuations.
There are four distinct moduli spaces, depending on whether $C$ blows up at 1) $\xi=0$, \emph{i.e.} at the BGM singularity; 2) $\xi=\pm 1$, \emph{i.e.} at the BGM horizons; 3) $\widetilde \xi:= \sqrt{\protect\overrightarrow{\widetilde K}^2}=0$, \emph{i.e.} at a membrane-like singularity outside the BGM horizons; and 4) $\widetilde \xi=1$ (also denoted by $\Delta=0$), \emph{i.e.} at a membrane-like singularity passing through the BGM horizon and singularity.}
\end{table}\label{Table}

\subsection{Diagonalizing the adjoint actions of $P$ and $B$}\label{Sec differenteigenvalues}

In what follows, we shall oscillator realize stargenfunctions $f_{\boldsymbol{\l}_L,\boldsymbol{\l}_R}$ with general complex left and right eigenvalues $\boldsymbol{\l}_L=(\lambda_{1L},\lambda_{2L})$ and $\boldsymbol{\l}_R=(\lambda_{1R},\lambda_{2R})$ of the number operators
\begin{equation}
w_{1}=\frac{i}{8}\left( B_{\underline{\alpha \beta }}-P_{\underline{%
\alpha \beta }}\right) Y^{\underline{\alpha }}Y^{\underline{\beta }}\text{ \
, \ \ }w_{2}=\frac{i}{8}\left( B_{\underline{\alpha \beta }}+P_{%
\underline{\alpha \beta }}\right) Y^{\underline{\alpha }}Y^{\underline{\beta 
}}\text{ \ ,}  \label{w12B}
\end{equation}
related to the Cartan pair $(iB,iP)$. 
To this end, we choose
\be B_{\underline{\alpha \beta }}=-(\G_{03})_{\underline{\alpha \beta }}\ ,\qquad P_{\underline{\alpha \beta }}=-(\G_{0'1})_{\underline{\alpha \beta }}\ ,\ee
and use the realization of the Dirac matrices given in Appendix \ref{App:conv}, to arrive at
\begin{gather}
a_{1}^+\ =\ \frac{1}{2}\left( y^{1}+\bar{y}^{\dot{1}}\right) \text{ , \ }%
a_{1}^-\ =\ \frac{i}{2}\left( y^{2}+\bar{y}^{\dot{2}}\right) \text{ ,}
\label{a1B} \\
a_{2}^+\ =\ \frac{i}{2}\left( y^{1}-\bar{y}^{\dot{1}}\right) \text{ , \ }%
a_{2}^-\ =\ \frac{1}{2}\left( y^{2}-\bar{y}^{\dot{2}}\right) \text{ .}
\label{a2B}
\end{gather}
These operators can be projected out from $Y_{\underline\alpha}$ using a spin-frame 
\be (u^{+\a},u^{-\a})\ ,\qquad  u^{+\a}u^{-}_{\a}\ = \ 1 \ , \ee
\emph{idem} their complex conjugates, as
\begin{equation}
a^{\pm }_{i}=\left( A^{\pm }_i\right) _{\underline{\alpha }}Y^{\underline{\alpha }}%
\text{ ,} \label{aAY}
\end{equation}
where 
\begin{gather}
A_{1\underline{\alpha }}^{+}\ =\ \frac{1}{2}\left( u^+_\a \, , \ \bar{u}^+_{\ad} \right) \text{ , \ }%
A_{1\underline{\alpha }}^-\ =\ -\frac{i}{2}\left(  u^-_\a \, , \ \bar{u}^-_{\ad}\right) \text{ ,}
\label{A1} \\
A_{2\underline{\alpha }}^+\ =\ \frac{i}{2}\left( u^+_\a \, , \ -\bar{u}^+_{\ad}\right) \text{ , \ }%
A_{2\underline{\alpha }}^-\ =\ -\frac{1}{2}\left( u^-_\a \, , \ -\bar{u}^-_{\ad}\right) \text{ .}
\label{A2}
\end{gather}
To proceed, we use  
\begin{eqnarray}
Y_{\underline{\alpha }}\star f\left( Y\right) &=&\left( Y_{\underline{\alpha 
}}+i\frac{\partial }{\partial Y^{\underline{\alpha }}}\right) f\left(
Y\right) \text{ ,} \\
f\left( Y\right) \star Y_{\underline{\alpha }} &=&\left( Y_{\underline{%
\alpha }}-i\frac{\partial }{\partial Y^{\underline{\alpha }}}\right) f\left(
Y\right) \text{ ,}
\end{eqnarray}
to derive   
\begin{eqnarray}
\left( Y_{\underline{\alpha }}Y_{\underline{\beta }}\right) \star f\left(
Y\right) &=&\left( Y_{\underline{\alpha }}Y_{\underline{\beta }}+iY_{%
\underline{\alpha }}\frac{\partial }{\partial Y^{\underline{\beta }}}+iY_{%
\underline{\beta }}\frac{\partial }{\partial Y^{\underline{\alpha }}}-\frac{%
\partial ^{2}}{\partial Y^{\underline{\alpha }}\partial Y^{\underline{\beta }%
}}\right) f\left( Y\right) \text{ ,}  \label{exprYYf} \\
f\left( Y\right) \star \left( Y_{\underline{\alpha }}Y_{\underline{\beta }%
}\right) &=&\left( Y_{\underline{\alpha }}Y_{\underline{\beta }}-iY_{%
\underline{\alpha }}\frac{\partial }{\partial Y^{\underline{\beta }}}-iY_{%
\underline{\beta }}\frac{\partial }{\partial Y^{\underline{\alpha }}}-\frac{%
\partial ^{2}}{\partial Y^{\underline{\alpha }}\partial Y^{\underline{\beta }%
}}\right) f\left( Y\right) \text{ .}  \label{exprfYY}
\end{eqnarray}
Thus, if $f(Y)=f(a^{+},a^{-})$, it follows that 
\begin{eqnarray}
\left( a^{+}a^{-}\right) \star f\left( a^{+},a^{-}\right) &=&\left(
a^{+}a^{-}+\frac{1}{2}a^{+}\frac{\partial }{\partial a^{+}}-\frac{1}{2}a^{-}%
\frac{\partial }{\partial a^{-}}-\frac{1}{4}\frac{\partial ^{2}}{\partial
a^{+}\partial a^{-}}\right) f\left( a^{+},a^{-}\right) \text{ ,}
\label{expraaf} \\
f\left( a^{+},a^{-}\right) \star \left( a^{+}a^{-}\right) &=&\left(
a^{+}a^{-}-\frac{1}{2}a^{+}\frac{\partial }{\partial a^{+}}+\frac{1}{2}a^{-}%
\frac{\partial }{\partial a^{-}}-\frac{1}{4}\frac{\partial ^{2}}{\partial
a^{+}\partial a^{-}}\right) f\left( a^{+},a^{-}\right) \text{ .}
\label{exprfaa}
\end{eqnarray}
Thus, the stargenvalue problem
\begin{eqnarray}
\left( a^{+}a^{-}\right) \star f_{\l_L,\l_R}\left( a^{+},a^{-}\right) &=&\lambda
_{L}f_{\l_L,\l_R}\left( a^{+},a^{-}\right) \text{ ,}  \label{originaleqL} \\
f_{\l_L,\l_R}\left( a^{+},a^{-}\right) \star \left( a^{+}a^{-}\right) &=&\lambda
_{R}f_{\l_L,\l_R}\left( a^{+},a^{-}\right) \text{ ,}  \label{originaleqR}
\end{eqnarray}
is equivalent to 
\begin{eqnarray}
\left( a^{+}a^{-}+\frac{1}{2}a^{+}\frac{\partial }{\partial a^{+}}-\frac{1}{
2 }a^{-}\frac{\partial }{\partial a^{-}}-\frac{1}{4}\frac{\partial ^{2}}{
\partial a^{+}\partial a^{-}}\right) f_{\l_L,\l_R}\left( a^{+},a^{-}\right) &=&\lambda
_{L}f_{\l_L,\l_R}\left( a^{+},a^{-}\right) \text{ ,}  \label{eqL} \\
\left( a^{+}a^{-}-\frac{1}{2}a^{+}\frac{\partial }{\partial a^{+}}+\frac{1}{
2 }a^{-}\frac{\partial }{\partial a^{-}}-\frac{1}{4}\frac{\partial ^{2}}{
\partial a^{+}\partial a^{-}}\right) f_{\l_L,\l_R}\left( a^{+},a^{-}\right) &=&\lambda
_{R}f_{\l_L,\l_R}\left( a^{+},a^{-}\right) \text{ .}  \label{eqR}
\end{eqnarray}
Adding and subtracting these equations, one finds
\begin{eqnarray}
\left( a^{+}\frac{\partial }{\partial a^{+}}-a^{-}\frac{\partial }{\partial
a^{-}}\right) f_{\l_L,\l_R}\left( a^{+},a^{-}\right) &=&\left( \lambda _{L}-\lambda
_{R}\right) f_{\l_L,\l_R}\left( a^{+},a^{-}\right) \text{ ,}  \label{eqdiff} \\
\left( 2a^{+}a^{-}-\frac{1}{2}\frac{\partial ^{2}}{\partial a^{+}\partial
a^{-}}\right) f_{\l_L,\l_R}\left( a^{+},a^{-}\right) &=&\left( \lambda _{L}+\lambda
_{R}\right) f_{\l_L,\l_R}\left( a^{+},a^{-}\right) \text{ .}  \label{eqsum}
\end{eqnarray}
The solutions to (\ref{eqdiff}) can be written equivalently as  
\begin{equation}
f_{\l_L,\l_R}\left( a^{+},a^{-}\right)\ =\ \left( a^{+}\right) ^{\lambda _{L}-\lambda
_{R}}g^{(+)}_{\l_L,\l_R}\left( a^{+}a^{-}\right)  \ = \ \left( a^{-}\right) ^{\lambda _{R}-\lambda
_{L}}g^{(-)}_{\l_L,\l_R}\left( a^{+}a^{-}\right) \text{ ,}  \label{soleqdiff}
\end{equation}
where 
\begin{equation}
2wg^{(+)}_{\l_L,\l_R}\left( w\right) -\frac{1}{2}\left( \lambda _{L}-\lambda _{R}+1\right)
g^{(+)\prime }_{\l_L,\l_R}\left( w\right) -\frac{1}{2}wg^{(+)\prime \prime }_{\l_L,\l_R}\left( w\right)
=\left( \lambda _{L}+\lambda _{R}\right) g^{(+)}_{\l_L,\l_R}\left( w\right)  \label{eqsumtrans}
\end{equation}
with $w=a^{+}a^{-}$; as each of Eqs. (\ref{eqL}) and (\ref{eqR}) is left invariant under the exchanges 
\be a^{+}\leftrightarrow a^{-}\ ,\qquad \lambda _{L}\leftrightarrow\lambda _{R}\ ,\label{exchange}\ee
the equation for $g^{(-)}_{\l_L,\l_R}$ can be obtained from \eq{eqsumtrans} by exchanging $\l_L\leftrightarrow \l_R$.

For generic eigenvalues\footnote{\label{Foot:kummer}
Substituting an Ansatz of the form $g^{(+)}_{\l_L,\l_R}(w)=e^{-2w}\tilde{g}_{\l_L,\l_R}(w)$ in \eq{eqsumtrans}, the latter is turned into the standard Kummer equation $z\tilde{g}_{\l_L,\l_R}''(z)+(b-z)\tilde{g}_{\l_L,\l_R}'(z)-a\tilde{g}_{\l_L,\l_R}=0$ with $z=4w$, $a=\frac12-\l_R$ and $b=\l_L-\l_R+1$ for $\tilde{g}_{\l_L,\l_R}(w)$, so the usual criteria for the construction of the two independent solutions apply. In particular, for $\lambda _{L}=\lambda _{R}$, the two terms in the solution \eq{gWsolnfull} degenerate
into one. In this situation, the second term should be replaced with $
\mathrm{\tilde{C}}e^{-2w}U\left( \frac{1}{2}-\lambda _{R},1+\lambda
_{L}-\lambda _{R},4w\right)$, where $U$ is the Tricomi confluent
hypergeometric function. The latter can be expressed as the linear combination $U(a,b,z)=\frac{\pi}{\sin(\pi b)}\left[\frac{1}{\G(b)\G(1+a-b)}{}_1F_1\left(a,b,z\right)-z^{1-b}\frac{1}{\G(a)\G(2-b)}{}_1F_1\left(1+a-b,2-b,z\right)\right]$ when $b$ is not an integer, but can be extended to any $b\in \mathbb{Z}$ \cite{AbramowitzStegun}.  Combinations of one the two solutions in \eq{gWsolnfull} and $U$ enable one to write a complete solution to \eq{eqsumtrans} also in the cases when $\l_L-\l_R=\pm1,\pm2,... $ in which one of the two terms in \eq{gWsolnfull} has simple poles. For the special case $\lambda _{L}=\lambda _{R}=\frac12$, ${}_1F_1\left(0,1,4w\right)=U\left(0,1,4w\right)=1$ and a second independent solution is given by the exponential integral  $\tilde{g}_{1/2,1/2}(w)=-\int_{-4w}^{\infty}\frac{e^{-t}}{t} dt$.}, 
the solution to (\ref{eqsumtrans}) can be given as 
\be
 g^{(+)}_{\l_L,\l_R}(w) 
 =   \mathrm{C}\,g^{(+,1)}_{\l_L,\l_R}(w) + \mathrm{\tilde{C}}\,g^{(+,2)}_{\l_L,\l_R}(w)  \ ,\ee
where $C$ and $\mathrm{\tilde{C}}$ are integration constants and
\bea
 g^{(+,1)}_{\l_L,\l_R}(w) & = & e^{-2w}{}_1F_1\left(\frac12-\l_R,\l_L-\l_R+1,4w\right)\nn \\
g^{(+,2)}_{\l_L,\l_R}(w)
&=& e^{-2w}(4w)^{\l_R-\l_L}{}_1F_1\left(\frac12-\l_L,\l_R-\l_L+1,4w\right)\text{ .}  \label{gWsolnfull}
\eea
The corresponding solution for $g^{(-)}_{\l_L,\l_R}$ is obtained from \eq{gWsolnfull} by performing the aforementioned exchange, whose action on the above basis elements is given by 
\be (a^+)^{\l_L-\l_R}g^{(+,2)}_{\l_L,\l_R} = (a^-)^{\l_R-\l_L}g^{(-,1)}_{\l_L,\l_R}\ .\ee
In what follows, we shall restrict the eigenvalues to
\begin{equation}
\l_L\in \Comp \ , \quad \lambda _{R}+\frac{1}{2}\in \mathbb{Z}^{+} \ ,\qquad {\rm or } \qquad \l_L-\frac12\in\mathbb{Z}^- \ , \quad \l_R\in \Comp \ ,  \label{rightlefteigenconstr}
\end{equation}
as this will lead to regular prescriptions that simplify the analysis of the spacetime dependence of the Weyl zero-form.
If $\l_R+\frac12\in\mathbb{Z}^+$, then the first confluent hypergeometric function in \eq{gWsolnfull} reduces to a generalized Laguerre polynomial, 
\begin{equation}
\l_R+\frac12\in\mathbb{Z}^+\ :\quad g^{(+,1)}_{\l_L,\l_R}(w) \ = \ \mathrm{c} \,e^{-2w}L_{\lambda _{R}-\frac{1}{2}}^{\lambda
_{L}-\lambda _{R}}\left( 4w\right)  \text{ .}  \label{gWsoln}
\end{equation}
These generalized polynomials capture $g^{(+,1)}_{\l_L,\l_R}(w)$ also  when $\l_L-\frac12\in\mathbb{Z}^-$ and $\l_R\in\Comp$, since by virtue of Kummer's transformation ${}_1F_1(a,b,z)=e^z{}_1F_1(b-a,b,-z)$, it follows that
\be \l_L-\frac12\in\mathbb{Z}^-\ :\quad g^{(1)}_{\l_L,\l_R}(w) \ = \ \mathrm{c} \,e^{-2w}L_{\lambda _{R}-\frac{1}{2}}^{\lambda
_{L}-\lambda _{R}}\left( 4w\right) \ = \ \mathrm{c}\, \frac{\sin(\l_R-\frac12)\pi}{\sin(\l_L-\frac12)\pi}\,e^{2w}L_{-\lambda _{L}-\frac{1}{2}}^{\lambda
_{L}-\lambda _{R}}\left( -4w\right)  \text{ .} \ee
Thus, for generic eigenvalues, we may take 
\begin{equation}
 f_{\l_L,\l_R}\left( a^{+},a^{-}\right) \ = \ \mathrm{c}\,e^{-2w}\left( a^{+}\right)
^{\lambda _{L}-\lambda _{R}}L_{\lambda _{R}-\frac{1}{2}}^{\lambda
_{L}-\lambda _{R}}\left( 4w\right) \text{ ,}\qquad \mathrm{\tilde{C}}=0\  , \label{fsoln}
\end{equation}
and generate the other branch by the exchange \eqref{exchange}.

Solving the eigenvalue equations without any assumption of real-analyticity in $Y$ implies in particular that the eigenvalues alone do not fully specify the function $f_{\l_L,\l_R}$, nor its algebraic properties. In fact, the total space of elements of the form \eq{soleqdiff}, that satisfy the eigenvalue equations, can be described as the overlap of different solution subspaces, whose precise form goes beyond the scope of the present paper, and that we shall study systematically in a future publication. Essentially, as we have seen Eq. \eq{eqsumtrans} admits two independent solutions that are functions of $w$ (plus two more independent solutions if we also admit distributions in $w$). 
The two independent functions $f^{(i)}_{\l_L,\l_R}=(a^+)^{\l_L-\l_R}g^{(i)}_{\l_L,\l_R}(w)$, $i=1,2$, account for this degeneracy, and are distinct by the fact that $f^{(1)}$ admits a closed contour integral presentation while $f^{(2)}$ needs an open contour presentation\footnote{For instance, admitting inverse powers of the oscillators in the realization of $f_{\l_L,\l_R}$ implies that an element like $f_{1/2,1/2}$ can be realized both by the Fock space lowest-weight state projector $f^{(1)}_{1/2,1/2}$ and by an element $f^{(2)}_{1/2,1/2}$ obtained by acting on the anti-Fock-space highest-weight state projector $f^{(1)}_{-1/2,-1/2}$ with inverse powers of the creation/annihilation operators, $(a^-)^{-1}\star f^{(1)}_{-1/2,-1/2}\star (a^+)^{-1}$. It is evident that $f^{(1)}_{1/2,1/2}$ and $f^{(2)}_{1/2,1/2}$  do \emph{not} coincide, as the latter is not annihilated from the left by $a^-$ and from the right by $a^+$. Indeed, as mentioned in Footnote \ref{Foot:kummer}, the eigenvalue equations \eq{eqsumtrans} for $\l_L=\l_R=\frac12$ is solved by $f_{1/2,1/2}=c_+ e^{-2w}+c_- e^{-2w}{\rm Ei}(4w)$, where ${\rm Ei}(x)=-\int_{-x}^{+\infty}\frac{e^{-t}}{t}dt$ is the exponential integral: while the Fock space lowest-weight projector corresponds to the exponential solution $e^{-2w}$, the element  $f^{(2)}_{1/2,1/2}$, generated by means of non-analytic functions of the oscillators, can be shown to be equal to the second, independent and non-elementary solution.} \cite{paper1prime}. 

In the following, we shall focus only on the simplest type of solutions that enable us to satisfy the periodicity condition in a non-trivial way and to study the possible resolution of singularities of the fluctuation fields in the higher-spin gravity setup. Such solutions admit a closed contour integral presentation for their ``diagonal'' factor $g_{\l_L,\l_R}(w)$, corresponding to a Laplace-like transform. We shall now turn to describing this integral transform, specifying the eigenfunctions that it can encode. 

\subsection{Regular presentation of the stargenfunctions}\label{Sec:regpres}

The stargenvalue equations give rise to non-polynomial functions $g(w)$ as well as complex powers of the oscillators. 
It is therefore important to specify a functional presentation for the eigenfunctions, both to make sense of the complex powers and because different presentation of the same non-polynomial function may have different star-product properties. 
For instance, it was shown in \cite{Iazeolla:2011cb} that in order to ensure that both Fock-space and anti-Fock-space elements (that are in general both required by reality conditions on the master fields) form an associative algebra, it is crucial to work with an integral presentation, with the prescription that all star products be worked out before evaluating the auxiliary integrals. 
The specific regular presentation that was used in that paper and its follow-ups  (see \cite{Iazeolla:2017vng,Aros:2017ror} for the use of this integral presentation for solutions of various physical interpretation), involving an integral around a ``small'' contour, is technically the simplest one (see \cite{Iazeolla:2008ix} for more general ones), and for this reason will be employed in the present paper to represent the factor $g_{\l_L,\l_R}(w)$. We shall moreover use a Mellin transform to account for the complex powers. Fixing this presentation will resolve the degeneracy in $f_{\l_L,\l_R}$ that we commented on above by limiting the choice of eigenfunction to a simple class.  We postpone the study of more general regular presentations to a future publication \cite{paper1prime}. 

Let us now show how one can solve the eigenvalue equation (\ref{eqsumtrans}) by means of a closed-contour Laplace-like transform 
\begin{equation}
g_{\l_L,\l_R}\left( w\right) \ = \ \oint_{C(\varsigma_0)} \frac{d\varsigma }{2\pi i}e^{-2\varsigma w}\tilde{g}_{\l_L,\l_R}%
\left( \varsigma \right) \text{ ,}
\end{equation}
where the factor of $2$ at the exponent has been inserted for future convenience and $C(\varsigma_0)$ is a closed contour encircling the point $\varsigma_0$ to be determined later, subject to the condition that the integrand must be single-valued along the integration path.
Eq. (\ref{eqsumtrans}) is then converted into 
\begin{equation}
\oint_{C(\varsigma_0)} \frac{d\varsigma }{2\pi i}\ e^{-2\varsigma w}\left[ 2w+\left(
\lambda _{L}-\lambda _{R}+1\right) \varsigma -2w\varsigma
^{2}-\left( \lambda _{L}+\lambda _{R}\right) \right] \tilde{g}_{\l_L,\l_R}\left(
\varsigma \right) =0\text{ .}
\end{equation}
Expressing $w$ in the square brackets in terms of derivatives of $%
e^{-2\varsigma w}$ w.r.t. $\varsigma $ and integrating by parts one turns the condition \eq{eqsumtrans} into a first-order differential condition on the transform $\tilde{g}_{\l_L,\l_R}\left(
\varsigma \right) $,
\begin{equation}
\oint_{C(\varsigma_0)} \frac{d\varsigma }{2\pi i}\ e^{-2\varsigma w}\left[ \left(1-
\varsigma ^{2}\right) \frac{\partial }{\partial \varsigma }+%
\left( \lambda _{L}-\lambda _{R}-1\right) \varsigma -\left( \lambda
_{L}+\lambda _{R}\right) \right] \tilde{g}_{\l_L,\l_R}\left( \varsigma \right) =0\text{ ,%
}
\end{equation}
which is solved by 
\begin{equation}
\tilde{g}_{\l_L,\l_R}\left( \varsigma \right) \ = \ {\cal N}\frac{\left( \varsigma+1
\right) ^{\lambda _{L}-\frac{1}{2}}}{\left(\varsigma-1 \right) ^{\lambda
_{R}+\frac{1}{2}}}\text{ ,}
\end{equation}
where ${\cal N}$ is a constant. Thus we obtain 
\begin{equation}
g_{\l_L,\l_R}\left( w\right) \ = \ {\cal N}_{\l_L,\l_R}\oint_{C(\varsigma_0)} \frac{d\varsigma }{2\pi i}\ 
\frac{\left(  \varsigma+1 \right) ^{\lambda _{L}-\frac{1}{2}}}{\left(\varsigma-1 \right) ^{\lambda _{R}+\frac{1}{2}}}\ e^{-2\varsigma w}\ ,\label{intrep}
\end{equation}
where ${\cal N}_{\l_L,\l_R}$ is a normalization constant, to be fixed by requiring closure of the associative algebra of $f_{\l_L,\l_R}$ elements\footnote{While for half-integer eigenvalues it is concretely possible to fix the normalization constants by requiring that $f_{\l_L,\l_R}\star f_{\l'_L,\l'_R}=\d_{\l_R,\l'_L}f_{\l_L,\l'_R}$, working with complex eigenvalues makes this issue subtler, and the simple integral realization that we use in this paper, while good enough for a linearized analysis, is not a satisfactory choice for such purpose. Indeed, such ``small contour'' integral presentation can only capture discrete eigenvalues, according to \eq{lpm}, and these cover only a portion of the full spectrum, as discussed in Section \ref{Sec:HSBTZ}. We expect in fact that the linearized solutions discussed in this paper can only be dressed into full solutions by starting from an enlarged set of states, with left and right complex eigenvalues. For this reason, we shall not fix the normalization constants in this paper, leaving this issue, as well as any other question related to the non-linear completion of these solutions, to a future work \cite{paper1prime} where we shall use a different contour-integral presentation which evades such restriction.}. Imposing that the contour encircle the point $\varsigma_0 =1$, \eq{intrep} gives an integral realization of the $g(w)$ function for elements \eq{soleqdiff} with $\l_L\in \Comp$ and $\lambda _{R}+\frac{1}{2}\in \mathbb{\ Z}^{+}$, provided the contour is small enough as not to cross the branch cut from $-1$ to $-\infty$ that arises from the numerator of the integrand when $\l_L -\frac{1}{2} \notin  \mathbb{ Z}$. Analogously,  choosing a small contour that encircles the point $\varsigma_0 =-1$ and does not cross the branch cut from $1$ to $+\infty$, \eq{intrep} gives an integral realization of the $g(w)$ function for elements with $\lambda _{L}-\frac{1}{2}\in \mathbb{\ Z}^{-}$ and $\l_R\in \Comp$ \footnote{Actually, when taking the product of two elements $f_{\l_L,\l_R}$ and $f_{\l'_L,\l'_R}$ for half-integer eigenvalues, the condition that they form an associative algebra in general requires that one can always deform one of the two closed contour to be infinitesimally close to $\varsigma_0 =1$ or $\varsigma_0 =-1$, in such a way that, even after the star-product is evaluated, $\varsigma_0 =\pm 1$ is still the only pole encircled by the contour (for details, see \cite{Iazeolla:2011cb,Aros:2017ror}). Thus, in practice, we shall always assume the contour in \eq{intrep} to be  ``sufficiently small'' and to encircle $\varsigma_0 =\pm 1$.}. 
We can therefore conclude that
\begin{equation}
g_{\l_L,\l_R}\left( w\right) \ = \ {\cal N}_{\l_L,\l_R}\oint_{C(\pm1)} \frac{d\varsigma }{2\pi i}\ 
\frac{\left(  \varsigma+1 \right) ^{\lambda _{L}-\frac{1}{2}}}{\left(\varsigma-1 \right) ^{\lambda _{R}+\frac{1}{2}}}\ e^{-2\varsigma w}\label{intreppm}
\end{equation}
gives an integral presentation of the function of $w$ accounting for the $w$-dependent factor of $f_{\l_L,\l_R}$ in \eq{soleqdiff}, with the limitation that 
\bea &\lambda _{R}+\frac{1}{2}\ \in \ \mathbb{ Z}^{+}  \ , \quad {\rm for} \ \ \varsigma_0 = 1& \ ,\nn\\
&\lambda_{L}-\frac{1}{2}\ \in \ \mathbb{ Z}^{-}  \ , \quad {\rm for} \ \ \varsigma_0 = -1 &\ .\label{lpm}\eea
We choose standard phase conventions around the branching points, with ${\rm Arg}(\varsigma)\in (-\pi,\pi]$ for the integrand when $\lambda _{R}+\frac{1}{2}\ \in \ \mathbb{ Z}^{+}  $ and $\l_L\in \Comp$, and ${\rm Arg}(\varsigma)\in [0,2\pi)$ when $\lambda_{L}-\frac{1}{2}\ \in \ \mathbb{ Z}^{-} $ and $\l_R\in \Comp$. The integral presentation \eq{intreppm} with \eq{lpm} indeed covers the cases \eq{gWsoln} with \eq{rightlefteigenconstr}, as anticipated.

The factor $\left( a^{+}\right) ^{\lambda _{L}-\lambda
_{R}} $ of \eq{soleqdiff} also can be given an integral representation, which is in fact crucial to encode complex left eigenvalues. One way of doing that is via a Mellin transform, %
\begin{equation}
\left( a^{+}\right) ^{\lambda _{L}-\lambda _{R}}=\int_{0}^{+\infty }d\tau 
\frac{\tau ^{\lambda _{R}-\lambda _{L}-1}}{\Gamma \left( \lambda
_{R}-\lambda _{L}\right) }e^{-\tau a^{+}}\text{ ,}  \label{apowerlambda}
\end{equation}%
where $\Gamma $ stands for the gamma function. 
The above integral only makes sense for $\mathrm{Re}\left( \lambda _{L}-\lambda_{R}\right) <0$ 
and $\mathrm{Re}\left( a^{+}\right) > 0$. In order to extend it to the rest of the 
parameter space of interest, we can analytically continue \eq{apowerlambda} with
\begin{equation}
\left( a^{+}\right)^{\lambda _{L}-\lambda _{R}} \ = \ \Gamma(1+\lambda _{L}-\lambda _{R})\int_\gamma \frac{d\tau}{2\pi i}\,{\tau^{\lambda _{R}-\lambda _{L}-1}}\,e^{\tau a^+} 
\ ,  \label{apwrlambdaHank}
\end{equation}
where $\gamma$ is a contour of Hankel type, represented in Figure \ref{Fig:hankel}.  

\begin{figure}
\centering
\includegraphics[scale=1]{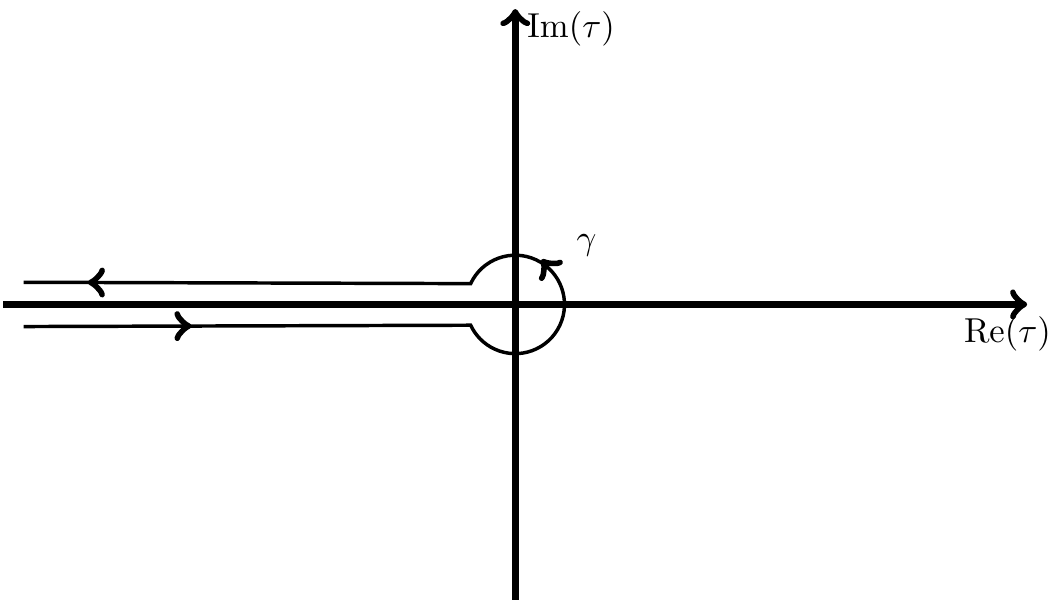}
\caption{The Hankel contour used in Eq.\eq{apwrlambdaHank}.}\label{Fig:hankel}
\end{figure}           

Such integral presentation is valid for any $\lambda _{L}-\lambda _{R} \neq - 1,- 2,...$ and $\mathrm{Re}( a^{+}) >0$.\footnote{This analytic continuation can be obtained for instance starting from the usual integral representation of the Gamma function via the Hankel contour, $\Gamma(z)=\frac{1}{2i\sin(\pi z)}\int_\gamma dt\,t^{z-1}\,e^{t}$, valid everywhere in the complex plane except at $z=0, -1, -2, ...$, rescaling the integration variable as $t = \tau x$, with $\mathrm{Re}(x)>0$, thus obtaining $x^{-z}=\frac{1}{2i\Gamma(z)\sin(\pi z)}\int_\gamma d\tau\,\tau^{z-1}\,e^{\tau x}$, and then using $\Gamma(z)\Gamma(1-z)=\frac{\pi}{\sin(\pi z)}$.} In practice, as we shall see, when evaluating the spacetime-dependent master field it will be possible to formally use the simpler presentation \eq{apowerlambda} in the relevant computations, and then analytically continue $\lambda _{L}-\lambda _{R}$ beyond the region $\mathrm{Re}\left( \lambda _{L}-\lambda_{R}\right) <0 $ after all star-products have been evaluated. 

Thus the solutions to the eigenvalue problem (\ref{fsoln}) that we shall focus on can be rewritten as%
\begin{equation}
f_{\lambda _{L},\lambda _{R}}\left( a_{+},a_{-}\right) \ = \ {\cal N}_{\lambda _{L},\lambda _{R}}%
\int_{0}^{+\infty }d\tau \frac{\tau ^{\lambda _{R}-\lambda _{L}-1}}{\Gamma
\left( \lambda _{R}-\lambda _{L}\right) }e^{-\tau a^{+}}\ \oint_{C(\pm1)}\frac{%
d\varsigma }{2\pi i}\ \frac{\left( \varsigma+1 
\right) ^{\lambda _{L}-\frac{1}{2}}}{\left( \varsigma -1\right) ^{\lambda
_{R}+\frac{1}{2}}}\, e^{-2\varsigma a^{+}a^{-}}\text{ ,}  \label{solnint}
\end{equation}%
with a (possibly redefined) normalization constant ${\cal N}_{\lambda _{L},\lambda _{R}}$, $\l_R+\frac{1}{2} \in \mathbb{Z}^+$ or $\l_L-\frac{1}{2} \in \mathbb{Z}^-$ according to \eq{lpm}, and the proviso that for a proper analytic continuation
one should use the Hankel contour integral \eq{apwrlambdaHank}. See Appendix \ref{Secfurthereigen} for further details on
the elements $f_{\lambda _{L},\lambda _{R}}$ in the regular presentation. 

As in this paper we are mainly concerned with the application of this formalism to the study of fluctuations over a BTZ-like background, for simplicity we shall limit ourselves to elucidating the main features of our construction by expanding the master fields on eigenfunctions of the form \eq{solnint} with $\l_R+\frac12 \in \mathbb{Z}^+$ accompanied by their hermitian conjugates, that the reality conditions will require (see Sections \ref{Sec conjf} and \ref{Sec:conj}). Elements admitting this type of ``small-contour'' integral transform correspond to eigenfunctions $f_{\l_L,\l_R}$ belonging to the subset of the $f^{(1)}_{\l_L,\l_R}$ that can be obtained as $(a^+)^{\l_L-\frac{1}{2}}\star f^{(1)}_{1/2,1/2}\star (a^-)^{\l_R-\frac{1}{2}} = (a^+)^{\l_L-\l_R}\star f^{(1)}_{\l_R,\l_R}\propto (a^+)^{\l_L-\l_R}e^{-2w}L^{\l_L-\l_R}_{\l_R-\frac{1}{2}}(4w)$. where $\l_L$ is at this level unconstrained and can have an imaginary part, while $\l_R$ is a positive half-integer. Constraints on $\l_L$ will arise from algebraic conditions and from imposing periodicity along the direction of identification. From now on we shall restrict our consideration to this class of eigenfunctions, and omit any of the superscripts used in this section to distinguish the various sectors of solutions to \eq{originaleqL}-\eq{originaleqR}. 

Note that, as elements like $f_{\l_L,\l_R}$ are in general non-analytic in $Y$ for $\l_L \in \Comp$, expanding the master-fields over such a basis seems incompatible with a physical interpretation of the expansion coefficients in terms of fields of various spins. However, we shall show in Section \ref{Sec zeroform-L} that this effect is peculiar to having started the construction with the master fields restricted at the unfolding point, and that reinstating the spacetime dependence via the gauge function $L$ in fact removes this problem, provided  that  the  star  products  with $L$ are performed prior to taking the limit back to the unfolding point.

Finally, it is useful to note that in the limit $\l_{L}-\l_{R}\to 0$, the integral presentation \eq{solnint} of $f_{\l_L,\l_R}$ smoothly reduces to that of a projector $f_{\l_R,\l_R}$ 
\begin{eqnarray}
&&f_{\lambda _{L},\lambda _{R}}\left( a^{+},a^{-}\right) \quad \xrightarrow[\l_{L}-\l_{R}\to 0]{} \quad  \oint_{C(\varepsilon)}\frac{%
d\varsigma }{2\pi i}\ \frac{\left( \varsigma+1 
\right) ^{\lambda _{R}-\frac{1}{2}}}{\left( \varsigma -1\right) ^{\lambda
_{R}+\frac{1}{2}}}\, e^{-2\varsigma w}\text{ ,}\label{diaglim}
\end{eqnarray}
where now $\varepsilon={\rm sign}(\l_R)$, in the sense that the divergence of each Gamma function at the denominator cancels exactly the one of the corresponding $\tau$-integral in the limit.  This is of course in agreement with the result of the limit taken on the non-integral presentation \eq{fsoln}.

The above results apply to each of the (commuting) $(a^{+}_1,a^{-}_1)$ and $(a^{+}_2,a^{-}_2)$ systems, so in the following sections we can directly use the above results by adding the labels ``1'' and ``2''.

\subsection{Reality properties of the eigenfunctions \label{Sec conjf}}

Later in this paper, we will discuss the reality condition imposed on
fields. To prepare for that discussion, we first investigate the reality properties of the 
eigenfunctions. 

Using the convention (\ref{a1B}) and (\ref{a2B}), we have for both the
\textquotedblleft 1\textquotedblright\ and the \textquotedblleft
2\textquotedblright\ system%
\begin{equation}
\left( a^{\pm }_i\right) ^{\dag }=\pm a^{\pm }_i\ \ \text{,}  \label{a-conj}
\end{equation}%
i.e.\ the creation and annihilation operators behave respectively like real and
imaginary numbers under hermitian conjugation.

The complex conjugate of (\ref{originaleqL}) and (\ref{originaleqR}) are: 
\begin{eqnarray}
f^{\dag }\left( a^{+},a^{-}\right) \star \left( a^{+}a^{-}\right)
&=&-\lambda _{L}^{\ast }f^{\dag }\left( a^{+},a^{-}\right) \text{ ,} \\
\left( a^{+}a^{-}\right) \star f^{\dag }\left( a^{+},a^{-}\right)
&=&-\lambda _{R}^{\ast }f^{\dag }\left( a^{+},a^{-}\right) \text{ ,}
\end{eqnarray}%
which shows that, due to the fact that $w^\dagger=-w$, the right (left)
eigenvalue of $f^\dagger$ is the opposite of the complex conjugate of the left
(right) eigenvalue of $f$.

Thus, the Hermitian conjugate of $f_{\lambda _{L},\lambda _{R}}$, with $\l_R+\frac12\in \mathbb{Z}^+$, is an element with left eigenvalue $\l'_L=-\l_R\in\mathbb{Z}^-+\frac12$ and complex right eigenvalue $\l'_R=-\l^\ast_L$, and as such admits a regular presentation as (see Eqs. \eq{solnint} with \eq{intreppm}-\eq{lpm})
\begin{equation}
f_{-\lambda_{R},-\lambda^\ast _{L}} \ = \ {\cal N}_{-\lambda_{R},-\lambda^\ast _{L}}\int_{0}^{+\infty }d\tau \frac{\tau ^{\lambda _{R}-\lambda _{L}^{\ast }-1}}{\Gamma\left( \lambda _{R}-\lambda
_{L}^{\ast }\right) }e^{-\tau a^{+}}\ \oint_{C(-1)}\frac{d\varsigma }{2\pi i}\
e^{-2\varsigma w}\frac{\left(\varsigma -1\right) ^{\lambda
_{L}^{\ast }-\frac{1}{2}}}{\left( \varsigma+1 \right) ^{\lambda _{R}+%
\frac{1}{2}}}\text{ .}  \label{fconj}
\end{equation}

Indeed, we can compare with the hermitian conjugate of $f_{\l_L,\l_R}$ from Eq. (\ref{solnint}) with $\l_R+\frac12\in\mathbb{Z}^+$, which reads
\be (f_{\l_L,\l_R})^{\dagger} \ = \  ({\cal N}_{\lambda _{L},\lambda _{R}})^\ast%
\int_{0}^{+\infty }d\tau \frac{\tau ^{\lambda _{R}-\lambda^\ast_{L}-1}}{\Gamma
\left( \lambda _{R}-\lambda^\ast_{L}\right) }e^{-\tau a^{+}}\ \oint_{C(1)}\frac{d\varsigma ^{\ast }}{2\pi i}\,\frac{\left( \varsigma ^{\ast }+1\right) ^{\lambda _{L}^{\ast }-%
\frac{1}{2}}}{\left( \varsigma ^{\ast }-1\right) ^{\lambda _{R}+%
\frac{1}{2}}}\, e^{2\varsigma ^{\ast
}w} \ , \ee
where a minus sign coming from the complex conjugation of the $i$ in the integration measure is compensated by an overall minus sign due to reversing the orientation of the contour. Changing integration variable as $\varsigma ^{\ast }=-\varsigma ^{\prime }$ and dropping the prime, 
\be (f_{\l_L,\l_R})^{\dagger} \ = \  ({\cal N}_{\lambda _{L},\lambda _{R}})^\ast e^{-i\pi(\l^\ast_L-\frac12)}(-1)^{\frac12-\l_R}
\int_{0}^{+\infty }\frac{d\tau\, \tau ^{\lambda _{R}-\lambda^\ast_{L}-1}}{\Gamma
\left( \lambda _{R}-\lambda^\ast_{L}\right) }e^{-\tau a^{+}}\ \oint_{C(-1)}\frac{d\varsigma}{2\pi i}\,\frac{\left( \varsigma -1\right)^{\lambda_L^\ast-\frac12}}{\left(\varsigma +1\right)^{\lambda_R+\frac12}}\, e^{-2\varsigma w} \ , \ee
where the phase factor was extracted  taking into account the phase conventions on (\ref{solnint}) when $\l_L\in \Comp$. Indeed the expression obtained above coincides with \eq{fconj} provided that\footnote{Note that the condition \eq{NastN} is indeed compatible with $( (f_{\l_L,\l_R})^{\dagger})^\dagger=f_{\l_L,\l_R}$, as it can be shown by repeating the reasoning that leads to \eq{NastN} for the case when $\l_R\in\Comp$, which results in ${\cal N}_{-\lambda^\ast_{R},-\lambda_{L}} \ = \ ({\cal N}_{\lambda _{L},\lambda _{R}})^\ast e^{i\pi(\l^\ast_R+\frac12)}(-1)^{\frac12+\l_L}$, and nesting the two formulas to get 
\bea &(({\cal N}_{\lambda _{L},\lambda _{R}})^\ast )^\ast=({\cal N}_{-\lambda_{R},-\lambda^\ast _{L}} e^{i\pi(\l^\ast_L-\frac12)}(-1)^{\l_R-\frac12})^\ast=({\cal N}_{-\lambda_{R},-\lambda^\ast _{L}})^\ast e^{-i\pi(\l_L-\frac12)}(-1)^{\l_R-\frac12}&\nn\\
&={\cal N}_{\l_L,\lambda_{R}} e^{-i\pi(-\l_L+\frac12)} e^{-i\pi(\l_L-\frac12)}(-1)^{2\l_R-1}= {\cal N}_{\l_L,\lambda_{R}} &\ .\nn\eea
}
\be {\cal N}_{-\lambda_{R},-\lambda^\ast _{L}} \ = \ ({\cal N}_{\lambda _{L},\lambda _{R}})^\ast e^{-i\pi(\l^\ast_L-\frac12)}(-1)^{\frac12-\l_R}
\ .\label{NastN}\ee

\subsection{$Sp(4;\Real)$-covariant notation for the eigenfunctions \label{Sec gravitational notation}}

Introducing the notation 
\begin{equation}
\boldsymbol{\lambda }:=\left\{ \lambda _{1L},\lambda _{1R},\lambda
_{2L},\lambda _{2R}\right\} \text{ ,\ \ }\boldsymbol{\lambda }_{1}:=\left\{
\lambda _{1L},\lambda _{1R}\right\} \text{ , \ }\boldsymbol{\lambda }%
_{2}:=\left\{ \lambda _{2L},\lambda _{2R}\right\} \text{ ,}
\end{equation}
we are now ready to expand the master field over the functions 
\bea & f_{\boldsymbol{\lambda }}\left( a_{1}^\pm ,a_{2}^\pm \right) \ = \ f_{\boldsymbol{\lambda }_1}\left( a_{1}^\pm\right)\star  f_{\boldsymbol{\lambda }_2}\left( a_{2}^\pm\right) \ = \ f_{\boldsymbol{\lambda }_1}\left( a_{1}^\pm\right) f_{\boldsymbol{\lambda }_2}\left( a_{2}^\pm\right)& \nn\\
& \ = \ {\rm c}_{\boldsymbol{\lambda }_1}{\rm c}_{\boldsymbol{\lambda }_2}\,e^{-2(w_1+w_2)}\left( a_1^{+}\right)
^{\lambda _{1L}-\lambda _{1R}}L_{\lambda _{1R}-\frac{1}{2}}^{\lambda
_{1L}-\lambda _{1R}}\left( 4w_1\right)\left( a_2^{+}\right)
^{\lambda _{2L}-\lambda _{2R}}L_{\lambda _{2R}-\frac{1}{2}}^{\lambda
_{2L}-\lambda _{2R}}\left( 4w_2\right) & \label{fexpol}
\eea
with regular presentation
\begin{eqnarray}
f_{\boldsymbol{\lambda }}\left( a_{1}^\pm ,a_{2}^\pm \right) & = & {\cal N}_{\boldsymbol{\lambda }_1}{\cal N}_{\boldsymbol{\lambda }_2}\int_{0}^{+\infty }d\tau _{1}\frac{\tau _{1}^{\lambda _{1R}-\lambda _{1L}-1}%
}{\Gamma \left( \lambda _{1R}-\lambda _{1L}\right) }e^{-\tau
_{1}a_{1}^+}\int_{0}^{+\infty }d\tau _{2}\frac{\tau _{2}^{\lambda
_{2R}-\lambda _{2L}-1}}{\Gamma \left( \lambda _{2R}-\lambda _{2L}\right) }%
e^{-\tau _{2}a_{2}^+}  \notag \\
&&\oint_{C(\pm1)}\frac{d\varsigma _{1}}{2\pi i}\frac{\left( \varsigma
_{1}+1\right) ^{\lambda _{1L}-\frac{1}{2}}}{\left(\varsigma _{1}-1\right)
^{\lambda _{1R}+\frac{1}{2}}}e^{-2\varsigma _{1}w_1}\oint_{C(\pm1)}\frac{%
d\varsigma _{2}}{2\pi i}\frac{\left( \varsigma
_{2}+1\right) ^{\lambda _{2L}-%
\frac{1}{2}}}{\left( \varsigma _{2}-1\right) ^{\lambda _{2R}+\frac{1}{2}}}%
e^{-2\varsigma _{2}w_2}\text{ .}
\end{eqnarray}
We recall that the second equality in \eq{fexpol} is due to the fact that the creation and
annihilation operators (\ref%
{a1B}) and (\ref{a2B}), commute under star-product. For $\l_L=\l_R\in\mathbb{Z}-\frac12$ one retrieves the projectors studied in \cite{Iazeolla:2011cb,Iazeolla:2012nf,Iazeolla:2017vng} and recalled as a special case in Section \ref{Section:recallproj}.

We can now rewrite the complete eigenfunctions $f_{\boldsymbol{\lambda }}$ in an $Sp(4;\Real)$-covariant notation, and break the latter into $SL(2;\Comp)$-covariant blocks when convenient. This will be useful to highlight the physical meaning of the various structures involved, and will facilitate the evaluation of the star products with the gauge function. We shall do it in general for an arbitrary family of solutions \eq{families}, and later specify to the case studied in the present paper. 

In order to shorten the expressions, let us  also introduce the notation
\be {\cal O}_{\boldsymbol{\lambda }_{i}}^{\varsigma_i} \ := \ \oint_{C(\pm1)}\frac{d\varsigma _{1}}{2\pi i}\frac{\left( \varsigma
_{1}+1\right) ^{\lambda _{1L}-\frac{1}{2}}}{\left(\varsigma _{1}-1\right)
^{\lambda _{1R}+\frac{1}{2}}}  \ . \ee
Then, ignoring for now the normalization constants, we can write
\begin{eqnarray}
 & f_{\boldsymbol{\lambda }}\left( a_{1}^\pm ,a_{2}^\pm \right) \ \equiv \ f_{\boldsymbol{\lambda }_1}\left( a_{1}^\pm\right) f_{\boldsymbol{\lambda }_2}\left( a_{2}^\pm\right) & \nn\\
& \displaystyle \propto \ {\cal O}_{\boldsymbol{\lambda }_{1}}^{\varsigma_1}{\cal O}_{\boldsymbol{\lambda }_{2}}^{\varsigma_2}%
\int_{0}^{+\infty }d\tau _{1}\frac{\tau _{1}^{\lambda _{1R}-\lambda _{1L}-1}%
}{\Gamma \left( \lambda _{1R}-\lambda _{1L}\right) }\int_{0}^{+\infty }d\tau _{2}\frac{\tau _{2}^{\lambda
_{2R}-\lambda _{2L}-1}}{\Gamma \left( \lambda _{2R}-\lambda _{2L}\right) }%
e^{-4\check K(\varsigma_1,\varsigma_2; Y)-\Theta(\t_1,\t_2) Y} & \ , \label{projkappa}
\end{eqnarray}
where, using matrix notation $A^{\underline{\a}} B_{\underline{\a}}  = : AB = ab+\bar a \bar b := a^\a b_\a + \bar a^{\ad}\bar b_{\ad} $, 
\bea
\check K(\varsigma_1,\varsigma_2; Y) & :=  & \frac12(s_1 w_1+s_2 w_2) \ = \ \frac{\varsigma_1+\varsigma_2}2 K_{(+)}+\frac{\varsigma_2-\varsigma_1}2  K_{(-)}  \notag\\
& = &  -\frac18 Y\check K(\varsigma_1,\varsigma_2)Y \ = \ -\frac18\left[ y\check\varkappa y + \yb\check{\bar \varkappa}\yb+2 y \check v \yb  \right] \ ,\label{Kcheck}
\eea
with
\bea
\check \varkappa_{\a\b} &:=  & \frac{\varsigma_1+\varsigma_2}2 \varkappa_{(+)\a\b}+\frac{\varsigma_2-\varsigma_1}2 \varkappa_{(-)\a\b}   \label{varkcheck}\\
\check v_{\a\bd} & :=  & \frac{\varsigma_1+\varsigma_2}2 v_{(+)\a\b}+\frac{\varsigma_2-\varsigma_1}2 v_{(-)\a\b} \ ,
\eea
idem $\check \varkb$, and 
\bea  \Theta Y & = & \theta y + \bar \theta \bar y \ ,  \nn\\
\displaystyle \Theta^{\underline{\a}} & = & \left(\theta^\a\ , \ \bar \theta^{\ad}\right) \ = \  -\left(\frac{\t_1+i\t_2}{2}\,u^{+\a} \, \ , \ \frac{\t_1-i\t_2}{2}\,\bar u^{+\ad}\right) \ . \label{Theta}\eea

As explained in Section \ref{Section:recallproj}, each one of the matrices 
\bea K_{(q)\underline{\a\b}} \ = \ \left(\begin{array}{cc}\vark_{(q)\a\b} & v_{(q)\a\bd} \\ 
\bar v_{(q)\ad\b} & \bar \vark_{(q)\ad\bd}\end{array}\right) \eea
is an $Sp(4,\Comp)$ Gamma matrix, so it is either block-diagonal ($v_{(q)\a\b}=0=\bar v_{(q)\ad\b}$), for the $\pi$-even generators, or it is block-off-diagonal (with $\varkappa_{(q)\a\b}=0=\bar \vark_{(q)\ad\bd }$) for $\pi$-odd generators. As we shall see later (see \cite{Iazeolla:2011cb} for more details), the star-products with the gauge function \eq{LrotPhi} will result in a conjugation of the $K_{(q)\underline{\a\b}}$ matrices by an $x$-dependent $Sp(4;\Real)$ matrix, giving rise to $K^L_{(q)\underline{\a\b}}$ matrices with all blocks non-vanishing: in particular, the off-diagonal blocks $v_{(q)}^L$ are the Killing vectors corresponding to the rigid isometry generator $K_{(q)}$ and the diagonal blocks $\varkappa_{(q)},\bar{\varkappa}_{(q)}$ the selfdual and anti-selfdual part of the corresponding Killing two-form.  In the case that we are studying in this paper, $K_{(+)}=iB$ and $K_{(-)}=iP$, and in particular, as is clear from \eq{BPrealization}, 
\bea \varkappa_{(iB)\a\b} &= & -i(\s_{03})_{\a\b}\ ,\qquad v_{(iB)\a\b}\ = \ 0\ ,\label{vkappaB}\\ 
\varkappa_{(iP)\a\b}&= &0\ ,\qquad v_{(iP)\a\bd}=-i(\s_{1})_{\a\bd} \ .\label{vkappaP}\eea
Note that choosing the integration contour to encircle the points $\varsigma_1 = \pm 1$, $\varsigma_2=\pm 1$, with signs correlated in such a way that $\varsigma_1\varsigma_2 = 1$, corresponds to choosing $iB$ as principal Cartan generator: that is, to fixing the lowest-weight state of the Fock space to be $4e^{-4iB}$ (and the corresponding anti-Fock space highest-weight state to be $4e^{4iB}$),  with the commuting generator $iP$ only appearing in the excited states (see \cite{Iazeolla:2011cb} for details). This also implies that the lowest-weight state, as well as the corresponding anti-Fock highest-weight state, have enhanced symmetry under an $\mso(2)_{B}\,\oplus\,\mso(2,1)_{\{M_{12},P_1,P_2\}}$ residual isometry algebra, any other diagonal state $f_{\l_{iL}=\l_{iR}}$ in the module is biaxially symmetric, with isometry algebra $\mso(2)_{B}\,\oplus\,\mso(2)_{P}$ \cite{Iazeolla:2011cb,Iazeolla:2012nf}, special non-diagonal states (such as those with identical real parts of the left and right eigenvalues and non-trivial momentum on $S^1$, that we shall examine in detail in Section \ref{Sec:scalar}) have only axial symmetry, while completely generic states $f_{\boldsymbol{\l}}$ do not preserve any isometry. These rigid symmetries of the $Y$-dependent expansion elements of $\Phi'(Y)$ are generically promoted to spacetime isometries preserved by the corresponding fluctuation modes of $\Phi^{(L)}(x,Y)$ via rotation with the gauge function $L$. However, due to the identification, the four-dimensional BTZ-like isometry algebra is reduced, compared to that of $AdS_4$, to $\mso(2)_{P}\,\oplus\, \msp(2)_{B,M_{02},M_{03}}$. As a consequence, the spacetime isometries preserved by the fluctuation modes actually reduce to the intersection of their ``local'' isometry algebra with that of the BTZ-like  background. For instance, the symmetries of the spacetime fluctuation based on the lowest-weight state $4e^{-4iB}$ will only be  $\mso(2)_{B}\,\oplus\,\mso(2)_{P}$.

Let us look at the generic element in the twisted sector in $\Phi' $ in terms of these building blocks. The star-product of \eq{projkappa} with $\k_y$ can be written as
\bea f_{\boldsymbol{\lambda }}\left( a_{1}^\pm ,a_{2}^\pm \right) \star \k_y &= & {\cal O}_{\boldsymbol{\lambda }_{1}}^{\varsigma_1}{\cal O}_{\boldsymbol{\lambda }_{2}}^{\varsigma_2}\int_{0}^{+\infty }d\tau _{1}\frac{\tau _{1}^{\lambda _{1R}-\lambda _{1L}-1}%
}{\Gamma \left( \lambda _{1R}-\lambda _{1L}\right) }\int_{0}^{+\infty }d\tau _{2}\frac{\tau _{2}^{\lambda
_{2R}-\lambda _{2L}-1}}{\Gamma \left( \lambda _{2R}-\lambda _{2L}\right) } \ \ \times\nn\\
&&\times \ \  \frac{1}{\sqrt{\check \vark^2}}\, \exp\left[-\frac1{2}(\ty-i\theta)\check \vark^{-1}(\ty-i\theta)+\frac{1}2\yb\check \varkb\yb-\bar \theta\yb\right] \ , \label{Phiprimecov}\eea
where we have defined the modified oscillators $\ty:=y-i\check v\yb$, and we recall that, $\vark_{(q)\a\b}$ being a symmetric $2\times 2$ matrix, $\vark^2_{(q)}:=\det \vark_{(q)}=\frac12\vark^{\a\b}_{(q)}\vark_{(q)\a\b}$ and $\vark^{-1}_{(q)\a\b}=-\frac{\vark_{(q)\a\b}}{\vark^2_{(q)}}$. It is clear from this expression that the elements of the twisted sector is, in fact, regular if $ \vark_{(q)\a\b}$ of the principal Cartan generator is non-singular, and otherwise non-real-analytic -- realizing, in fact, a delta-function in $Y$ space, as we shall show in Appendix \ref{Sec:singularity} (see also \cite{Iazeolla:2011cb} for the diagonal case). Distributional master-fields in the twisted sector are then smoothened  into regular functions almost everywhere once the $x$-dependence is reinstated via the gauge function $L$, but maintain the delta-function behaviour on the spacetime surface where $(\vark^{L}_{(q)})^2(x)=0$. This is the case for the twisted sector of the family ${\cal M}(E;J)$, investigated in \cite{Iazeolla:2011cb,Iazeolla:2017vng}, where the singular behaviour occurs in $r=0$ and results in a spherically-symmetric black-hole-like behaviour in the tower of Weyl tensors encoded in $\Phi^{(L)}$. For the $\pi$-even principal Cartan generators $J$ and $iB$, instead, $\vark_{(J)}^2=\vark_{(iB)}^2=1$, so regular and twisted sector are, in fact, degenerate, as anticipated. This, however, does not imply that the $L$-rotated elements of the twisted sector are regular, as this depends on whether $(\vark^{L}_{(q)})^2(x)$ has a region where it vanishes or not (which ultimately depends on whether the corresponding Killing vector has positive defined norm or not). While solutions based on $J$ as principal Cartan generator will always be regular, as we shall see in this paper the solutions based on $iB$ possess a singular surface, and we shall study the smoothening of the corresponding singularity in  Appendix \ref{Sec:singularity}. 

For the family ${\cal M}(iB;iP)$ regular and twisted sector are equivalent (the lowest/highest weight elements $e^{\mp 4iB}$ are in fact eigenstates of $\k_y$). However, the integral presentation that we use selects the twisted sector in $\Phi'$, in the sense that, as we shall explain in Appendix \ref{App:trouble}, the elements of the regular sector, once transformed by means of the gauge function $L$, will give rise to integrands that are incompatible with the small-contour integral representation employed in this paper, at least in some spacetime region\footnote{This is to be contrasted with the situation for families of solutions based on $\pi$-odd principal Cartan generators, where the regular and twisted sectors give rise to very different solution spaces (such as massless particle and black-hole states, as studied in \cite{Iazeolla:2017vng}) and no incompatibility arises.}. For this reason we shall discard them in the following. Thus, we shall expand $\Phi'$ on the regular sector, i.e.
\bea \Phi' \ = \ \sum_{\boldsymbol{\lambda }} \n_{\boldsymbol{\lambda }}f_{\boldsymbol{\lambda }}(Y)  \star \kappa_y   \ ,\label{Phitwist} \eea
and we shall now turn our attention to imposing constraints on the states allowed in such expansion.

\subsection{Identification conditions}\label{Sec ident}

Fluctuation fields over the four-dimensional BTZ-like background need to be left invariant by a full spatial transvection along the $S^1$ cycle.  We shall now impose this condition on the Weyl zero-form master field.

The Weyl zero-form (see Section \ref{SecVasilieveqs}) transforms as 
\begin{equation}
\Phi^{(L)} \quad \longrightarrow \quad (\gamma ^{(L)})^{-1}\star \Phi^{(L)}\star \pi
( \gamma ^{(L)}) \text{ ,}\label{Ltransf}
\end{equation}
where in $L$ gauge  
\begin{equation}
\gamma ^{(L)}	\ = \	L^{-1}\star \gamma ^{\prime }\star L\text{ ,}
\end{equation}
and $\gamma ^{\prime }$ induces the corresponding transformation on the rigid $Y$-space element $\Phi'$,
\begin{equation}
\Phi ^{\prime } \quad \longrightarrow \quad \gamma ^{\prime -1}\star \Phi ^{\prime }\star \pi
\left( \gamma ^{\prime }\right) \text{ .}
\end{equation}
A finite transvection generated by $P:=P_{0'1}$ is therefore implemented via  
\begin{equation}
\gamma ^{\prime } \ =  \ e_{\star }^{-\frac{i}{8}\sqrt{M}\varphi P_{\underline{%
\alpha \beta }}Y^{\underline{\alpha }}Y^{\underline{\beta }}}=e_{\star }^{%
\frac{1}{2}\sqrt{M}\varphi \left( w_{1}-w_{2}\right) }\text{ ,}
\label{gamma is P}
\end{equation}
as explained in Appendix \ref{Sec transformations}, the $K_{\underline{\alpha \beta }}$ of which is here specified to $iP_{\underline{\alpha \beta }}$, and $\varphi $ is rescaled for convenience of later discussion. Thus the
BTZ-like periodicity conditions are imposed by
\begin{equation}
\Phi ^{\prime } \ = \ \gamma ^{\prime -1}\star \Phi ^{\prime }\star \pi \left(
\gamma ^{\prime }\right) |_{\varphi =\varphi _{0}}\text{ ,} 
\end{equation}
where $\sqrt{M}\varphi _{0}$ represents the circumference of the $S^1$ cycle of the BGM background. We can choose  $%
\varphi _{0}=2\pi $. Note that $\pi \left( \gamma ^{\prime }\right) =\gamma
^{\prime -1}$.

Imposing the identification condition on \eq{Phitwist} amounts to imposing it on each factor $f_{\boldsymbol{\l}}$. The transformation of $f_{\boldsymbol{\l}}$ can be written as
\begin{eqnarray}
f_{\boldsymbol{\l}} &\longrightarrow &\gamma ^{\prime -1}\star f_{\boldsymbol{\l}}\star \gamma
^{\prime }  \notag \\
&=&e_{\star }^{-\frac{1}{2}\sqrt{M}\varphi \left( w_{1}-w_{2}\right) }\star
\left( f_{\boldsymbol{\l}} \right) \star e_{\star }^{\frac{1}{2}\sqrt{M}\varphi \left(
w_{1}-w_{2}\right) }  \notag \\
&=&e^{\frac{1}{2}\sqrt{M}\varphi \left[ -\left( \lambda _{1L}-\lambda
_{2L}\right) +\left( \lambda _{1R}-\lambda _{2R}\right) \right] }f_{\boldsymbol{\l}} %
\text{ ,}
\end{eqnarray}
and requiring that the transformation is periodic in $%
\varphi $ amounts to imposing the condition 
\begin{equation}
\left[ -\left( \lambda _{1L}-\lambda _{2L}\right) +\left( \lambda
_{1R}-\lambda _{2R}\right) \right] \in i\mathbb{R}\text{ .}
\end{equation}
Since we assume that $\lambda _{1,2\ R}+\frac{1}{2}\in \mathbb{Z}^{+}$,
this condition reduces to  
\begin{equation}
\mathrm{Re}\left( \lambda _{1L}-\lambda _{2L}\right) =\left( \lambda
_{1R}-\lambda _{2R}\right) \text{ .}\label{idreal}
\end{equation}
Furthermore, if we require that the transformation for $\varphi =2\pi $ be an
identity, we need to further impose
\begin{equation}
\mathrm{Im}\left[ \frac{\sqrt{M}}{2}\left( \lambda _{1L}-\lambda
_{2L}\right) \right] \in \mathbb{Z}\text{ .}\label{idim}
\end{equation}

We can therefore expand $\Phi ^{\prime }$ over states $f_{\boldsymbol{\l}}$ compatible with the BTZ-like identification by restricting the eigenvalues to those satisfying \eq{idreal}-\eq{idim}, and we write 
\begin{equation}
\Phi ^{\prime } \ = \ \Phi ^{\prime }_{{\rm discrete}}=\sum_{\substack{ \text{All valid }  \\ \text{values of }%
\boldsymbol{\lambda }}}\left[ \nu _{\boldsymbol{\lambda }}f_{\boldsymbol{\lambda }_{1}}\left(
a_{1}^\pm\right) f_{\boldsymbol{%
\lambda }_{2}}\left( a_{2}^\pm \right) \star \kappa _{y}\right] \ +\ \text{conj ,}
\label{Phi-prime sum}
\end{equation}
with the assumption that
\begin{equation}
\lambda _{1R}+\frac{1}{2}\in \mathbb{Z}^{+}\text{ \ and\ \ }\lambda _{2R}+%
\frac{1}{2}\in \mathbb{Z}^{+}
\end{equation}
in the first term of the sum. The notation \textquotedblleft conj\textquotedblright\ for the second term in the sum stands for the conjugate terms required by the reality conditions, which we now turn to determining. We recall that this expansion only captures one subsector of the full spectrum, the one distinguished by discrete eigenvalues. Had we started from the more complicated setup envisaged in Section \ref{Sec:HSBTZ}, even after imposing the identification condition we would still be left with one unconstrained complex eigenvalue, and one of the sums in \eq{Phi-prime sum} would then be substituted by an integral over the latter. The discrete spectrum contains simpler states, and the linearized analysis that we undertake in this paper does not require the introduction of the full spectrum of states. For this reason we shall limit our expansion to the discrete spectrum, and this is the reason for the subscript on $\Phi'$ used in \eq{Phi-prime sum}. 

\subsection{The conjugate terms}\label{Sec:conj}

There are more kinematic conditions to impose on the Weyl zero-form that further constrain the eigenvalues: the bosonic projection  and the reality conditions \eq{bosandreal} . 


Satisfying the bosonic projection condition
\begin{equation}
\Phi ^{\prime }=\pi \bar{\pi}\left( \Phi ^{\prime }\right)  \label{bos cond}
\end{equation}
amounts to imposing that
\begin{equation}
f_{\boldsymbol{%
\lambda }_{1}}f_{\boldsymbol{%
\lambda }_{2}} \ = \ \pi \bar{\pi}\left( f_{\boldsymbol{%
\lambda }_{1}}f_{\boldsymbol{%
\lambda }_{2}}\right) \text{ ,}  \label{bos f1f2}
\end{equation}%
which means that $f_{\lb}$ is invariant when the sign of $Y^{\underline{\alpha 
}}$ is flipped. To this end, it is convenient to represent the elements $f_{\lb}$ with $\l_{iR}\in\mathbb{Z}^+-\frac12$ as 
\begin{equation}
f_{\lb_1}f_{\lb_2} \ \propto \ \left( a_{1}^+\right) ^{\lambda _{1L}-\frac{1}{2}}\star
\left( a_{2}^+\right) ^{\lambda _{2L}-\frac{1}{2}}\star e^{-\frac{1}{2}iB_{%
\underline{\alpha \beta }}Y^{\underline{\alpha }}Y^{\underline{\beta }%
}}\star \left( a_{1}^-\right) ^{\lambda _{1R}-\frac{1}{2}}\star \left(
a_{2}^-\right) ^{\lambda _{2R}-\frac{1}{2}}\text{ ,}
\end{equation}
where we have written explicitly the lowest weight state $f_{1/2,1/2}=4e^{-\frac{1}{2}iB_{%
\underline{\alpha \beta }}Y^{\underline{\alpha }}Y^{\underline{\beta }}}$. Then
\begin{equation}
\pi \bar{\pi}\left( f_{\lb_1}f_{\lb_2}\right) \ \propto \ \left( -a_{1}^+\right)
^{\lambda _{1L}-\frac{1}{2}}\star \left( -a_{2}^+\right) ^{\lambda _{2L}-%
\frac{1}{2}}\star e^{-\frac{1}{2}iB_{\underline{\alpha \beta }}Y^{\underline{%
\alpha }}Y^{\underline{\beta }}}\star \left( -a_{1}^-\right) ^{\lambda _{1R}-%
\frac{1}{2}}\star \left( -a_{2}^-\right) ^{\lambda _{2R}-\frac{1}{2}}\text{ .}
\label{pipibarf1f2}
\end{equation}
Therefore, (\ref{bos f1f2}) implies that\footnote{As usual, since $\lambda _{1,2\ L}$ are complex, we obtain the overall phase \eq{minus} by extracting the $-1$ from the first two factors in \eq{pipibarf1f2} within standard branch cut conventions, assigning phases in such a way that  ${\rm Arg}(-a^{+}_i)\in(-\pi,\pi]$, $i=1,2$ and taking into account the reality conditions \eq{a-conj} and the assumption ${\rm Re}(a^+_i) > 0$ used in defining \eq{apowerlambda} and \eq{apwrlambdaHank}.} 
\begin{equation}
\left( e^{i\pi}\right) ^{\lambda _{1L}+\lambda _{2L}+\lambda _{1R}+\lambda _{2R}} \ = \ 1
\text{ ,}\label{minus}
\end{equation}
i.e., under the assumption that $\lambda _{1,2\ R}+\frac{1}{2}\in \mathbb{Z}%
^{+}$, \ 
\begin{equation}
\mathrm{Im}\left( \lambda _{1L}+\lambda _{2L}\right) =0\text{ \ \ \ \ and \
\ \ \ }\frac{1}{2}\left[ \mathrm{Re}\left( \lambda _{1L}+\lambda
_{2L}\right) +\lambda _{1R}+\lambda _{2R}\right] \in \mathbb{Z}\text{ \ .}\label{pipibaronl}
\end{equation}%

Now we consider the reality condition. Using $\pi^2=1$,
it is clear that any $\Phi ^{\prime }$ written as the combination $\Phi ^{\prime
}=C^{\prime }+\pi \left( C^{\prime \dag }\right) $ satisfies 
\begin{equation}
\Phi ^{\prime }=\pi \left( \Phi ^{\prime \dag }\right) \ . \label{real cond}
\end{equation}
Therefore, the \textquotedblleft conj\textquotedblright\ terms in (\ref%
{Phi-prime sum}) are obtained by applying the $\pi $-automorphism to the Hermitian conjugate of
the first term, i.e.,
\begin{eqnarray}
&&\pi \left\{ \left[ \nu _{\boldsymbol{\lambda }}f_{\boldsymbol{\lambda }_{1}}\left( a_{1}^\pm 
\right) f_{\boldsymbol{\lambda }_2}\left( a_{2}^\pm \right) \star \kappa _{y}\right] ^{\dag }\right\}  \notag \\
&=&\left( \nu _{\boldsymbol{\lambda }}\right) ^{\ast }\ \left[ f_{\boldsymbol{\lambda }_{1}}\left( a_{1}^\pm 
\right)  f_{\boldsymbol{\lambda }_2}\left( a_{2}^\pm \right) ) \star \kappa _{y}\star \bar{\kappa}_{\bar{y}}\right]
^{\dag }\star \kappa _{y}\text{ .}  \label{firstconj}
\end{eqnarray}
Using that
\begin{equation}
f_{ \boldsymbol{\lambda }_{1}}\left( a_{1}^\pm\right) f_{\boldsymbol{\lambda }_{2}}\left( a_{2}^\pm %
\right) \propto \left( a_{1}^+\right) ^{\lambda
_{1L}-\frac{1}{2}}\star \left( a_{2}^+\right) ^{\lambda _{2L}-\frac{1}{2}%
}\star e^{-\frac{1}{2}iB_{\underline{\alpha \beta }}Y^{\underline{\alpha }%
}Y^{\underline{\beta }}}\star \left( a_{1}^-\right) ^{\lambda _{1R}-\frac{1}{2%
}}\star \left( a_{2}^-\right) ^{\lambda _{2R}-\frac{1}{2}}\text{ ,}
\end{equation}%
\begin{equation}
\left( a_{1}^-\right) ^{\lambda _{1R}-\frac{1}{2}}\star \left( a_{2}^-\right)
^{\lambda _{2R}-\frac{1}{2}}\star \kappa _{y}\star \bar{\kappa}_{\bar{y}%
}=\left( -1\right) ^{\lambda _{1R}+\lambda _{2R}-1}\kappa _{y}\star \bar{%
\kappa}_{\bar{y}}\star \left( a_{1}^-\right) ^{\lambda _{1R}-\frac{1}{2}%
}\star \left( a_{2}^-\right) ^{\lambda _{2R}-\frac{1}{2}}\text{ ,}
\end{equation}%
\begin{equation}
e^{-\frac{1}{2}iB_{\underline{\alpha \beta }}Y^{\underline{\alpha }}Y^{%
\underline{\beta }}}\star \kappa _{y}\star \bar{\kappa}_{\bar{y}}\ = \ e^{-\frac{1%
}{2}iB_{\underline{\alpha \beta }}Y^{\underline{\alpha }}Y^{\underline{\beta 
}}}\text{ ,}
\end{equation}%
and that $\lambda _{1,2\ R}+\frac{1}{2}\in \mathbb{Z}^{+}$, we derive that (\ref%
{firstconj}) is equal to
\bea
& \pi \left\{ \left[ \nu _{\boldsymbol{\lambda }}f_{\boldsymbol{\lambda }_{1}}\left( a_{1}^\pm 
\right) f_{\boldsymbol{\lambda }_2}\left( a_{2}^\pm \right) \star \kappa _{y}\right] ^{\dag }\right\} \ = \ \left( \nu _{\boldsymbol{\lambda }}\right) ^{\ast }\ \left[ f_{\boldsymbol{\lambda }_{1}}\left( a_{1}^\pm 
\right) f_{\boldsymbol{\lambda }_2}\left( a_{2}^\pm \right) \right] ^{\dag }\star \kappa _{y} & \nn\\
&  = \ \left( \nu _{\boldsymbol{\lambda }}\right) ^{\ast } f_{-\lambda _{1R},-\lambda _{1L}^{\ast }}\left( a_{1}^\pm 
\right) f_{-\lambda _{2R},-\lambda _{2L}^{\ast }}\left( a_{2}^\pm \right)\star \kappa _{y} &\ , 
\eea
where in the last equality we have used the results of Section \ref{Sec conjf}. Note that the first condition for the bosonic projection in Eq. \eq{pipibaronl} also ensures that 
\be \kappa_y\bar \kappa_{\yb}\star  f_{\boldsymbol{\lambda }_{1}}\left( a_{1}^\pm 
\right)  f_{\boldsymbol{\lambda }_2}\left( a_{2}^\pm \right) \ = \ \pm f_{\boldsymbol{\lambda }_{1}}\left( a_{1}^\pm 
\right)  f_{\boldsymbol{\lambda }_2}\left( a_{2}^\pm \right)\ , \label{kkbarf}
\ee 
which is a necessary condition for the element $\k_y\bar \k_{\yb}\star f_{\boldsymbol{\lambda }_1}f_{\boldsymbol{\lambda }_2}$, that appears in the reality condition \eq{firstconj}, to be an admissible element of an associative algebra, since $\k_y$ and $\bar \k_{\yb}$ are unimodular\footnote{The analogous condition involving the star multiplication with $\kappa_y\bar \kappa_{\yb}$ from the right is trivially satisfied with our choice of $\l_{iR}+\frac12\in \mathbb{Z}^+$. The condition \eq{kkbarf} also constrains the real parts of the left eigenvalues as ${\rm Re}(\l_{1L}+\l_{2L})\in \mathbb{Z}$, which, again due to $\l_{iR}+\frac12\in \mathbb{Z}^+$, is anyway implied by the second condition in \eq{pipibaronl}. Note however that imposing \eq{kkbarf} together with its right analogue would restrict ${\rm Re}(\l_{1L}+\l_{2L})\in \mathbb{Z}$ and ${\rm Re}(\l_{1R}+\l_{2R})\in \mathbb{Z}$ even without fixing $\l_{iR}+\frac12\in \mathbb{Z}^+$.} (see Eqs. \eq{ky21}-\eq{kby21}). 

We are finally ready to give the Weyl zero-form that we are going to focus on in the remainder of the paper,
\begin{eqnarray}
\Phi ^{\prime } \ = \ \Phi ^{\prime }_{{\rm discrete}} 
   \ = \ &\displaystyle\sum_{\substack{ \text{All valid }  \\ \text{values of }%
\boldsymbol{\lambda }}}&\left[ \ \nu _{\boldsymbol{\lambda }}f_{\lambda _{1L},\lambda _{1R},\lambda _{2L},\lambda _{2R}}\left( a_{1}^\pm
,a_{2}^\pm \right)\right.\notag \\
& & \displaystyle + \ \left. \left( \nu _{\boldsymbol{\lambda }}\right) ^{\ast }f_{-\lambda _{1R},-\lambda _{1L}^{\ast },-\lambda
_{2R},-\lambda _{2L}^{\ast }}\left( a_{1}^\pm ,a_{2}^\pm\right)\right] \star \kappa _{y} \text{ ,} \label{phiprimefinal}
\end{eqnarray}
where the elements $f_{\lambda _{a},\lambda _{b},\lambda _{c},\lambda
_{d}}$ entering the expansion of the Weyl zero-form admit the integral presentation
\begin{eqnarray}
\hspace{-0.5cm}f_{\lambda _{a},\lambda _{b},\lambda _{c},\lambda
_{d}}\left( a_{1}^\pm ,a_{2}^\pm\right) 
&=&\int_{0}^{+\infty }d\tau _{1}\frac{\tau _{1}^{\lambda _{b}-\lambda _{a}-1}%
}{\Gamma \left( \lambda _{b}-\lambda _{a}\right) }e^{-\tau
_{1}a_{1}^+}\int_{0}^{+\infty }d\tau _{2}\frac{\tau _{2}^{\lambda
_{d}-\lambda _{c}-1}}{\Gamma \left( \lambda _{d}-\lambda _{c}\right) }%
e^{-\tau _{2}a_{2}^+}  \notag \\
&&\oint_{C(\pm 1)}\frac{d\varsigma _{1}}{2\pi i}\frac{\left( \varsigma
_{1}+1\right) ^{\lambda _{a}-\frac{1}{2}}}{\left( \varsigma _{1}-1\right)
^{\lambda _{b}+\frac{1}{2}}}e^{-2\varsigma _{1}a_{1}^+a_{1}^-}\oint_{C(\pm 1)}\frac{%
d\varsigma _{2}}{2\pi i}\frac{\left( \varsigma _{2}+1\right) ^{\lambda _{c}-%
\frac{1}{2}}}{\left( \varsigma _{2}-1\right) ^{\lambda _{d}+\frac{1}{2}}}%
e^{-2\varsigma _{2}a_2^+ a_2^-}\text{ .}\label{fgen}
\end{eqnarray}
and where we recall that:
\begin{itemize}
\item for simplicity of the integral presentation we set 
\be \l_{iR}+\frac12\in\mathbb{Z}^+ \ ,\qquad  i=1,2\,;  \ee
and, as a consequence of the identification constraints and of the bosonic projection condition (see  \eq{idreal}-\eq{idim} and \eq{pipibaronl}), both the real and the imaginary part of the left eigenvalues are quantized, and in particular 
\be {\rm Re}(\l_{iL})-\frac12\in \mathbb{Z} \ ,\qquad  {\rm with} \quad {\rm Re}(\l_{1L})-\l_{1R} \ = \ {\rm Re}(\l_{2L})-\l_{2R}\label{ident}\ee
and 
\be {\rm Im}(\l_{1L})\ = \ -{\rm Im}(\l_{2L})\in \frac{\mathbb{Z}} {\sqrt{M}}\ ,\label{imbos}\ee 
from which it follows that 
\be  \l_{1L}+\l_{2L}\ =  (\l_{1R}+\l_{2R}){\rm mod \,}2 \ ; \label{rebos}\ee

\item the contour encircles $\pm 1$ according to the pole of the integrand, i.e., it encircles $+1$ for the elements encoded in the first term in the sum \eq{phiprimefinal} and $-1$ for those in the second one;

\item the previously used normalization constants ${\cal{N}}_{{\boldsymbol \l}_i}$ have been absorbed into the deformation parameters $\nu_{\boldsymbol{\l}} $.

\end{itemize}


\section{Fluctuation fields in spacetime}\label{Sec:fluct}


Having determined the Weyl zero-form encoding fluctuation fields of all integer spins over a 4D BTZ-like background at the unfolding point, we shall now spread this initial data over a spacetime chart and reinstate the $x$-dependence  via the star products with the gauge function $L$ (as in \eq{LrotPhi}). We shall examine the resulting behaviour of the individual spacetime fields, and resolve their apparent singularities at the level of their embedding into the master fields living on the full $(x,Y)$-space.


\subsection{The Weyl zero-form in $L$-gauge}
\label{Sec zeroform-L}


\paragraph{Gauge function.} We choose $L$ to be \cite{Bolotin:1999fa}
\begin{equation}
L\left( x;y,\bar{y}\right) \ =\ \frac{2h}{1+h}\exp \left( \frac{i }{%
	1+h}x^{a}(\sigma _{a})^{\alpha \dot{\alpha}}y_{\alpha }\bar{y}_{\dot{\alpha}%
}\right) \text{ ,}
\end{equation}%
where $\sigma_a $ are the Van der Waerden symbol (see Appendix \ref{App:conv} for
explicit realizations), and 
\begin{equation}
h=\sqrt{1-\eta _{ab}x^{a}x^{b}}\text{ \ , \ \ }\eta _{ab}=%
\mathrm{diag}(-1,1,1,1)\text{ \ .}
\end{equation}
With this choice the $AdS_4$ background one-form connection is given by
\begin{equation}
\O_{\mu } \ = \ L^{-1}\star dL \ =\ -\frac{i}{2}e_{\mu }^{\alpha \dot{\alpha}}y_{\alpha }\bar{y}_{\dot{%
		\alpha}}-\frac{i}{4}\left( \omega _{\mu }^{\alpha \beta }y_{\alpha }y_{\beta
}+\bar{\omega}_{\mu }^{\dot{\alpha}\dot{\beta}}\bar{y}_{\dot{\alpha}}\bar{y}%
_{\dot{\beta}}\right) \text{ \ ,}
\end{equation}%
where%
\begin{equation}
e_{\mu }^{\alpha \dot{\alpha}}=- h^{-2}\delta _{\mu }^{a}\left(
\sigma _{a}\right) ^{\alpha \dot{\alpha}}\text{\ \ \ ,\ \ \ \ }\omega _{\mu
}^{\alpha \beta }=- h^{-2}\delta _{\mu }^{a}x^{b}\left( \sigma
_{ab}\right) ^{\alpha \beta }\text{ \ and \ }\bar{\omega}_{\mu }^{\dot{\alpha%
	}\dot{\beta}}=-h^{-2}\delta _{\mu }^{a}x^{b}\left( \bar{\sigma}%
_{ab}\right) ^{\dot{\alpha}\dot{\beta}}\text{ \ ,}
\end{equation}%
are the vierbeins and spin-connection and the $x^a$ are stereographic
coordinates, which are related to the embedding coordinates (\ref{embcoord4D}%
) by%
\begin{equation}
x^{a}\ =\ \frac{X^{a}}{1+| X^{0^{\prime }}|}.
\end{equation}
We refer to Appendix \ref{App:conv} for  more details on the relation of stereographic coordinates with the other coordinate systems that we use in this paper. 

\paragraph{Spacetime-dependent Weyl zero-form.} The evaluation of \eq{LrotPhi} is facilitated by noting that the adjoint action of $L$ on a $Z$-independent symbol $f$ amounts to a rotation of the $Y$ oscillators, \emph{viz.}
\begin{equation}
L^{-1}\star f\left( Y_{\underline{\alpha }}\right) \star L \ = \ f\left( L_{%
\underline{\alpha }}{}^{\underline{\beta }}Y_{\underline{\beta }}\right) 
\text{ ,}
\end{equation}
where $L_{\underline{\a}}{}^{\underline{\b}}(x)$ is the $x$-dependent $Sp(4)$ matrix
\begin{equation}
L_{\underline{\alpha }}{}^{\underline{\beta }} \ = \ \frac{1}{h}\left(
\begin{array}{cc}
\delta_{\alpha }{}^{\beta } &  x_{\alpha }{}^{\dot{\beta}} \\ 
\bar{x}_{\dot{\alpha}}{}^{\beta } & \delta_{\dot{\alpha}}{}^{\dot{%
\beta}}%
\end{array}%
\right) \text{ .}  \label{Lmatrix}
\end{equation}
In order to compute  $\Phi^{(L)}$, it is therefore useful to write the eigenfunctions $f_{\boldsymbol{\l}}$ in the $Sp(4,\Real)$-covariant notation of Section \ref{Sec gravitational notation}. The $L$-rotation of the $Y$ oscillators induces a spacetime-dependent transformation of all the structures contracted with them, the polarization spinor $\Theta_{\underline{\a}}$ and the $K_{\underline{\a\b}}$ matrix (with its $\vark_{\a\b}$, $\varkb_{\ad\bd}$ and $v_{\a\bd}$ blocks) entering Eqs. \eq{Kcheck}-\eq{Theta}. Ultimately, their transformations all descend from the $L$-rotation induced on the spin-frame basis spinors $u^\pm_\a$ and $\bar u^\pm_{\ad}$. We shall henceforth denote with a label $(L)$ the corresponding $L$-transformed quantities, and refer to the so-transformed master fields as being in $L$-gauge. 

Thus, 
\begin{eqnarray}
 \Phi ^{(L)} \ = \  \Phi ^{(L)}_{{\rm discrete}}\ = \ L^{-1}\star \Phi ^{\prime }_{{\rm discrete}}\star \pi \left( L\right)
\ = \  \sum_{\substack{ \text{All valid }  \\ \text{values of }\boldsymbol{%
\lambda }}}\nu _{\boldsymbol{\lambda }}f_{\boldsymbol{\lambda }%
}^{L}\star \kappa _{y}+\text{ \ conj}^{(L)} \text{ ,}
\label{PhiL}
\end{eqnarray}
where 
\bea f_{\boldsymbol{\lambda }%
}^{L} & = & f^L_{\boldsymbol{\lambda }}\left( x,Y \right) \ = \ L^{-1}\star f_{\boldsymbol{\lambda }}\left( a_{1}^\pm ,a_{2}^\pm \right)  \star  L \nn\\
& = & {\cal O}_{\boldsymbol{\lambda }_{1}}^{\varsigma_1}{\cal O}_{\boldsymbol{\lambda }_{2}}^{\varsigma_2}\int_{0}^{+\infty }d\tau _{1}\frac{\tau _{1}^{\lambda _{1R}-\lambda _{1L}-1}%
}{\Gamma \left( \lambda _{1R}-\lambda _{1L}\right) }\int_{0}^{+\infty }d\tau _{2}\frac{\tau _{2}^{\lambda
_{2R}-\lambda _{2L}-1}}{\Gamma \left( \lambda _{2R}-\lambda _{2L}\right) }e^{-4\check K^L(\varsigma_1,\varsigma_2; x,Y)-\Theta^L(\t_1,\t_2;x) Y}  \ , \label{fL}\eea 
where we have defined
\bea
\check K^L(\varsigma_1,\varsigma_2; Y) \ :=  \ \frac{\varsigma_1+\varsigma_2}2 K^L_{(+)}+\frac{\varsigma_2-\varsigma_1}2  K^L_{(-)} 
\ = \   \ -\frac18\left[ y\check\varkappa^L y + \yb\check{\bar \varkappa}^L\yb+2 y \check v^L \yb  \right] \ ,
\eea 
with 
\bea &\displaystyle K^L_{(q)} \ = \ -\frac18 Y^{\underline{\a}}K^L_{(q)\underline{\a}}{}^{\underline{\b}}Y_{\underline{\b}} \ ,& \nn\\
&\displaystyle K^L_{(q)\underline{\a}}{}^{\underline{\b}} \ = \ -\left(L^T K_{(q)} L\right)_{\underline{\a}}{}^{\underline{\b}} \ = \ \left(\begin{array}{cc}\vark^L_{(q)\a\b} & v^L_{(q)\a\bd} \\ 
\bar v^L_{(q)\ad\b} & \bar \vark^L_{(q)\ad\bd}\end{array}\right) \ ,&\eea
where the matrix $L$ is given in \eq{Lmatrix}, and
\be  \Theta^{L\underline{\a}}(x;\t_1,\t_2)  \ = \ (\Theta(\t_1,\t_2)L(x) )^{\underline{\a}} \ .\ee 
In our specific case $K_{(+)}=iB$ and $K_{(-)}=iP$, so
\bea
\check \varkappa_{\a\b}^L &=  & \frac{\varsigma_1+\varsigma_2}2 \varkappa_{(iB)\a\b}^L+\frac{\varsigma_2-\varsigma_1}2 \varkappa_{(iP)\a\b}^L   \ , \label{chvarkL}\\
\check v_{\a\bd}^L & =  & \frac{\varsigma_1+\varsigma_2}2 v_{(iB)\a\bd}^L+\frac{\varsigma_2-\varsigma_1}2 v_{(iP)\a\bd}^L \ ,\label{chvL}
\eea
with $v^L_{(M)\a\bd}$ being the Killing vector associated to the  $\msp(4;\Comp)$ (complexified) isometry generator $M$ and $\vark^L_{(M)\a\b}$ ($\varkb^L_{(M)\ad\bd}$) being the (anti-)selfdual part of the corresponding Killing two-form. Using \eq{vkappaB}-\eq{vkappaP}, their explicit expressions in global, embedding coordinates are
\bea
\varkappa_{(iB)\a\b}^L & = & -i(\s_{03})_{\a\b}+\frac{2i}{1-X_{0'}} \,X_{[0}X^a(\s_{3]a})_{\a\b} \ ,\label{varkiB}\\
v_{(iB)\a\bd}^L & = & 2i X_{[0}(\s_{3]})_{\a\bd} \ ,\label{viB}\\
\varkappa_{(iP)\a\b}^L & = & i X^a (\s_{a1})_{\a\b} \ ,\label{varkiP}\\
v_{(iP)\a\bd}^L & = & -i(\s_{1})_{\a\bd}+\frac{2i}{1-X_{0'}} \,X_{[1}X^a(\s_{a]})_{\a\b} \ .\label{viP}
\eea
It will be useful in the following to write all van der Waerden symbols in terms of a spin-frame $(u^{+\a},u^{-\a})$, $u^{+\a}u^{-}_{\a}=1$, idem their complex conjugates (see Appendix \ref{App:conv} for details). As a consequence, \eq{varkiB}-\eq{viP} can be rewritten as
\bea
\varkappa_{(iB)\a\b}^L & = & -i\left[1+\frac{X_3^2-X_0^2}{1-X_{0'}}\right](u^{+}_\a u^{-}_\b+u^{-}_\a u^{+}_\b) \notag\\
&&-\frac{i}{1-X_{0'}} \left[(X_0+X_3)(X_1-iX_2)u^+_\a u^+_\b+(X_0-X_3)(X_1+iX_2)u^-_\a u^-_\b\right] \ , \label{varkiB2}\\
v_{(iB)\a\bd}^L & = & i (X_{0}+X_3)u^+_\a \ub^+_{\bd}-i (X_{0}-X_3)u^-_\a \ub^-_{\bd} \ , \label{viB2}\\
\varkappa_{(iP)\a\b}^L & = & -X_2(u^{+}_\a u^{-}_\b+u^{-}_\a u^{+}_\b) -i (X_{0}+X_3)u^+_\a u^+_{\b}-i (X_{3}-X_0)u^-_\a u^-_{\b}  \\
v_{(iP)\a\bd}^L & = &  i\left[1+\frac{X^2-X_1^2}{1-X_{0'}}\right](u^{+}_\a \ub^{-}_{\bd}+u^{-}_\a \ub^{+}_{\bd}) \notag\\
&&+\frac{iX_1}{1-X_{0'}} \left[(X_0+X_3)u^+_\a \ub^+_{\bd}+(X_0-X_3)u^-_\a \ub^-_{\bd}+iX_2(u^+_\a \ub^-_{\bd}-u^-_\a\ub^+_{\bd})\right] \ .
\eea
It is easy to show that 
\bea
(\vark_{(iB)}^{L})^2 & := & \frac12 \vark_{(iB)}^{L\a\b}\vark_{(iB)\a\b}^L \ = \ \det\vark_{(iB)}^L \ = \ 1-X_0^2+X_3^2  \ =: \ \Delta^2 \ , \\
(v^{L}_{(iB)})^2 & := & \frac12 v_{(iB)}^{L\a\bd}v_{(iB)\a\bd}^L \ = \ \det v_{(iB)}^L \ = \ X_0^2-X_3^2 \ ,
\eea
while
\bea
(\vark_{(iP)}^{L})^2 & = &  -X^\n X_\n  \ = \ 1-X_{0'}^2+X_1^2 \ , \qquad \n = 0,2,3 \ , \\
(v^{L}_{(iP)})^2 & = & 1 +X^\n X_\n \ = \ X_{0'}^2-X_1^2 \ =: \  \frac{\xi^2}{M} \ . 
\eea
Moreover, 
\bea
\theta^L_\a  =  \frac{1}{2^{\frac32}\sqrt{1-X_{0'}}}\left\{\left[(X_1-iX_2)(\t_1-i\t_2) -(1-X_{0'})(\t_1+i\t_2)\right]u^+_\a+(X_3-X_0)(\t_1-i\t_2)u^-_\a \right\} \ .
\label{thetaL}\eea
We also note that, once all constraints \eq{ident}-\eq{rebos} on the eigenvalue that we allow in the expansion of the Weyl zero-form have been taken into account, Eq. \eq{fL}, in particular the $\tau$-integrals, can be rewritten in the more suggestive form
\bea   f^L_{\boldsymbol{\lambda }} & = & \frac{1}{\Gamma \left(- {\rm Re}(\D\lambda)-ip\right) \Gamma \left(-{\rm Re}( \D\lambda)+ip\right)} \,{\cal O}_{\boldsymbol{\lambda }_{1}}^{\varsigma_1}{\cal O}_{\boldsymbol{\lambda }_{2}}^{\varsigma_2}\nn\\
& & \times \ \ \int_{0}^{+\infty }d\tau _{1} \int_{0}^{+\infty }d\tau _{2} \frac{1}{(\tau _{1}\tau _{2})^{{\rm Re}(\D\lambda)+1}}\left(\frac{\t_2}{\t_1}\right)^{ip} e^{-4\check K^L(\varsigma_1,\varsigma_2; x,Y)-\Theta^L(\t_1,\t_2;x) Y} \ ,\label{fcond}\eea
where we have defined $p:={\rm Im}(\l_{1L})=-{\rm Im}(\l_{2L})$, $\D\lambda:=\l_{1L}-\l_{1R}=\l_{2L}-\l_{2R}$ (the last equality following from the identification condition \eq{ident}). 

In terms of the $L$-rotated quantities above introduced, the star product of \eq{fL} with $\k_y$ reads
\bea  f^L_{\boldsymbol{\lambda }}\star \k_y &= & {\cal O}_{\boldsymbol{\lambda }_{1}}^{\varsigma_1}{\cal O}_{\boldsymbol{\lambda }_{2}}^{\varsigma_2}\int_{0}^{+\infty }d\tau _{1}\frac{\tau _{1}^{\lambda _{1R}-\lambda _{1L}-1}%
}{\Gamma \left( \lambda _{1R}-\lambda _{1L}\right) }\int_{0}^{+\infty }d\tau _{2}\frac{\tau _{2}^{\lambda
_{2R}-\lambda _{2L}-1}}{\Gamma \left( \lambda _{2R}-\lambda _{2L}\right) } \nn\\
&&\times \ \  \frac{1}{\sqrt{(\check \vark^L)^2}}\, \exp\left[-\frac1{2}(\ty^L-i\theta^L)(\check \vark^L)^{-1}(\ty^L-i\theta^L)+\frac{1}2\yb\check \varkb^L\yb-\bar \theta^L\yb\right] \ , \label{Phicov}\eea
with the modified oscillators $\ty^L:=y-i\check v^L\yb$ and where we recall that $(\vark^L_{(q)})^{-1}_{\a\b}=-\frac{\vark^L_{(q)\a\b}}{(\vark^L_{(q)})^2}$. This is the expression of the generic term in the expansion of our Weyl zero-form $\Phi^{(L)}$ in \eq{PhiL}. In particular, the generating function of the scalar and the (self-dual part of the) generalized Weyl tensor fields is
\bea  {\cal C}(x,y) & = & \Phi^{(L)} \big|_{\yb=0} \ = \  \sum_{\substack{ \text{All valid }  \\ \text{values of }\boldsymbol{%
\lambda }}}\left(\nu _{\boldsymbol{\lambda }}f_{\boldsymbol{\lambda }%
}^{L}\star \kappa_{y}+\text{ \ conj}^{(L)}\right) \big|_{\yb=0} \ ,\label{calCm}\eea
where 
\bea f_{\boldsymbol{\lambda }%
}^{L}\star \kappa_{y}\big|_{\yb=0}  & = & {\cal O}_{\boldsymbol{\lambda }_{1}}^{\varsigma_1}{\cal O}_{\boldsymbol{\lambda }_{2}}^{\varsigma_2}\int_{0}^{+\infty }d\tau _{1}\frac{\tau _{1}^{\lambda _{1R}-\lambda _{1L}-1}%
}{\Gamma \left( \lambda _{1R}-\lambda _{1L}\right) }\int_{0}^{+\infty }d\tau _{2}\frac{\tau _{2}^{\lambda
_{2R}-\lambda _{2L}-1}}{\Gamma \left( \lambda _{2R}-\lambda _{2L}\right) } \nn\\
&&\times \ \ \frac{1}{\sqrt{(\check \vark^L)^2}}\, \exp\left[-\frac1{2}(y-i\theta^L)(\check \vark^L)^{-1}(y-i\theta^L)\right] \label{calC} \ .
\eea
\paragraph{Real-analyticity at the horizon.} While the expression for $\Phi^{(L)}$ \eq{Phicov} obtained above may seem, at a first glance, essentially identical to its $x$-independent counterpart at the unfolding point on the horizon \eq{Phiprimecov},  a closer look reveals an important difference: the $Y$-independent term at the exponent --- bilinear in $\theta^L_\a$, that is, blinear in the $\t_i$ --- is here actually non-vanishing, whereas its counterpart in \eq{Phiprimecov} is in fact trivial. Indeed, as $\check \vark_{\a\b}$ reduces to $\vark_{(iB)\a\b}$ (see Eqs. \eq{varkcheck} and \eq{vkappaB}-\eq{vkappaP}) and $\theta_\a$ is collinear with one of the eigenspinors of $\vark_{(iB)\a\b}$ (see Eqs. \eq{Theta},\eq{vkappaB} and \eq{D.2.0}), clearly the quadratic term in $\theta$ at the exponent of \eq{Phiprimecov} vanishes. However, the situation changes once the star products with the gauge function displace the Weyl zero-form from the unfolding point at the horizon: $\check \vark^L_{\a\b}(x)$ and $\theta^L_\a(x)$ acquire extra, spacetime-dependent terms that complicate their spinorial structure, giving rise to non-trivial, $Y$-independent terms bilinear in $\t_i$.  \\
\noindent This is crucial in order for the linearized Weyl zero-form to be considered a proper generating function of fluctuation fields. Indeed, the scalar field ($s=0$) and the (self-dual part of the) spin-$s$ linearized Weyl tensor $C_{\a(2s)}$ ($s = 1,2,3,\dots$) are extracted from $\Phi^{(L)}$ via 
\begin{equation}
\left.
C_{\a(2s)}(x) \ = \ \frac{\partial}{\partial y^{\alpha_1}}
\cdots
\frac{\partial}{\partial y^{\alpha_{2s}}}
\Phi^{(L)}
\right|_{Y=0} \ = \ \left.\frac{\partial}{\partial y^{\alpha_1}}
\cdots
\frac{\partial}{\partial y^{\alpha_{2s}}} {\cal C}\right|_{\yb=0}
\end{equation}
(analogously, with the roles of $y$  and $\yb$ interchanged for the anti-self-dual part of the Weyl tensors). Therefore, in order for $\Phi^{(L)}$ to contain the propagating degrees of freedom in the coefficients of its $Y$ expansion it is crucial that it be real-analytic in $Y=0$. In this respect, the expansion \eq{phiprimefinal}-\eq{fgen} that we start from at the unfolding point is problematic, since, as we have seen, whenever complex eigenvalues are involved, as it is in this case necessary in order to have non-trivial momentum on $S^1$. This leads to complex powers of the oscillators, which reflect themselves, in the integral presentation, into  ill-defined $\tau$-integrals in the limit $Y\to 0$ (\ref{fgen}). However, as we have commented on above, displacing the Weyl zero-form away from the unfolding point by means of the star products with the gauge function leads to an integrand of schematic form
\begin{equation}
e^{O\left( \tau ^{2}\right) +O\left( \tau y\right) +O\left( y^{2}\right) }%
\text{ ,}
\end{equation}
which, after taking the derivatives w.r.t.\ $y$-coordinates, gives 
\begin{equation}
e^{O\left( \tau ^{2}\right) +O\left( \tau y\right) +O\left( y^{2}\right) }
\text{ Polynomial}(\tau,y)
\text{ .}
\end{equation}
It is the appearance of the non-trivial $Y$-independent terms bilinear in $\t_i$ at the exponent that helps the convergence of the Mellin transforms and restores analyticity in $Y$ (at least for generic spacetime points), as we shall show with examples in the next Subsection.

\paragraph{Resolution of membrane-like singularities.} On specific surfaces the Weyl zero-form (as well as each spin-$s$ component field) may have an analytic curvature singularity. As clear from \eq{Phicov}, this happens where the Killing two-form becomes degenerate, that is, where $(\check \vark^L)^2=0$, which, with our restriction on the eigenfunctions as in \eq{ident}-\eq{rebos}, reduces to the surface $\D^2\equiv 1+X_3^2-X_0^2=0$. Singularities of this type  were already studied in the context of spherically-symmetric higher-spin black holes in \cite{Iazeolla:2011cb,Iazeolla:2012nf,Iazeolla:2017vng} and, as we shall see, it is possible to generalize the conclusions of those papers to our present case of fluctuations of type \eq{Phicov} over a BTZ-like background: what happens is that, as anticipated in Section \ref{Sec:resolution}, the embedding of such curvature singularities in a higher-spin covariant theory ---  where higher-spin symmetries force the appearance of one such singular Weyl tensor field for every component of an infinite-dimensional multiplet, all packed as coefficients of the expansion of the Weyl zero-form onto an infinite-dimensional, non-commutative fibre algebra $Y$ --- effectively trades the space-time singularities of the component fields for a delta-function-like behaviour in $Y$ of the corresponding master field. In practice, the quantity  $\sqrt{\D^2}$ enters the formula \eq{Phicov} as the vanishing parameter of a delta sequence: away from the surface $\D^2=0$ the Weyl zero-form is a smooth Gaussian function of the oscillators, while it approaches a Dirac delta function on $Y$-space in the $\D^2\to 0$  limit.  However, unlike the delta function on a commutative space, the delta function in the non-commutative $Y$-space, thought of as a symbol for an element of a star product algebra, is smooth. In other words, the mapping of the spacetime curvature singularities to a distribution in the fibre has the advantage that the latter type of singularity can be handled better, as the resulting distributions have good star-product properties and can be considered elements of an associative algebra\footnote{Furthermore, it is to some extent possible to consider a delta function of the oscillators as a bounded function (which would give an even stronger meaning to the notion of resolution of curvature singularities) in the sense that, on a non-commutative space, a change in the ordering prescription can turn a delta function into a smooth symbol (e.g., an exponential \cite{Iazeolla:2011cb}). Changes of ordering are formally part of the gauge transformation that leave the classical observables of the Vasiliev system invariant (with important subtleties that are currently being studied \cite{Vasiliev:2015mka,DeFilippi:2019jqq}), so in this sense the above resolution of curvature singularities would amount to saying that the latter are an artifact of the ordering choice for the infinite-dimensional symmetry algebra governing the Vasiliev system.}, see \cite{Iazeolla:2011cb,Iazeolla:2017vng,Aros:2017ror}.  \\
\noindent We shall spell out the details of the limit in Appendix \ref{Sec:singularity} for the simplest non-trivial choice of eigenvalues (all real parts of left and right eigenvalues take the lowest-weight value $1/2$) . Qualitatively, the result will be that
\be  \lim_{\D^2\to 0} \Phi^{(L)} \ \propto \  f(X) {\cal O}_{\boldsymbol{\l}_1} {\cal O}_{\boldsymbol{\l}_2} \d^2(\hat y) \ .  \ee
where $f(X)$ is a function of the spacetime coordinates and $\hat y:=\lim_{\D^2\to 0}\ty^L$. We defer to Appendix \ref{Sec:singularity}  the precise result and more comments. 

Thus, from the considerations above we expect that the Weyl zero-form can be analytically continued through the horizon (to which the unfolding point belongs), and that the membrane-like curvature singularities in $\Delta=0$ are resolved at the master-field level in the sense specified above. 

\paragraph{Limit to the singularity of the BGM background.} Furthermore, one can observe that $\D^2 |_{X^2=0}=\xi^2/M$. 
The analysis of the membrane-like singularity therefore suggests that also $\xi=0$ is a regular point, in the sense that the master field is given here by a well-defined regular prescription.
Therefore, we expect that the master field configuration can be continued through the singularity, thus exploring the full background manifold $AdS_3\times_\xi S^1$.

In what remains, we shall turn our attention to extracting and studying the behaviour of the Lorentz scalar fluctuation field.

\subsection{The scalar field}\label{Sec:scalar}

\paragraph{Choice of quantum numbers.} For simplicity, we shall begin by studying the scalar field from the simplest non-trivial choice of eigenvalues that our kinematical conditions allow: that is, 
\begin{equation}
C(x) \ := \ \ f^L_{\boldsymbol{\lambda }}\star \k_y|_{Y=0} + {\rm c.c.}\ , \label{scal}
\end{equation} 
with
\begin{equation}
\lambda _{1L}=\frac{1}{2}+i\,\frac{n}{\sqrt{M}}\ \ ,\ \ \ \lambda _{2L}=\frac{1%
}{2}-i\,\frac{n}{\sqrt{M}}\ \ ,\ \ \ \lambda _{1R}=\lambda _{2R}=\frac{1}{2}\
\ ,\ \ \ (n\in \mathbb{Z)}\text{ ,}
\label{eigenchoice}
\end{equation}
The complex conjugate, denoted with ${\rm c.c.}$ is extracted by means of the identical projection on the conjugate term ${\rm conj}^{(L)}$ of \eq{PhiL} (or, equivalently, \eq{calCm}). 

Recalling the realization of $iP$ in terms of number operators \eq{w12B}, it is evident that, operating on an element $f_{\boldsymbol{\l}}$ with eigenvalues \eq{eigenchoice}, the twisted adjoint action of $iP$ (according to \eq{twadjeig}) on it extracts the eigenvalue $\frac{\l_{2L}-\l_{1L}}{2}-\frac{\l_{2R}-\l_{1R}}{2}=-ip$; that is, the non-trivial imaginary part of the left eigenvalues gives rise to the oscillating dependence from the coordinate on the $S^1$, as we shall see in \eq{C1Schw}-\eq{C2Schw}. On the other hand, having chosen identical real parts for the left and right eigenvalues and recalling \eq{w12B} and \eq{twadjeig}, the eigenvalue of $iB$ results $\frac{\l_{2L}+\l_{1L}}{2}-\frac{\l_{2R}+\l_{1R}}{2}=0$, and as a consequence the fields obtained from a Weyl zero-form with \eq{eigenchoice} do not break the $U(1)_{B}$ symmetry. This is also why it makes sense to use the Schwarzschild-like coordinates, not adapted to the action of the Killing vector corresponding to $iB$, to write the fluctuation fields for this choice of eigenvalues. Note that a left-right asymmetric choice of real parts of the eigenvalues would give rise to exponentially growing/decreasing quasi-normal modes in the coordinate dual to $B$.

\paragraph{Performing parametric integrals in global coordinates.}

The contour integrals in (\ref{Phicov}) can be evaluated immediately, and simply set $\varsigma_1=\varsigma_2=1$. Projecting onto $Y=0$, \eq{scal} is reduced to
\bea & \displaystyle C(x) \ := \ \ f^L_{\boldsymbol{\lambda }}\star \k_y|_{Y=0} + {\rm c.c.} &\nn\\
& \displaystyle= \  \int_{0}^{+\infty }d\tau _{1}\frac{\tau _{1}^{-ip-1}}{\Gamma \left(
-ip\right) }\ \int_{0}^{+\infty }d\tau _{2}\frac{\tau _{2}^{ip-1}}{\Gamma
\left( ip\right) } \frac{1}{\sqrt{(\vark_{iB}^L)^2}}\, \exp\left[-\frac1{2(\vark_{(iB)}^L)^2 }\theta^L \vark^L_{(iB)} \theta^L\right] +{\rm c.c.} & \label{scalC} \ ,
\eea
where we have defined $p := \frac{n}{\sqrt{M}}$. The exponent is a quadratic form in the $\t_i$, which can be computed by substituting \eq{varkiB2} and \eq{thetaL} (alternatively, see Appendix \ref{Sec:singularity} for an ``adapted'' spin-frame, in which the basis spinors are chosen as the eigenspinors of $\vark_{(iB)}^L$), 
\be \exp\left[-\frac1{2(\vark_{(iB)}^L)^2 }\theta^L \vark^L_{(iB)} \theta^L\right] \ = \ e^{-i\left( c_{1}\tau _{1}{}^{2}+c_{2}\tau _{1}\tau
_{2}+c_{3}\tau _{2}{}^{2}\right) } \ee
where, in embedding coordinates,
\begin{gather}
c_{1}\ := \ \left( X^{0^{\prime }}-X^{1}\right) \mathrm{A} \ ,\qquad
c_{2}\ := \ 2X^{2}\mathrm{A}\ ,\qquad c_{3}\ := \ \left( X^{0^{\prime }}+X^{1}\right) 
\mathrm{A} \ \ ,  \notag \\
\mathrm{A} \ := \ \frac{X^{0}+X^{3}}{8%
\D^2 }\text{ \ ,}
\label{Mellinparam}
\end{gather}
and where we recall that $\D^2\equiv (\vark^L_{(iB)})^2=1+X^2_3-X_0^2=X_{0'}^2-X_1^2-X_2^2$. Finally, the remaining integrals over $\tau _{1}$ and $\tau _{2}$ can be computed using the following formula (see Appendix \ref{App:integralformula} for the details of the derivation, and Appendix \ref{App:03divergence} for a succinct analysis of the extraction of component fields on the seemingly problematic surface $X^0+X^3=0$):
\begin{eqnarray}
&&\int_{0}^{+\infty }d\tau _{1}\frac{\tau _{1}^{-ip-1}}{\Gamma \left(
-ip\right) }\ \int_{0}^{+\infty }d\tau _{2}\frac{\tau _{2}^{ip-1}}{\Gamma
\left( ip\right) }e^{-i\left( c_{1}\tau _{1}{}^{2}+c_{2}\tau _{1}\tau
_{2}+c_{3}\tau _{2}{}^{2}\right) }  \notag \\
& = & \left(\frac{c_1}{c_3}\right)^{\frac{ip}{2}}\cosh\left[p\arcsin\left(\sqrt{1-\frac{c_2^2}{4c_1 c_3}}\right)\right] \ .
\label{Mellinscalarsin}
\end{eqnarray}
Thus, the scalar field profile can be written as 
\bea 
C(X) & = & 
 \left( \frac{X^{0^{\prime }}-X^{1}}{X^{0^{\prime }}+X^{1}}%
\right) ^{\frac{in}{2\sqrt{M}}}\frac{\cosh \left\{ \frac{n}{\sqrt{M}}\arcsin %
\left[ \sqrt{\frac{M\D^2}{\xi^2}}\right] \right\} } {\sqrt{\D^2}}+{\rm c.c.}\ ,\label{C1C2}
\eea
where we recall $\xi^2 := M(v^L_{(iP)})^2=M((X^{0'})^2-(X^1)^2)$. The first factor in \eq{C1C2} guarantees periodicity along the direction of identification, as will become manifest in Schwarzschild-like coordinates. As the norm of the identification Killing vector $P\equiv M_{0'1}$ is everywhere positive on the BTZ-like background, $\xi^2\equiv M(v^L_{(iP)})^2=-M v_{(iP)}^\m v_{(iP)\m}>0$ everywhere, too, and therefore the behaviour of the second factor in \eq{C1C2} is essentially determined by the sign of $\D^2$:  
\bea 
C(X) & = &   \left( \frac{X^{0^{\prime }}-X^{1}}{X^{0^{\prime }}+X^{1}}%
\right) ^{\frac{in}{2\sqrt{M}}}\frac{\cosh \left\{ \frac{n}{\sqrt{M}}\arcsin %
\left[ \frac{\sqrt{M}\D}{\xi}\right] \right\}}{\D}+{\rm c.c.}\ ,\qquad \D^2>0 \ ,\label{C1C2bis1}\\
C(X) & = &   -i\left( \frac{X^{0^{\prime }}-X^{1}}{X^{0^{\prime }}+X^{1}}%
\right) ^{\frac{in}{2\sqrt{M}}}\frac{\cos \left\{ \frac{n}{\sqrt{M}}{\rm arcsinh} %
\left[ \frac{\sqrt{M}|\D|}{\xi}\right] \right\} }{|\D|}+{\rm c.c.}\ ,\qquad \D^2<0 \ , \label{C1C2bis2}
\eea
where we are defining $\D:=\sqrt{\D^2}$ and $\xi:=\sqrt{\xi^2}$. As $\D^2=\xi^2/M-(X^2)^2$, for $\D^2>0$ the argument of $\arcsin$ is real and bounded from above, $ 0\leq\sqrt{\frac{M\D^2}{\xi^2}}\leq 1$, and as a consequence, according to \eq{C1C2bis1} the scalar field diverges for $\D^2\to 0^+$ and decays essentially as $1/\D$ for large enough $\D$.  On the other hand, for $\D^2<0$, the argument of ${\rm arcsinh}$ is real and unbounded, $0 \leq\sqrt{\frac{M|\D^2|}{\xi^2}}<+\infty$, and consequently, as $|\D^2|$ increases, the scalar field in the region $\D^2< 0$ decreases towards zero from the divergence in $\D^2\to 0^-$ with fastly suppressed oscillations.  

\paragraph{Alternative expression in Schwarzschild coordinates.}
For $\frac{c_2}{\sqrt{c_1 c_3}}>0$, that is, for $X^2>0$ ($\sin\theta>0$ in Schwarzschild-like coordinates), one can rewrite the result in \eq{Mellinscalarsin} as
\begin{eqnarray}
&&\int_{0}^{+\infty }d\tau _{1}\frac{\tau _{1}^{-ip-1}}{\Gamma \left(
-ip\right) }\ \int_{0}^{+\infty }d\tau _{2}\frac{\tau _{2}^{ip-1}}{\Gamma
\left( ip\right) }e^{-i\left( c_{1}\tau _{1}{}^{2}+c_{2}\tau _{1}\tau
_{2}+c_{3}\tau _{2}{}^{2}\right) }  \notag \\
&=&\left( \frac{c_{1}}{c_{3}}\right) ^{\frac{ip}{2}}\cosh \left[ p\arccos
\left( \frac{c_{2}}{2\sqrt{c_{1}c_{3}}}\right) \right] \text{ ,}
\label{Mellinscalar}
\end{eqnarray}
and therefore, in embedding coordinates,
\begin{equation}
C(X) \ = \ \left( \frac{X^{0^{\prime }}-X^{1}}{X^{0^{\prime }}+X^{1}}%
\right) ^{\frac{in}{2\sqrt{M}}}\frac{\cosh \left\{ \frac{n}{\sqrt{M}}\arccos %
\left[ \frac{\sqrt{M}X^{2}}{\sqrt{\xi^{2}}}\right] \right\} }{\sqrt{\D^2}} +{\rm c.c.} \label{C1C2arccos}
\end{equation}
Then for $r>\sqrt{M}$ it can be further converted into  
\begin{equation}
C \ = \ e^{-in\phi }\frac{\cosh \left\{ \frac{n}{\sqrt{M}}\arccos \left[ 
\sqrt{1-\frac{M}{r^{2}}}\cosh \left( \sqrt{M}t\right) \sin \left(
\theta \right) \right] \right\} }{\sqrt{\frac{r^{2}}{M}-\left( \frac{%
r^{2}}{M}-1\right) \cosh ^{2}\left(\sqrt{M}t\right) \sin
^{2}\left( \theta \right) }}+{\rm c.c.}\text{ ,}\label{C1Schw}
\end{equation}
and for $r<\sqrt{M}$
\begin{equation}
C \ = \ e^{-in\phi }\frac{\cosh \left\{ \frac{n}{\sqrt{M}}\arccos \left[ 
\sqrt{\frac{M}{r^{2}}-1}\sinh \left( \sqrt{M}t\right) \sin \left(
\theta \right) \right] \right\} }{\sqrt{\frac{r^{2}}{M}-\left( 1-\frac{%
r^{2}}{M}\right) \sinh ^{2}\left( \sqrt{M}t\right) \sin
^{2}\left( \theta \right) }}+{\rm c.c.}\text{ .}\label{C2Schw}
\end{equation}%
The scalar field is manifestly periodic in $\phi $, as expected, and one can check
that it indeed satisfies the Klein-Gordon equation
\begin{equation}
\left( \Box +2\right) C =0 \ .
\end{equation}

\paragraph{Horizon limit.}
We can check that the coefficients of the $Y$ expansion of the master field  \eq{Phicov} with eigenvalues \eq{eigenchoice} in the scalar sector are all well-defined at the horizon (i.e., in the limit $r\to\sqrt{M}$), as anticipated. For instance, in this limit the scalar field converges to 
\begin{equation}
    \lim_{r \rightarrow  \sqrt{M}} C = e^{-i n\phi} \cosh\left(\frac{n}{\sqrt{M}}\right)\ ,
\end{equation}
where for notational simplicity we omit the ${\rm c.c.}$ term.  This result can be actually obtained on either of the patch of coordinates for $r>   \sqrt{M}$ or $ r <  \sqrt{M}$. The directional derivatives also remain well-defined at the horizon.  Let us first consider the limit on the outer patch. In order to define the directional derivatives we can introduce the local frame 
\begin{eqnarray*}
  e^0 = \left( r^{2}-M\right)^{\frac{1}{2}} dt,  & &  e^2 = \left( r^{2}-M\right)^{\frac{1}{2}} \cosh\left( \sqrt{M}t\right) d\theta \\
   e^1 = \left( r^{2}- M\right)^{-\frac{1}{2}} dr  &\textrm{and}& e^3 = r  d\phi,
\end{eqnarray*}
where the direction along (1) can be identify with the {\it radial} direction. In this way,  
\begin{equation}\label{RadialDerivative}
  \lim_{r \rightarrow  \sqrt{M}} \left( \nabla_{(1)} \right) C = - n \frac{e^{-i n\phi}}{M } \sinh\left(\frac{n}{2\sqrt{M}}\right) \sin(\theta)\cosh(\sqrt{M}t).
\end{equation}
Analogously, 
\begin{equation}\label{TimelDerivative}
  \lim_{r \rightarrow  \sqrt{M}} \left( \nabla_{(0)} \right) C =  - n e^{-i n\phi}\sinh\left(\frac{n}{\sqrt{M}}\right) \sinh(\sqrt{M}t).
\end{equation}
and
\begin{equation}\label{ThetaDerivative}
  \lim_{r \rightarrow  \sqrt{M}} \left( \nabla_{(2)} \right) C =   n \frac{e^{-i n\phi}}{2}\sinh\left(\frac{n}{\sqrt{M}}\right) \cos(\theta)\cosh(\sqrt{M}t).
\end{equation}
It is direct to check that the same limit can be obtained from the patch $r<\sqrt{M}$, with the proviso that of course the coordinates are different and that $ \sinh(\sqrt{M}t)$ and $ \cosh(\sqrt{M}t)$ are interchanged (compare Eq.(\ref{embed-HP}) and Eq.(\ref{embed-HPSecond})).

\paragraph{Singular limits.} As can be seen directly from \eq{C1C2bis2}, the scalar field has a membrane-like singularity as $\Delta\rightarrow 0$.
Approaching the singularity of the BGM background, \emph{i.e.} in the limit $\xi\to 0$, the scalar field remains bounded but becomes indefinite, as it starts oscillating with a diveregent frequency.
However, as discussed in Section \ref{Sec zeroform-L}, the linearized Weyl zero-form master field remains well-defined in both of these limits in the sense that its limiting values are fiber space distributions with a regular presentation.
More precisely, the resolution of the membrane-like singularity of the scalar field is in terms of a fiber space delta function, as shown in Appendix \ref{Sec:singularity}.
The resolution of the scalar field singularity at $\xi=0$, on the other hand, is in terms of a more general distribution, as discussed at the end of Section \ref{Sec zeroform-L}.

\section{Conclusions and outlook}\label{Sec:concl}

In this paper we have studied massless fluctuations of all integer spins over the four-dimensional uplift of the eternal spinless BTZ black hole --- i.e., over the eternal spinless BGM black hole. The latter, like its 3D counterpart, can be represented as a flat connection, and as such it is also a vacuum solution of the full 4D Vasiliev equations of higher-spin gravity.  For this reason, we have constructed fluctuation fields as solutions to the Vasiliev equations linearized around the BGM spacetime. In doing so, we have made use of the unfolded formulation, thereby constructing both the background solution and the fluctuations by means of spacetime-dependent gauge functions and fiber space elements containing the local data that reconstruct the spacetime fields around any regular point. 

In fact, as we showed in Section \ref{Sec:BTZ}, writing the background solution by means of a globally defined gauge function facilitates its extension to the full topologically extended eternal spinless BGM black hole, consisting  of  two  eternal  BGM  black  holes glued  together  across  their singularities. The natural question, that we addressed next, is whether fluctuation fields can be thought of as smooth at the BGM singularity (as well as on other submanifold, such as the horizon), which seems impossible in the usual gravitational analysis.

We have showed that higher-spin gravity provides interesting mechanisms for resolving classical singularities in gravity.
These mechanisms rely on an interplay between differential and operator algebras.
The former can be used to treat fluctuations on manifolds with degenerate metrics.
Moreover, the unfolded machinery requires the introduction of infinitely many form fields, which can be packaged into master fields taking their values in operator algebras.
In the presence of higher spin symmetry, these operators become density matrices on non-commutative symplectic manifolds.
As we have seen in this paper, the linearized master fields can be continued across horizons and singularities (and other surfaces), where individual Lorentz tensorial fields have fatal singularities: typically, the singular behaviour of individual fields on such surfaces manifests itself in the fact that the master fields become delta function in the fiber coordinates. As the latter are non-commutative variables, however, the master fields remain well-defined as symbols of operator algebra  elements, and in that sense the limit to the horizon or the singularity is uneventful. In that sense, the fluctuations do explore the full topologically extended eternal spinless BGM black hole. 

In order for such mechanisms to survive at the fully non-linear level, one has to show that the aforementioned operator algebras admit a well-defined quantum star product.
We shall address this important issue, which involves the composition of operators in the image of the Wigner-Ville map applied to wave-functions that are not $L^2$, in a future publication \cite{paper1prime}.
Preliminary results show that Holder duality \cite{math_people} as well as the particle/black hole duality \eq{calad} may play an important role in constructing these algebras.

The formalism also leads to a natural twistor space regularization of the self-energy for Coulomb--like solutions on $AdS_4$, which clearly deserves further scrutiny.
Another physically interesting feature that emerges naturally from our construction is the appearance of quasi-normal modes on the BGM black-hole background. They arise essentially as a result of the fact that if the adjoint action of ${\rm ad}^\star_K$ of the oscillator realization $K$ of the identification Killing vector $\overrightarrow{K}$ has an integer spectrum, then the spectrum of ${\rm ad}^\star_{\widetilde K}$, where $\overrightarrow{\widetilde{K}}$ is a dual Killing vector, contains imaginary parts. The latter are responsible for the appearance of exponentially growing/decaying modes, which we interpret as quasi-normal modes. Our construction based on the unfolded formulation provides a systematic way of obtaining them analytically, and may therefore prove useful to study quasi-normal modes and the properties of their dual thermal states in greater detail.

As a final remark, we would like to stress that, to our best understanding, it is possible to consider gauge functions generated by identifications along general conjugacy classes as formal solutions to Vasiliev's equations including higher spin fluctuations.
Moreover, the issue of horizons is sidestepped simply due to the fact that the physical observables of higher spin gravity theory are of a quite different type than those used in ordinary gravity.
In particular, one may label the BHTZ-like higher spin geometries using the holonomy of the flat Vasiliev connection along the identification circle.

\section*{Acknowledgements} 
We would like to thank I. Araya, C. Arias, E. Bergshoeff, R. Bonezzi, N. Boulanger, A. Campoleoni, E. Cheung, D. de Filippi, G. de Nittis, F. Diaz, V. E. Didenko, A. Faraggi, D. Grumiller, M. Henneaux, Y.-P. Hu, A. Kehagias, R. Olea, F. Rojas, M. Romo, A. Sagnotti, E. Sezgin, M. Taronna, M. Valenzuela, B. Vallilo and M. A. Vasiliev for valuable discussion.
This work was partially funded by grants FONDECYT 1151107, 1140296. 
PS, RA and YY would like to thank DPI20140115 for some financial support. 
The work of CI was supported in part by the Russian Science Foundation grant 
14-42-00047 in association with the Lebedev Physical Institute 
in Moscow. 
The initial stage of YY's work was supported by the FONDECYT Project 3150692.
Finally, CI, PS and YY would like to thank the Erwin Schrödinger Institute in Vienna for hospitality during the program `Higher Spins and Holography' where this work was completed.

\appendix{}


\section{Spinor conventions and $AdS_4$ background}\label{App:conv}


We use conventions in which $SO(3,2)$ generators $M_{AB}$ with $A,B=0,1,2,3,0'$ obey
\be [M_{AB},M_{CD}]\ =\ 4i\eta_{[C|[B}M_{A]|D]}\ ,\qquad
(M_{AB})^\dagger\ =\ M_{AB}\ ,\label{sogena}\ee
which can be decomposed using $\eta_{AB}~=~(\eta_{ab};-1)$, with $a,b=0,1,2,3$ as
\be [M_{ab},M_{cd}]_\star\ =\ 4i\eta_{[c|[b}M_{a]|d]}\ ,\qquad
[M_{ab},P_c]_\star\ =\ 2i\eta_{c[b}P_{a]}\ ,\qquad [P_a,P_b]_\star\ =\
i\lambda^2 M_{ab}\ ,\label{sogenb}\ee
where $M_{ab}$ generate the Lorentz subalgebra $\mso(3,1)$, and $P_a=\l M_{0'a}$ with $\l=l^{-1}$ being the inverse $AdS_4$ radius related to the cosmological constant via $\Lambda=-3\l^2$. The Lorentz metric $\eta_{ab}$ is taken as ${\rm diag}(-+++)$. Decomposing further under the maximal compact subalgebra, the $AdS_4$ energy generator $E=P_0=\l M_{0'0}$ and the spatial $\mso(3)$ rotations are generated by $M_{rs}$ with $r,s=1,2,3$.
In terms of the oscillators $Y_{\underline\a}=(y_\a,\yb_{\ad})$, their realization is taken to be
\be M_{AB}~=~ -\frac{1}{8}  (\G_{AB})_{\underline{\a\b}}\,Y^{\underline\a}\star Y^{\underline\b}\ ,\label{MAB}\ee
 \be
 M_{ab}\ =\ -\frac18 \left[~ (\s_{ab})^{\a\b}y_\a\star y_\b+
 (\sb_{ab})^{\ad\bd}\bar y_{\ad}\star \yb_{\bd}~\right]\ ,\qquad P_{a}\ =\
 \frac{\l}4 (\s_a)^{\a\bd}y_\a \star \yb_{\bd}\ ,\label{mab}
 \ee
using Dirac matrices obeying $(\Gamma_A)_{\underline{\a}}{}^{\underline{\b}}(\Gamma_B C)_{\underline{\b\g}}=
\eta_{AB}C_{\underline{\a\g}}+(\Gamma_{AB} C)_{\underline{\a\g}}$, and van der Waerden symbols obeying
 \be
  (\s^{a})_{\a}{}^{\ad}(\sb^{b})_{\ad}{}^{\b}~=~ \eta^{ab}\d_{\a}^{\b}\
 +\ (\s^{ab})_{\a}{}^{\b} \ ,\qquad
 (\sb^{a})_{\ad}{}^{\a}(\s^{b})_{\a}{}^{\bd}~=~\eta^{ab}\d^{\bd}_{\ad}\
 +\ (\sb^{ab})_{\ad}{}^{\bd} \ ,\label{so4a}\ee\be
 \frac{1}{2} \e_{abcd}(\s^{cd})_{\a\b}~=~ i (\s_{ab})_{\a\b}\ ,\qquad \frac{1}{2}
 \e_{abcd}(\sb^{cd})_{\ad\bd}~=~ -i (\sb_{ab})_{\ad\bd}\ ,\label{so4b}
\ee
\be ((\s^a)_{\a\bd})^\dagger~=~
(\sb^a)_{\ad\b} ~=~ (\s^a)_{\b\ad} \ , \qquad ((\s^{ab})_{\a\b})^\dagger\ =\ (\sb^{ab})_{\ad\bd} \ .\ee
and raising and lowering spinor indices according to the
conventions $A^\a=\epsilon^{\a\b}A_\b$ and $A_\a=A^\b\epsilon_{\b\a}$ where
\be \e^{\a\b}\e_{\g\d} \ = \ 2 \d^{\a\b}_{\g\d} \ , \qquad
\e^{\a\b}\e_{\a\g} \ = \ \d^\b_\g \ ,\qquad (\e_{\a\b})^\dagger \ = \ \e_{\ad\bd} \ .\ee
In order to avoid cluttering the expression with many spinor indices, in the paper we also use the matrix notations 
\bea & A^{\underline{\a}} B_{\underline{\a}} \ = :\  AB \ = \ ab+\bar a \bar b \ := \ a^\a b_\a + \bar a^{\ad}\bar b_{\ad} \ ,&\\
& aMb \ := \ a^\a M_{\a}{}^\b b_\b \ , \qquad aN\bar{b} \ := \ a^\a N_{\a}{}^{\bd} \bar{b}_{\bd} \ . &
\eea
The van der Waerden symbols can be realized in a given spin-frame 
\be U~=~(u^\pm_\a,{\bar u}^\pm_{\ad})\ ,\qquad \bar u^\pm_{\ad}~=~(u^\pm_\a)^\dagger\ ,\qquad u^{+\a}u^-_\a=1=\bar u^{+\ad}\bar u^-_{\ad}\ ,\label{D.1}\ee
\be \e_{\a\b} \ = \ (u^- u^+- u^+ u^-)_{\a\b}\ ,\qquad \e^{0123}~=~1\ ,\ee
as
\bea \s_0|_U& = & -u^+\bar u^+- u^-\bar u^- \ , \qquad \s_1|_U~=~-u^+\bar u^-- u^-\bar u^+ \ ,\label{sigmaupm}\\[5pt]
\s_2|_U & = & i(u^+\bar u^-- u^-\bar u^+) \ ,\qquad \s_3|_U \ = \ u^+\bar u^+- u^-\bar u^-\ ,\eea
\be \s_{01}|_U ~=~ u^+ u^+- u^- u^-\ , \quad \s_{02}|_U ~=~ -i (u^+ u^++ u^- u^-) \ ,\quad  \s_{03}|_U ~=~ u^+ u^-+ u^- u^+ \label{D.2.0}\ee\be
\s_{12}|_U~=~ -i\s_{03}|_U\ ,\qquad \s_{23}|_U~=~ -i\s_{01}|_U\ ,\qquad \s_{31}|_U~=~i\s_{02}|_U\ ,\label{D.2}\ee
with $\bar\s_{ab}|_U$ given by complex conjugates. Realizing the spin-frame as $u^+_\a = \left(\begin{array}{c} 1 \\ 0 \end{array}\right)$, $u^-_\a = \left(\begin{array}{c} 0 \\ -1 \end{array}\right)$, the van der Waerden symbols take the form
\begin{equation}
\left( \sigma ^{0}\right) _{\alpha \dot{\alpha}}=\left( 
\begin{array}{cc}
1 & 0 \\ 
0 & 1%
\end{array}
\right) \text{\ , \ }\left( \sigma ^{1}\right) _{\alpha \dot{\alpha}}=\left( 
\begin{array}{cc}
0 & 1 \\ 
1 & 0%
\end{array}
\right) \text{\ , \ }\left( \sigma ^{2}\right) _{\alpha \dot{\alpha}}=\left( 
\begin{array}{cc}
0 & -i \\ 
i & 0%
\end{array}
\right) \text{\ , \ }\left( \sigma ^{3}\right) _{\alpha \dot{\alpha}}=\left( 
\begin{array}{cc}
1 & 0 \\ 
0 & -1%
\end{array}
\right) \text{ ,}
\label{explicitvdw}
\end{equation}
with gamma matrices 
\begin{equation}
\left( \Gamma ^{a}\right) _{\underline{\alpha }}^{\ \ \underline{\beta }
}=\left( 
\begin{array}{cc}
0 & \left( \sigma ^{a}\right) _{\alpha }^{\ \ \dot{\beta}} \\ 
\left( \bar{\sigma}^{a}\right) _{\dot{\alpha}}^{\ \ \beta } & 0%
\end{array}
\right) \text{ ,}
\end{equation}
and
\begin{equation}
\left( \Gamma _{ab}\right) _{\underline{\alpha \beta }}=\left( 
\begin{array}{cc}
\left( \sigma _{ab}\right) _{\alpha \beta } & 0 \\ 
0 & \left( \bar{\sigma}_{ab}\right) _{\dot{\alpha}\dot{\beta}}%
\end{array}
\right) \text{ .}
\end{equation}
In particular, the $SO(3,2)$ generators that define the families of solutions \eq{families} are realized as
\begin{align}
E_{\underline{\alpha \beta }}\ & =\ -\left( \Gamma _{0}\right) _{\underline{
\alpha \beta }}=\left( 
\begin{array}{cccc}
0 & 0 & 1 & 0 \\ 
0 & 0 & 0 & 1 \\ 
1 & 0 & 0 & 0 \\ 
0 & 1 & 0 & 0%
\end{array}
\right) \text{ \ ,\ \ \ }J_{\underline{\alpha \beta }}\ =\ -\left( \Gamma
_{12}\right) _{\underline{\alpha \beta }}=\left( 
\begin{array}{cccc}
0 & -i & 0 & 0 \\ 
-i & 0 & 0 & 0 \\ 
0 & 0 & 0 & i \\ 
0 & 0 & i & 0%
\end{array}
\right) \text{ \ ,}  \label{EJrealization} \\
B_{\underline{\alpha \beta }}\ & =\ -\left( \Gamma _{03}\right) _{\underline{
\alpha \beta }}=\left( 
\begin{array}{cccc}
0 & 1 & 0 & 0 \\ 
1 & 0 & 0 & 0 \\ 
0 & 0 & 0 & 1 \\ 
0 & 0 & 1 & 0%
\end{array}
\right) \text{ \ , \ \ }P_{\underline{\alpha \beta }}\ =\ -\left( \Gamma
_{1}\right) _{\underline{\alpha \beta }}=\left( 
\begin{array}{cccc}
0 & 0 & 0 & -1 \\ 
0 & 0 & -1 & 0 \\ 
0 & -1 & 0 & 0 \\ 
-1 & 0 & 0 & 0%
\end{array}
\right) \text{ \ .}  \label{BPrealization}
\end{align}

The $\mso(3,2)$-valued connection
 \be
  \O~:=~-i \left(\frac12 \omega^{ab} M_{ab}+e^a P_a\right) ~:=~ \frac1{2i}
 \left(\frac12 \omega^{\a\b}~y_\a \star y_\b
 +  e^{\a\dot\b}~y_\a \star {\bar y}_{\dot\b}+\frac12 \bar{\omega}^{\dot\a\dot\b}~{\bar y}_{\dot\a}\star {\bar y}_{\dot\b}\right)\
 ,\label{Omega}
 \ee
  \be
 \o^{\a\b}~=~ -\frac{1}{4}(\s_{ab})^{\a\b}~\o^{ab}\ , \qquad \omega_{ab}~=~\frac{1}{2}\left( (\s_{ab})^{\a\b} \o_{\a\b}+(\bar\s_{ab})^{\ad\bd} \bar\o_{\ad\bd}\right)\ ,\ee
 \be e^{\a\dot\a}~=~ \frac{\lambda}{2}(\s_{a})^{\a \dot\a}~e^{a}\ , \qquad e_a~=~ -\l^{-1} (\s_a)^{\a\ad} e_{\a\ad}\ ,\label{convert}\ee
and field strength
\begin{align} {\cal R}~&:=~ d\O+\O\star \O~:=~-i \left(\frac12 {\cal R}^{ab}M_{ab}+{\cal R}^a P_a\right)\\ ~&:=~ \frac1{2i}
 \left(\frac12 {\cal R}^{\a\b}~y_\a \star y_\b
 +  {\cal R}^{\a\dot\b}~y_\a \star {\bar y}_{\dot\b}+\frac12 \bar{\cal R}^{\dot\a\dot\b}~{\bar y}_{\dot\a}\star {\bar y}_{\dot\b}\right)\
 ,\label{calRdef}\end{align}
\begin{align}
 {\cal R}^{\a\b}\ &=\ -\frac{1}{4}(\s_{ab})^{\a\b}~{\cal R}^{ab}\ ,
 \qquad {\cal R}_{ab}~=~\frac{1}{2}\left( (\s_{ab})^{\a\b} {\cal R}_{\a\b}+(\bar\s_{ab})^{\ad\bd} \bar{\cal R}_{\ad\bd}\right)\ ,\\
 {\cal R}^{\a\dot\a}\ &=\ \frac{\lambda}{2}(\s_{a})^{\a \dot\a}~{\cal R}^{a}\ ,
 \qquad {\cal R}_a~=~ -\l^{-1} (\s_a)^{\a\ad} {\cal R}_{\a\ad}\ .\end{align}
In these conventions, it follows that
 \be
 {\cal R}_{\a\b}~=~ d\o_{\a\b} -\o_{\a}{}^{\g}\wedge\o_{\g\b}-
 e_{\a}{}^{\cd}\wedge\bar e_{\cd\b}\ ,\qquad
 {\cal R}_{\a\dot\b}~=~  de_{\a\bd}+ \o_{\a\g}\wedge
 e^{\g}{}_{\bd}+\bar{\o}_{\bd\dd}\wedge e_{\a}{}^{\dd}\
 ,\ee\be
 {\cal R}^{ab}~=~ R_{ab}+\lambda^2
 e^a\wedge e^b\ ,\qquad R_{ab}~:=~d\o^{ab}+\o^a{}_c\wedge\o^{cb}\ ,\ee\be
 {\cal R}^a~=~ T^a ~:=~d e^a+\o^a{}_b\wedge e^b\ ,
 \label{curvcomp} \ee
where $R_{ab}:=\frac12 e^c e^d R_{cd,ab}$ and $T_a:=e^b e^c T^a_{bc}$ are the Riemann and torsion two-forms.
The metric $g_{\mu\nu}:=e^a_\mu e^b_{\nu}\eta_{ab}$. The $AdS_4$ vacuum solution $\O_{(0)}=e_{(0)}+\o_{(0)}$ obeying $d\O_{(0)}+\O_{(0)}\star\O_{(0)}=0$, with Riemann tensor $ R_{(0)\m\n,\rho\s}=
 -\lambda^2 \left( g_{(0)\mu\rho} g_{(0)\nu\sigma}-
  g_{(0)\nu\rho} g_{(0)\mu\sigma} \right)$ and vanishing torsion, can be expressed as $\O_{(0)}=L^{-1}\star dL$ where the gauge function $L\in SO(3,2)/SO(3,1)$. The stereographic coordinates $x^a$ are related to the coordinates $X^A$ of the five-dimensional embedding space with metric
$ds^2  =dX^A dX^B\eta_{AB}$,
in which $AdS_4$ is embedded as the hyperboloid
$X^A X^B \eta_{AB}=  -\frac{1}{\l^2} = -l^2$,
as
\bea &  \displaystyle  x^a \ = \ \frac{X^a}{1+\sqrt{1+\l^2 X^a X_a}} \ = \ \frac{X^{a}}{1+l^{-1}X^{0^{\prime }}}\text{ \ \ for \ \ }X^{0^{\prime
}}>0 & \text{ ,} \\
& \displaystyle X^a \ = \ \frac{2x^a}{1-\l^2 x^2}\ , \qquad a \ = \ 0,1,2,3 & \ .\label{stereo-embed}\eea
%

The familiar global spherical coordinates $(t,r,\theta,\phi)$ in which the metric reads
\bea  ds^2 \ = \ -(1+\l^2r^2)dt^2+\frac{dr^2}{1+\l^2 r^2}+r^2(d\theta^2+\sin^2\theta d\phi^2) \ ,\label{metricglob}\eea
are related locally to the embedding coordinates by
\bea & X_0 \ = \ \sqrt{\l^{-2}+r^2}\sin t \ , \qquad X_{0'} \ = \ \sqrt{\l^{-2}+r^2}\cos t \ , & \nn\\[5pt]
& X_1 \ = \ r\sin\theta\cos\phi \ , \quad  X_2 \ = \ r\sin\theta\sin\phi \ , \quad X_3 \ = \ r\cos\theta \ ,& \eea
providing a one-to-one map if $t\in [0,2\pi)$, $r\in[0,\infty)$, $\theta\in[0,\pi]$ and $\phi\in[0,2\pi)$ defining the single cover of $AdS_4$. This manifold can be covered by two sets of stereographic coordinates, $x^\mu_{(i)}$, $i=N,S$, related by the inversion $x^\m_N = -x^\m_S/(\l x_S)^2$ in the overlap region $\l^2 (x_N)^2, \l^2 (x_S)^2  <  0$, and the transition function $T_N^S=(L_N)^{-1}\star L_S\in SO(3,1)$. The map $x^\mu \rightarrow -x^\m/(\l x)^2$ leaves the metric invariant, maps the future and past time-like cones into themselves and exchanges the two space-like regions $0<\l^2 x^2< 1$ and $\l^2 x^2 > 1$ while leaving the boundary $\l^2 x^2 =1$ fixed. It follows that the single cover of $AdS_4$ is formally covered by taking $x^\mu\in \Real^{3,1}$.

\noindent For simplicity, we set $l=\l^{-1}=1$ in the body of the paper.

\section{Finite transformations of the Cartan generators}\label{Sec transformations}

In this appendix, we use a simple calculation to illustrate in the star-product language the finite transformations corresponding to the Cartan generators. 
We investigate the simple example:
\begin{equation}
\gamma ^{-1}\star Y^{\underline{\alpha }}\star \gamma \text{ ,}
\end{equation}%
where 
\begin{equation}
\gamma \left( \varphi K\right) \ =\ e_{\star }^{-\frac{1}{4}\varphi K_{%
\underline{\alpha \beta }}Y^{\underline{\alpha }}Y^{\underline{\beta }}}\ =\ 
\mathrm{sech}^{2}\left( \frac{\varphi }{2}\right) e^{-\frac{1}{2}\tanh
\left( \frac{\varphi }{2}\right) K_{\underline{\alpha \beta }}Y^{\underline{%
\alpha }}Y^{\underline{\beta }}}\text{ ,}
\end{equation}%
and its star-inverse 
\begin{equation}
\gamma ^{-1}\left( \varphi K\right) \ =\ e_{\star }^{\frac{1}{4}\varphi K_{%
\underline{\alpha \beta }}Y^{\underline{\alpha }}Y^{\underline{\beta }}}\ =\ 
\mathrm{sech}^{2}\left( \frac{\varphi }{2}\right) e^{\frac{1}{2}\tanh \left( 
\frac{\varphi }{2}\right) K_{\underline{\alpha \beta }}Y^{\underline{\alpha }%
}Y^{\underline{\beta }}}\text{ .}
\end{equation}%
$\varphi $ is any real or imaginary number. Using the property $K_{%
\underline{\alpha \beta }}K^{\underline{\beta \gamma }}=\delta _{\underline{%
\alpha }}{}^{\underline{\gamma }}$, we can derive 
\begin{equation}
\gamma ^{-1}\left( \varphi K\right) \star Y^{\underline{\alpha }}\star
\gamma \left( \varphi K\right) =\left[ \cosh \left( \varphi \right) \delta _{%
\underline{\beta }}{}^{\underline{\alpha }}-i\ \sinh \left( \varphi \right)
K_{\underline{\beta }}{}^{\underline{\alpha }}\right] Y^{\underline{\beta }}%
\text{ .}
\end{equation}%
If we replace $\varphi $ with $i\varphi $, we obtain 
\begin{equation}
\gamma ^{-1}\left( i\varphi K\right) \star Y^{\underline{\alpha }}\star
\gamma \left( i\varphi K\right) =\left[ \cos \left( \varphi \right) \delta _{%
\underline{\beta }}{}^{\underline{\alpha }}+\sin \left( \varphi \right) K_{%
\underline{\beta }}{}^{\underline{\alpha }}\right] Y^{\underline{\beta }}%
\text{ .}
\end{equation}%
From the above formulas we can see that, for $\varphi \in \mathbb{R}$, $%
\gamma \left( i\varphi E\right) $ and $\gamma \left( i\varphi J\right) $ are
periodic transformations, and $\gamma \left( i\varphi B\right) $ and $\gamma
\left( i\varphi P\right) $ are non-periodic transformations, which well-correspond to their (non-)compact nature that we expect from $AdS_4$ isometries.


\section{Further comments on the eigenfunctions \label{Secfurthereigen}}


In this Appendix we shall show explicitly how the eigenfunctions (\ref%
{solnint}) arise starting from the integral presentation of the projectors \eq{projF}. For the latter, i.e., for the case $\lambda _{L}=\lambda _{R}$ and $\lambda
_{L,R}+\frac{1}{2}\in \mathbb{Z}^{+}$, it was established in \cite{Iazeolla:2011cb} that different eigenfunctions are related by 
creation and annihilation operators, with $f_{\frac{1}{2},\frac{1}{2}}\left(
a^{+},a^{-}\right) $ being the ground state, i.e.\ $a^{+}\star f_{\frac{1}{2},%
\frac{1}{2}}\left( a^{+},a^{-}\right) =f_{\frac{1}{2},\frac{1}{2}}\left(
a^{+},a^{-}\right) \star a^{+}=0$. Moreover, diagonal elements with different half-integer eigenvalues are orthogonal with respect to the star product, and form an associative algebra (which can be extended by the corresponding twisted sector, see \cite{Iazeolla:2017vng}).

Things are much more complicated, however, for general complex eigenvalues.
In this paper we have not yet constructed a well-defined quantum system, and in particular we shall defer to a forthcoming paper the study of their algebraic properties under star product \cite{paper1prime}.
Below we will only qualitatively show that different eigenfunctions can be
brought from one to another by using creation and annihilation operators
with complex powers, which can be in their turn realized by means of the integral transform \eq{apowerlambda}.

We will first show that, starting from an eigenfunction with equal
eigenvalues $\lambda _{L}+\frac{1}{2}=\lambda _{R}+\frac{1}{2}\in \mathbb{Z}%
^{+}$, and by acting the creation operator on the left, we obtain an
eigenfunction with a different left eigenvalue $\lambda _{L}\in \mathbb{C}$,
i.e.\ 
\begin{equation}
f_{\lambda _{L},\lambda _{R}}\left( a^{+},a^{-}\right) \propto \left(
a_{+}\right) ^{\lambda _{L}-\lambda _{R}}\star f_{\lambda _{R},\lambda
_{R}}\left( a^{+},a^{-}\right) \text{ .}  \label{eigenrelation}
\end{equation}

To show that the r.h.s.\ of (\ref{eigenrelation}) produces the eigenfunction
with $\lambda _{L}$, we use the results from Section \ref{Sec
differenteigenvalues}. We substitute (\ref{apowerlambda}) and 
\begin{equation}
f_{\lambda _{R},\lambda _{R}}\left( a^{+},a^{-}\right) \propto \oint_{C(1)}%
\frac{d\varsigma }{2\pi i}\frac{\left(
\varsigma+1 \right) ^{\lambda _{R}-\frac{1}{2}}}{\left(\varsigma -1\right)
^{\lambda _{R}+\frac{1}{2}}}\ e^{-2\varsigma a^{+}a^{-}}
\end{equation}%
into the r.h.s.\ and evaluate the star-product between the $Y$-dependent factors of the integrands: 
\begin{eqnarray}
e^{-\tau a^{+}}\star e^{-2\varsigma a^{+}a^{-}}  
\ = \ e^{-\tau \left( 1+\varsigma \right) a^{+}}e^{-2\varsigma a^{+}a^{-}}\text{
.}
\end{eqnarray}%
Using this result, we have%
\begin{eqnarray}
&&\left( a^{+}\right) ^{\lambda _{L}-\lambda _{R}}\star f_{\lambda
_{R},\lambda _{R}}\left( a^{+},a^{-}\right)   \notag \\
&\propto &\oint_{C(1)}\frac{d\varsigma }{2\pi i}\frac{%
\left(\varsigma+1 \right) ^{\lambda _{R}-\frac{1}{2}}}{\left( \varsigma-1
\right) ^{\lambda _{R}+\frac{1}{2}}}\,e^{-2\varsigma a^{+}a^{-}} \int_{0}^{+\infty }d\tau \frac{\tau
^{\lambda _{R}-\lambda _{L}-1}}{\Gamma \left( \lambda _{R}-\lambda
_{L}\right) }\ e^{-\tau \left( 1+\varsigma \right) a^{+}}  \notag \\
&=&\oint_{C(1)}\frac{d\varsigma }{2\pi i}\frac{\left(
\varsigma +1\right) ^{\lambda _{R}-\frac{1}{2}}}{\left( \varsigma -1\right)
^{\lambda _{R}+\frac{1}{2}}}\,e^{2\varsigma a^{+}a^{-}} \left[ \left( 1+\varsigma \right) a^{+}\right]
^{\lambda _{L}-\lambda _{R}}  \notag \\
& = &\left( a^{+}\right) ^{\lambda _{L}-\lambda _{R}}\oint_{C(1)}\frac{%
d\varsigma }{2\pi i}\frac{\left( \varsigma+1
\right) ^{\lambda _{L}-\frac{1}{2}}}{\left( \varsigma-1 \right) ^{\lambda
_{R}+\frac{1}{2}}} \,e^{-2\varsigma a^{+}a^{-}} \notag \\
&\propto &f_{\lambda _{L},\lambda _{R}}\left( a^{+},a^{-}\right) \text{ .}
\end{eqnarray}%

On the other hand, following the discussion around (\ref{exchange}), if
we assume that $\lambda _{L}+\frac{1}{2}\in \mathbb{Z}^{+}$ always holds, we
can similarly derive the relation between different complex right eigenvalues:%
\begin{eqnarray}
&&f_{\lambda _{L},\lambda _{L}}\left( a^{+},a_{-}\right) \star \left(
a^{-}\right) ^{\lambda _{R}-\lambda _{L}}  \notag \\
&\propto &\left( a^{-}\right) ^{\lambda _{R}-\lambda _{L}}\oint_{C(1)}\frac{%
d\varsigma }{2\pi i}\frac{\left( \varsigma+1
\right) ^{\lambda _{R}-\frac{1}{2}}}{\left(\varsigma -1\right) ^{\lambda
_{L}+\frac{1}{2}}}\,e^{-2\varsigma a^{+}a^{-}}  \notag \\
&\propto &f_{\lambda _{L},\lambda _{R}}\left( a^{+},a^{-}\right) \ .
\end{eqnarray}

The above discussion is still far from a systematic study to build up a
quantum system. In particular, as stressed in Section \ref{Sec:regpres}, allowing both left and right eigenvalues to take complex values  involves alternative choices of 
contour other than the small circle around $\pm 1$. Moreover, such different contour-integral presentation has also the advantage of giving well-defined star-product properties between eigenfunctions with arbitrary complex eigenvalues.  This goes beyond the
scope of this paper, and we will continue our report on this issue in a forthcoming
work \cite{paper1prime}.

\section{Analysis of membrane-like curvature singularities}\label{Sec:singularity}

In this Appendix we shall study the limit $\D^2\to 0$, corresponding to an analytic singularity for every individual fluctuation fields extracted from the generating function \eq{PhiL}, at the level of Weyl zero-form master field --- that is, in terms of the behaviour of the latter in the full $(x,Y)$-space.\\
\noindent It is instructive to first study the limit in the diagonal case, i.e., for $\l_{iL}=\l_{iR}=\l\in\mathbb{Z}-1/2 $, $i=1,2$, that is (see Eq. \eq{diaglim}) for elements \eq{Phicov} of the form
\bea f^L_{\l,\l;\l,\l}\star \k_y  &= &  {\cal O}_{\l,\l}^{\varsigma_1} {\cal O}_{\l,\l}^{\varsigma_2} \frac{1}{\sqrt{(\check \vark^L)^2}}\, \exp\left[-\frac1{2}\ty^L(\check \vark^L)^{-1}\ty^L+\frac{1}2\yb\check \varkb^L\yb \right] \ , \label{Phicovdiag}\eea
where 
\be  {\cal O}_{\l,\l}^{\varsigma_i}  \ := \ \oint_{C(\e)}\frac{ d\varsigma_i}{2\pi i }\frac{(\varsigma_i+1)^{\l-\frac12}}{ (\varsigma_i-1)^{\l+\frac12}} \ ,\label{O11}\ee
with  $\e={\rm sign}(\l)$, and where the quantities in the integrand were defined in Section \ref{Sec zeroform-L}.
We shall also restrict our discussion to the simplest non-trivial choice of eigenvalues \eq{eigenchoice}, that in the diagonal limit $n=0$ reduces to studying the Weyl zero-form resulting from the lowest-weight element $\l_{iL}=\l_{iR}=\l=\frac{1}{2}$ only. In this case the contours encircle the points $\varsigma_1=1=\varsigma_2$, and as a consequence $\check \vark^L_{\a\b}$ and $\check v^L_{\a\bd}$ \eq{chvarkL}-\eq{chvL} reduce to $\vark^L_{(iB)\a\b}$  and $v^L_{(iB)\a\bd}$. As the dependence on $iP$ disappears, modulo a redefinition of the normalization factor (which we shall ignore here) we can simplify the notation by substituting the two contour integrals with a single one \cite{Iazeolla:2011cb}, 
\bea f^L_{\frac{1}{2},\frac{1}{2};\frac{1}{2},\frac{1}{2}}\star \k_y  &= &  {\cal O}_{1;1}^{\varsigma}
 \frac{1}{\varsigma\sqrt{( \vark^L)^2}}\, \exp\left[-\frac{1}{2\varsigma}\ty^L (\vark^L)^{-1} \ty^L+\frac{\varsigma}{2}\yb\varkb^L\yb \right] \ , \label{Phicovdiag1}
 \eea
where now $\ty^L:=y-i\varsigma v^L\yb$,
\be   {\cal O}_{1;1}^{\varsigma} \ := \ \oint_{C(1)}\frac{ d\varsigma}{2\pi i }\frac{\varsigma+1}{ \varsigma-1} \ ,\ee
and for notational simplicity we are now omitting the label $(iB)$, which is henceforth understood everywhere unless specified otherwise. We recall that, as discussed in Section \ref{Sec differenteigenvalues}, keeping the contour integral is part of the regular presentation of our eigenfunctions: we evaluated them in Section \ref{Sec:scalar} purely for the purpose of looking at the spacetime dependence of the component fields, but whenever the behaviour on the non-commutative $Y$ space is of relevance, as will be in the interpretation of the $\D^2\to 0$ limit, we should keep them as they are an integral part of the definition of the eigenfunctions from the point of view of their algebraic behaviour. 

In the $\D^2\to 0$ limit $\vark^L_{\a\b}$ and $\bar \vark^L_{\ad\bd}$ become degenerate,  and, as we shall now show,  the integrand takes the form of a delta-sequence 
\be \lim_{\e\to 0}\frac1{\e} e^{\frac{i}{2\e}\hat y D\hat y}\ = \ 2 \pi \delta^2(\hat{y}) \ ,\label{delta0}
\ee
where $\hat y_\a := (y+S\yb)_\a$, with $S_{\a\bd}$ a van der Waerden symbol coming from the $\D^2\to 0$ limit of $-i\varsigma v^L_{\a\bd}$, and $D_{\a\b}:=b^+_\a b_\b^-+b^-_\a b_\b^+$, in terms of an $x$-dependent spin-frame that we shall now introduce.

To study the limit precisely, it is convenient to perform a local $SL(2,\Comp)$ transformation to rewrite  $\vark^L_{\a\b}$  and $v^L_{\a\bd}$ on a common ``adapted'' spin-frame $(b^{+\a},b^{-\a})$, $b^{+\a}b^{-}_{\a}=1$ on which $\varkappa_{\a\b}^L $ takes the canonical form 
\bea
\varkappa_{\a\b}^L & = & i\sqrt{1+X_3^2-X_0^2}(b^{+}_\a b^{-}_\b+b^{-}_\a b^{+}_\b) \ =: \  i\D D_{\a\b}\ .
\eea
This condition determines the matrix of the Lorentz transformation only up to a free complex parameter $w$ as
\bea b^{+}_{\a} & = &  \frac{\D-Q}{2\D w} \,u^+_\a + \frac{(X_3-X_0)(X_1+iX_2)}{2\D w (1-X_{0'})}\,u^-_\a \ , \label{sl2c1}\\
b^{-}_{\a} & = & - \frac{(X_3+X_0)(X_1-iX_2)}{(\D-Q) (1-X_{0'})}w\, u^+_\a+w\, u^-_\a \ , \label{sl2c2}
 \eea
where $Q:=1+\frac{X_3^2-X_0^2}{1-X_{0'}}$. 

Performing this local Lorentz transformation corresponds to choosing a basis of the tangent space in which all spin-$s$ Weyl tensors extracted from \eq{Phicovdiag1} are manifestly of Petrov-type D \cite{Iazeolla:2011cb}. Retracing the analysis of the spherically-symmetric higher-spin black holes of \cite{Iazeolla:2011cb,Iazeolla:2017vng}, it is convenient to use the free parameter $w$ in order to realize the Killing vector $v^L_{\a\bd}$ on the adapted spin-frame in a canonical form, i.e., in terms of a single van der Waerden symbol. However, a novel feature arises here from the fact that both the norm of the Killing vector field and the determinant of its Killing two-form are not positive-definite: $v^L$ is spacelike for $|X_3|>|X_0|$ and there $\det \vark^L$ is also positive;  while $v^L$ is timelike for $|X_0|>|X_3|$ and in this region $\det \vark^L$ is positive as long as $X^2_3 < X_0^2<X^2_3+1$ and negative when $X_0^2>X^2_3+1$. As a consequence, the specific canonical form for $v^L$ changes in these three different spacetime regions: in particular, one can take 
\bea & X_3^2>X_0^2 &: \qquad \qquad w \ = \ |w| \ = \ \sqrt{\frac{\D-Q}{2\D}}\left(\frac{X_3-X_0}{X_3+X_0}\right)^{1/4} \ ,\\
 && v^L_{\a\bd} \ = \ i\sqrt{X_3^2-X_0^2}(b^+_\a \bar b^+_{\bd}+b^-_\a \bar b^-_{\bd}) \ ,  \eea
\bea & X^2_3 < X_0^2<X^2_3+1 &: \qquad \qquad w \ = \ |w| \ = \ \sqrt{\frac{Q-\D}{2\D}}\left(\frac{X_0-X_3}{X_0+X_3}\right)^{1/4} \ ,\\
 && v^L_{\a\bd} \ = \ -\sqrt{X_3^2-X_0^2}(b^+_\a \bar b^+_{\bd}-b^-_\a \bar b^-_{\bd}) \ ,  \label{w2}\eea
\bea & X_0^2>X^2_3+1 &: \qquad \qquad w \ = \ e^{i\pi/4}\left( \frac{\D-Q}{\D+Q}\,\frac{X_1+iX_2}{X_1-iX_2}\right)^{1/4} \ ,\\
 && v^L_{\a\bd} \ = \ -\sqrt{X_3^2-X_0^2}(b^+_\a \bar b^-_{\bd}+b^-_\a \bar b^+_{\bd}) \ .  \label{w3}\eea

Of the above three regions, the relevant ones for the study of the singularity on $(\vark^L)^2=0$ are clearly $X^2_3 < X_0^2<X^2_3+1$ and $X_0^2>X^2_3+1 $. It is then easy to show that in the limit $1+X^2_3-X^2_0 \to 0$ the integrand in \eq{Phicovdiag1} becomes
\be
\frac{1}{\varsigma\sqrt{( \vark^L)^2}}\,\exp\left[-\frac1{2\varsigma}\ty^L (\vark^L)^{-1} \ty^L+\frac{\varsigma}2\yb\varkb^L\yb \right] \quad \xrightarrow[\D\to 0]{} \quad \lim_{\e\to 0}\frac1{\e} e^{\frac{i}{2\e} (y+S\yb)D(y+S\yb) }\ = \ 2 \pi \delta^2(\hat y) \ ,\label{delta1}
\ee
where
\bea
 \hat y \ := \ \lim_{\D\to 0} \ty^L_\a \ = \  (y+S\yb)_\a \ , \qquad S_{\a\bd} \ := \  \left\{\begin{array}{cc}    -\varsigma(\s_3)_{\a\bd} & \qquad {\rm for} \quad X^2_3 < X_0^2<X^2_3+1  \ , \\ 
 \varsigma(\s_1)_{\a\bd} & \qquad {\rm for} \quad X_0^2>X^2_3+1  \ . \end{array}\right. \label{yts}
\eea
where the realization of the van der Waerden symbols in terms of a spin-frame\footnote{We note that while the entries of the $SL(2,\Comp)$ matrix \eq{sl2c2} with $w$ given by \eq{w2}-\eq{w3} separately scale like $\D^{-1/2}$, its determinant remains finite and equal to $1$ everywhere, including in the limit $\D\to 0$. This implies that $b^{+\a}b^-_{\a}=1$ also for $\D=0$, i.e., that $b^\pm_{\a}$ give a good spin-frame everywhere, thus in particular enabling to split $\ty^L_\a$ into $\ty^{L\pm}=b^{\pm\a}\ty^L_\a$ components which remain non-commuting, in such a way that, in particular, $[\hat y^-,\hat y^+]_\star=2i(1-s^2)$, which is in turn crucial to defining a proper non-commutative two-dimensional delta function.} has been given in Appendix \ref{App:conv}. As a consequence, 
\begin{eqnarray}
\lim_{\D \to 0} f^L_{\ft12,\ft12;\ft12,\ft12} \star \k_y
& \propto & {\cal O}_{1;1}^{\varsigma}\,2\pi \delta^2(\hat{y}) \ ,\label{deltareg}
\end{eqnarray}
where $\hat{y}=\hat{y}(\varsigma)$ is given in \eq{yts}.

A number of observations are now in order. 
First, it is interesting to note that, differently from the spherically-symmetric black-hole-like solutions where this singular behaviour was first observed, in this case the singular, delta-sequence limit is not obtained at the unfolding point $x^a=0$ (a point on the horizon of the gravitational background, in this paper) where the master-field \eq{Phicovdiag1} is, instead, regular. This is because, as evident from the discussion above, such distributional behaviour that characterizes the curvature singularities at the master-field level is strictly connected to the points at which the Killing two-form is degenerate. The latter was strictly vanishing at the unfolding point for the black-hole solutions (and for all solutions based on a $\pi$-odd Cartan principal generator -- that is, on $E$ and $iP$ up to $SO(3,2)$ transformations) whereas the Killing two-form of the solutions studied in the present paper is clearly non-degenerate for $x^a=0$, and the only reason that it can have zeroes outside the horizon is due to the fact that the corresponding Killing vector has indefinite norm. 

As already concluded in \cite{Iazeolla:2011cb,Iazeolla:2017vng,Iazeolla:2017dxc} for the spherically-symmetric solutions, this delta-function-like limit indicates that, even though at  $(\vark^L)^2\to 0$ every Weyl tensor diverges, the Weyl zero-form remains well-defined at $X_0^2-X_3^2=1$ as an operator. Indeed, a delta function of noncommutative variables has well-defined star product composition properties (and, in fact, is part of the associative algebra to which the exact solutions studied in \cite{Iazeolla:2017vng,Aros:2017ror} belong). In this sense, thought of as a symbol for an element of a star product algebra, such a master field remains smooth in the $(\vark^L)^2\to 0$  limit.  

We stress that the integral presentation \eq{deltareg} is crucial to the interpretation of the resulting distribution as an associative algebra element, and, therefore, to the above interpretation of the Weyl zero-form in the $\D^2\to 0$ limit. In fact, using contour integrals to represent Fock-space endomorphisms by means of $\mso(2,3)$ enveloping-algebra elements (with the obvious prescription to take all star products before performing the contour integrals) is the core of the regular presentation scheme that was essential to the solution-building method presented in \cite{Iazeolla:2011cb,Iazeolla:2017vng,Aros:2017ror}). In the case at hand, performing the auxiliary contour integrals first would lead to the delta function $\delta^2(\hat y\left|_{\varsigma=1}\right.)$ which has divergent star product with itself as $\hat y\left|_{\varsigma=1}\right.$ are abelian oscillators. On the other hand, using the regular presentation ensures that the element \eq{deltareg} has good star product properties \cite{Iazeolla:2017vng,Aros:2017ror}. In this sense, the regular presentation can be thought of as way of regulating the star products of non-polynomial elements by introducing auxiliary, complex integration variables to achieve a sort of point-splitting procedure in ${\cal Y}$ space, that gets rid of divergent terms and keeps the finite part of the star products above in a way which is compatible with associativity (at least within the Fock-space projectors with identical left and right eigenvalues and its dual space, obtained via star-multiplication with $\k_y$) \cite{Iazeolla:2017vng}.

However, note that, differently from the cases treated in detail in \cite{Iazeolla:2011cb}, in this case the Weyl zero-form admits a delta-function limit of modified oscillators $\hat y_\a$ that are specific to each side of the surface of apparent singularity $(\vark^L)^2=0$, which would correspond to a discontinuity in the component fields at $(\vark^L)^2=0$ if the component field description would make any sense there. The Weyl zero-form stays anyway regular in the sense above as a master field, which is the only suitable description in the strong coupling region. 

This concludes the discussion of the diagonal limit.



 
 \vspace{0.5cm} 
 
Such a resolution of the curvature singularity that arises in the limit $(\vark^L)^2\to 0$ can be shown, in fact, to still take place when $\l_{iL}-\l_{iR}$ is non-vanishing. For definiteness, let us first focus in greater detail on the choice $\l_{1L}=\frac12+ip$, $\l_{2L}=\frac12-ip$, $\l_{1R}=\frac12=\l_{2R}$, which is the case treated in greater detail at the end of Section \ref{Sec:scalar}. The master field in \eq{Phicov}  reads in this case
\bea
&\displaystyle f^L_{\ft12+ip,\ft12;\ft12-ip,\ft12}\star\k_y \ = \ {\cal O}_{\ft12+ip,\ft12}^{\varsigma_1} {\cal O}_{\ft12-ip,\ft12}^{\varsigma_2} \frac{1}{\G(ip)\G(-ip)}\int_0^\infty\frac{d\t_1}{\t_1}\int_0^\infty\frac{d\t_2}{\t_2}\left(\frac{\t_2}{\t_1}\right)^{ip}&\nn\\ 
&\displaystyle \times \ \ \ \left[ \frac{1}{\varsigma\sqrt{(\vark^L)^2}}\, e^{-\frac1{2\varsigma}(\ty^L-i\theta^L)(\vark^L)^{-1}(\ty^L-i\theta^L)-\frac{\varsigma}2\yb\varkb^L\yb-\bar \theta^L\yb+O(\varsigma_2-\varsigma_1)}+O(\varsigma_2-\varsigma_1) \right]&\ , \label{limitfull0}  
\eea
with $\theta^L$ given in \eq{thetaL},  $\varsigma:=\frac{\varsigma_1+\varsigma_2}{2}$, and where we denote with $O(\varsigma_2-\varsigma_1)$ all the terms weighted by the combination $\varsigma_2-\varsigma_1$ (i.e., carrying the dependence on $iP$), that vanish once one evaluates the contour integrals. Indeed, such terms will have no effect on the result, since with the choice $\l_{1L}=\frac12+ip$, $\l_{2L}=\frac12-ip$, $\l_{1R}=\frac12=\l_{2R}$  the basic effect of the two contour integrations, featuring a simple pole, is just to set $\varsigma_1=\varsigma_2=1$. All relevant quantities are therefore projected onto the $iB$ sector and therefore, for the sake of brevity, we shall henceforth omit the evanescent terms altogether. 
 
Now, away from the surface $\D^2=0$ we can use the $SL(2,\Comp)$ transformation \eq{sl2c1}-\eq{sl2c2} to write all quantities on the adapted spin-frame $b^\pm_\a$. Let us first approach the limit from the region $X^2_3<X^2_0<X^2_3+1$. Inverting the transformation \eq{sl2c1}-\eq{sl2c2} with \eq{w2}, we get  
\bea
\theta^L_\a \ = \ \frac{1}{\sqrt{\D}} \left[(F(X) \t_1+G(X) \t_2)b^+_\a +(L(X)\t_1+K(X)\t_2)b^-_\a \right]\ , \label{U}
\eea
where $F,G,L,K$ are complex functions of the embedding coordinates given by 
\bea 
\displaystyle F & = & \frac{1}{4\sqrt{1-X_{0'}}}\left(\frac{X_0-X_3}{X_0+X_3}\right)^{1/4}\left\{(X_1-iX_2)\left[\sqrt{Q-\D}+\frac{X_0^2-X_3^2}{\sqrt{Q-\D}(1-X_{0'})}\right]\right. \nn\\
\displaystyle && \qquad \qquad\qquad \qquad  \left.- (1-X_{0'})\sqrt{Q-\D}\right\} \ , \\
G & = & \frac{-i}{4\sqrt{1-X_{0'}}}\left(\frac{X_0-X_3}{X_0+X_3}\right)^{1/4}\left\{(X_1-iX_2)\left[\sqrt{Q-\D}+\frac{X_0^2-X_3^2}{\sqrt{Q-\D}(1-X_{0'})}\right]\right. \nn\\
\displaystyle && \qquad \qquad\qquad \qquad  \left.+ (1-X_{0'})\sqrt{Q-\D}\right\} \ ,\\
L & = & \frac{X_0-X_3}{4\sqrt{1-X_{0'}}}\left(\frac{X_0+X_3}{X_0-X_3}\right)^{1/4}\left\{\frac{X_1+iX_2}{(1-X_{0'})\sqrt{Q-\D}}\left[X_1-iX_2-1+X_{0'}\right]\right. \nn\\
\displaystyle && \qquad \qquad\qquad \qquad  \left.+\sqrt{Q-\D}\right\} \ ,\\
K & = &  -i\frac{X_0-X_3}{4\sqrt{1-X_{0'}}}\left(\frac{X_0+X_3}{X_0-X_3}\right)^{1/4}\left\{\frac{X_1+iX_2}{(1-X_{0'})\sqrt{Q-\D}}\left[X_1-iX_2+1-X_{0'}\right]\right. \nn\\
\displaystyle && \qquad \qquad\qquad \qquad  \left.+\sqrt{Q-\D}\right\} \ .
 \eea
Substituting in \eq{limitfull0}, the master field takes the form
\bea   
f^L_{\ft12+ip,\ft12;\ft12-ip,\ft12}\star\k_y & = & {\cal O}_{\ft12+ip,\ft12}^{\varsigma_1} {\cal O}_{\ft12-ip,\ft12}^{\varsigma_2}\frac{1}{\G(ip)\G(-ip)}  \int_0^\infty\frac{d\t_1}{\t_1}\int_0^\infty\frac{d\t_2}{\t_2}\left(\frac{\t_2}{\t_1}\right)^{ip} \nn\\
&& \times \ \  \frac{1}{\varsigma\D} \exp\left[ {\frac{i}{2\varsigma\D}(\ty^L-i\theta^L)D(\ty^L-i\theta^L)-\frac{i\varsigma\D}2\yb\bar{D}\yb-\bar \theta^L\yb}\right] \  ,
\eea
and rescaling the integration variables of the Mellin transforms as $\t_i \to \t'_i:=\frac{\t_i}{\sqrt{\D}}$, and then omitting the primes on $\t_i$, we get 
\bea   
f^L_{\ft12+ip,\ft12;\ft12-ip,\ft12}\star\k_y & =  &{\cal O}_{\ft12+ip,\ft12}^{\varsigma_1} {\cal O}_{\ft12-ip,\ft12}^{\varsigma_2}\frac{1}{\G(ip)\G(-ip)}  \int_0^\infty\frac{d\t_1}{\t_1}\int_0^\infty\frac{d\t_2}{\t_2}\nn\\
& & \ \ \left(\frac{\t_2}{\t_1}\right)^{ip} \frac{1}{\varsigma\D} e^{\frac{i}{2\varsigma\D}(\ty^L-i\sqrt{\D}\theta^L)D(\ty^L-i\sqrt{\D}\theta^L)-\frac{i\varsigma\D}2\yb\bar{D}\yb-\sqrt{\D}\bar \theta^L\yb} \ .
\eea

Let us now take the limit $\D\to 0$. Then
\bea   
\lim_{\D\to 0} f^L_{\ft12+ip,\ft12;\ft12-ip,\ft12}\star\k_y & =  &   {\cal O}_{\ft12+ip,\ft12}^{\varsigma_1} {\cal O}_{\ft12-ip,\ft12}^{\varsigma_2}\frac{2\pi}{\G(ip)\G(-ip)}\nn\\
& \times & \int_0^\infty\frac{d\t_1}{\t_1}\int_0^\infty\frac{d\t_2}{\t_2}\left(\frac{\t_2}{\t_1}\right)^{ip}\,\d^2 (\hat y-i\theta^{\prime L}) e^{-\bar \theta^{\prime L}\yb}\ , \label{limitfull}
\eea
where $\theta^{\prime L}_\a$ is defined by
\bea
\theta^{\prime L}_\a & := & \left.\sqrt{\D}\theta^L_\a\right|_{\D= 0} \ \equiv \ \left[(F(X) \t_1+G(X) \t_2)b^+_\a +(L(X)\t_1+K(X)\t_2)b^-_\a \right]_{\D= 0} \nn\\
& = &  (f(X) \t_1+g(X) \t_2)b^+_\a +(l(X)\t_1+k(X)\t_2)b^-_\a  \ , 
\eea
with
\bea  
 & \displaystyle l \ = \ -\frac14 \sqrt{\frac{X_3-X_0}{X_{0'}}}(X_{0'}+X_1+iX_2) \ , \qquad  k \ = \ \frac{-i}4 \sqrt{\frac{X_3-X_0}{X_{0'}}}(-X_{0'}+X_1+iX_2) \ , & \\
\displaystyle & f \ = \ -l^\ast \ , \qquad g \ = \ -k^\ast \ , &
\eea
that is,
\bea
\theta^{\prime L}_\a \ = \  -(l^\ast(X) \t_1+k^\ast(X) \t_2)b^+_\a +(l(X)\t_1+k(X)\t_2)b^-_\a \ . 
\eea

Substituting in \eq{limitfull}, splitting the two-dimensional delta as ($\psi^\pm:=b^{\pm\a}\psi_\a$)
\bea  &\d^2 (\hat y-i\theta^{\prime L}) \ = \ \d(\hat y^{+}-i\theta^{\prime L+})\d(\hat y^{-} -i\theta^{\prime L-}) \ , & \nn\\
& \theta^{\prime L+}  \ = \ l\t_1+k\t_2 \ , \qquad  \theta^{\prime L-}  \ = \ l^\ast\t_1+k^\ast\t_2 \ = \ \bar \theta^{\prime L+} \ ,&\eea
the Weyl zero-form master field in the limit $\Delta=0$ takes the form
\bea   
&\displaystyle \lim_{\D\to 0} f^L_{\ft12+ip,\ft12;\ft12-ip,\ft12}\star\k_y \ = \   {\cal O}_{\ft12+ip,\ft12}^{\varsigma_1} {\cal O}_{\ft12-ip,\ft12}^{\varsigma_2}\, \frac{2\pi}{\G(ip)\G(-ip)}  \int_0^\infty\frac{d\t_1}{\t_1}\int_0^\infty\frac{d\t_2}{\t_2}&\nn \\ 
&\displaystyle \times\ \  \left(\frac{\t_2}{\t_1}\right)^{ip} \,\d(\hat y^{+}-i\theta^{\prime L+})\d(\hat y^{-} -i\theta^{\prime L-}) \,e^{(l \bar y^+-l^\ast \bar y^-)\t_1+(k \bar y^+-k^\ast \bar y^-)\t_2}&\  .
\eea
Representing the delta functions in Fourier transform,
\bea  
\d(\hat y^{\pm}-i\theta^{\prime L\pm}) \ = \ \int_{-\infty}^{+\infty}\frac{dw}{2\pi} \, e^{i(\hat y^{\pm}-i\theta^{\prime L \pm})w} \ ,
 \eea
we have
\bea   
& \displaystyle \lim_{\D\to 0} f^L_{\ft12+ip,\ft12;\ft12-ip,\ft12}\star\k_y \ = \  {\cal O}_{\ft12+ip,\ft12}^{\varsigma_1} {\cal O}_{\ft12-ip,\ft12}^{\varsigma_2} \,\frac{2\pi}{\G(ip)\G(-ip)}  \int_0^\infty\frac{d\t_1}{\t_1^{1+ip}}\int_0^\infty d\t_2\,\t_2^{ip-1}  & \nn\\
& \displaystyle \times\ \  \int_{-\infty}^{+\infty}\frac{dw}{2\pi} \int_{-\infty}^{+\infty}\frac{dw'}{2\pi} e^{i(\hat y^{+}-i(l\t_1+k\t_2))w+i(\hat y^{-}-i(l^\ast\t_1+k^\ast\t_2))w'}\,e^{(l \bar y^+-l^\ast \bar y^-)\t_1+(k \bar y^+-k^\ast \bar y^-)\t_2}& \  . 
\eea
The two $\tau_i$-integrals are now disentangled, and give
\bea 
\int_0^\infty d\t_1\,\t_1^{-ip-1} \,e^{[l(w+\bar y^+)+l^\ast(w'-\bar y^-)]\t_1} & = &  \G(-ip) [l(w+\bar y^+)+l^\ast(w'-\bar y^-)]^{ip} \ ,\\
\int_0^\infty d\t_2\,\t_2^{ip-1} \,e^{[k(w+\bar y^+)+k^\ast(w'-\bar y^-)]\t_2} & = & \frac{\G(ip)}{[k(w+\bar y^+)+k^\ast(w'-\bar y^-)]^{ip} } \ ,
 \eea
provided that ${\rm Re}[l(w+\bar y^+)+l^\ast(w'-\bar y^-)]<0$, ${\rm Re}[k(w+\bar y^+)+k^\ast(w'-\bar y^-)]<0$.
Then,
\bea   
&\displaystyle \lim_{\D\to 0} f^L_{\ft12+ip,\ft12;\ft12-ip,\ft12}\star\k_y \ = \ {\cal O}_{\ft12+ip,\ft12}^{\varsigma_1} {\cal O}_{\ft12-ip,\ft12}^{\varsigma_2}\, \frac1{2\pi }&\nn\\
&\displaystyle\times\ \   \int_{-\infty}^{+\infty}dw \int_{-\infty}^{+\infty}dw' 
\,e^{i\hat y^{+}w+i\hat y^{-}w'}\,\left[\frac{l(w+\bar y^+)+l^\ast(w'-\bar y^-)}{k(w+\bar y^+)+k^\ast(w'-\bar y^-)}\right]^{ip} &\ , \label{limitFourier}
\eea
which indeed reduces to the limit of the diagonal case \eq{deltareg} for $p= 0$, 
\bea \left. \lim_{\D\to 0} f^L_{\ft12+ip,\ft12;\ft12-ip,\ft12}\star\k_y\right|_{p=0} & = & {\cal O}_{\ft12+ip,\ft12}^{\varsigma_1} {\cal O}_{\ft12-ip,\ft12}^{\varsigma_2} \, \frac1{2\pi } \int_{-\infty}^{+\infty}dw \int_{-\infty}^{+\infty}dw' 
e^{i\hat y^+w+i\hat y^-w'} \nn\\
& = &  {\cal O}_{\ft12+ip,\ft12} {\cal O}_{\ft12-ip,\ft12} \, 2\pi\, \d^2(\hat y) \ . \eea
The $w'$-integral in \eq{limitFourier} can be evaluated by first simplifying the integrand using the fact that $lk^\ast-l^\ast k =0$ for $\Delta=0$,
\bea & \displaystyle\int_{-\infty}^{+\infty}dw' 
e^{i\hat y^-w'}\,\left[\frac{l(w+\bar y^+)+l^\ast(w'-\bar y^-)}{k(w+\bar y^+)+k^\ast(w'-\bar y^-)}\right]^{ip} & \nn\\
& \displaystyle = \ \int_{-\infty}^{+\infty}dw' 
e^{i\hat y^-w'}\,\frac1{(k^\ast)^{ip}}\left[\frac{l k^\ast (w+\bar y^+) -l^\ast k (w+\bar y^+)}{k(w+\bar y^+)+k^\ast(w'-\bar y^-)}+l^\ast \right]^{ip}  &\nn\\
& \displaystyle = \ 2\pi\left(\frac{l^\ast}{k^\ast}\right)^{ip}\d(\hat y^-)  \ .&
\eea
Substituting in \eq{limitFourier} and evaluating the $w$-integral, we finally get
\bea \lim_{\D\to 0} f^L_{\ft12+ip,\ft12;\ft12-ip,\ft12}\star\k_y& = &   {\cal O}_{\ft12+ip,\ft12}^{\varsigma_1} {\cal O}_{\ft12-ip,\ft12}^{\varsigma_2}\, 2\pi\,\left(\frac{l^\ast}{k^\ast}\right)^{ip} \d^2(\hat y)\nn\\ & = &  {\cal O}_{\ft12+ip,\ft12}^{\varsigma_1} {\cal O}_{\ft12-ip,\ft12}^{\varsigma_2} \, 2\pi\, \left[-i\frac{X_{0'}+X_1-iX_2}{X_{0'}-X_1+iX_2}\right]^{ip} \d^2(\hat y) \ . \eea

\vspace{1cm}

The same procedure can be repeated approaching the limit $\D^2\to 0$ ``from below'', i.e., from the region $X_0^2>X_3^2+1$. In this case one uses the transformation \eq{sl2c1}-\eq{sl2c2} with \eq{w3} to obtain 
\bea
\theta^L_\a \ = \ \frac{1}{\D} \left[(F(X) \t_1+G(X) \t_2)b^+_\a +(L(X)\t_1+K(X)\t_2)b^-_\a \right]\ , \label{U2}
\eea
with coefficients
\bea 
\displaystyle F & = & e^{i\pi/4} \D\left(\frac{\D-Q}{\D+Q}\,\frac{X_1+iX_2}{X_1-iX_2}\right)^{1/4}\left[(X_1-iX_2)\left(1-\frac{X^2_0-X^2_3}{(\D-Q)(1-X_{0'})}\right)\right. \nn\\ 
\displaystyle && \hspace{4cm} - (1-X_{0'})\Big] \ , \\
\displaystyle G & = & -ie^{i\pi/4}\D \left(\frac{\D-Q}{\D+Q}\,\frac{X_1+iX_2}{X_1-iX_2}\right)^{1/4}\left[(X_1-iX_2)\left(1-\frac{X^2_0-X^2_3}{(\D-Q)(1-X_{0'})}\right)\right. \nn\\ 
\displaystyle && \hspace{4cm} - (1-X_{0'})\Big] \ , \\
\displaystyle L & = & e^{-i\pi/4} \left(\frac{\D+Q}{\D-Q}\,\frac{X_1-iX_2}{X_1+iX_2}\right)^{1/4}\frac{X_0-X_3}{2}\left[\frac{X^2_1+X^2_2}{1-X_{0'}}-X_1-iX_2+Q-\D\right] \ , \\
\displaystyle K & = & -i e^{-i\pi/4} \left(\frac{\D+Q}{\D-Q}\,\frac{X_1-iX_2}{X_1+iX_2}\right)^{1/4}\frac{X_0-X_3}{2}\left[\frac{X^2_1+X^2_2}{1-X_{0'}}+X_1+iX_2+Q-\D\right]  \ ,
 \eea
to be substituted in 
\bea   
 f^L_{\ft12+ip,\ft12;\ft12-ip,\ft12}\star \k_y & = & {\cal O}_{\ft12+ip,\ft12}^{\varsigma_1} {\cal O}_{\ft12-ip,\ft12}^{\varsigma_2} \, \frac{1}{\G(ip)\G(-ip)}  \int_0^\infty\frac{d\t_1}{\t_1}\int_0^\infty\frac{d\t_2}{\t_2}\left(\frac{\t_2}{\t_1}\right)^{ip}\nn \\
 && \times \ \  \frac{1}{\varsigma\D} e^{\frac{i}{2\varsigma\D}(\ty^L-i\theta^L)D(\ty^L-i\theta^L)-\frac{i\varsigma\D}2\yb\bar{D}\yb-\bar \theta^L\yb} \  .
\eea
Again, to take the limit it is useful to rescale the integration variables as $\t_i \to \t'_i= \frac{\t_i}{\D}$, in terms of which (again omitting the primes after the change of variables) 
\bea   
f^L_{\ft12+ip,\ft12;\ft12-ip,\ft12}\star\k_y & = & {\cal O}_{\ft12+ip,\ft12}^{\varsigma_1} {\cal O}_{\ft12-ip,\ft12}^{\varsigma_2} \, \frac{1}{\G(ip)\G(-ip)}  \int_0^\infty\frac{d\t_1}{\t_1}\int_0^\infty\frac{d\t_2}{\t_2}\left(\frac{\t_2}{\t_1}\right)^{ip} \nn\\
&& \times \ \ \frac{1}{\varsigma\D} e^{\frac{i}{2\varsigma\D}(\ty^L-i\D \theta^L)D(\ty^L-i\D \theta^L)-\frac{i\varsigma\D}2\yb\bar{D}\yb-\D\bar \theta^L\yb} \ .
\eea
Let us now take the limit $\D\to 0$. Then
\bea   
\lim_{\D\to 0} f^L_{\ft12+ip,\ft12;\ft12-ip,\ft12}\star\k_y & = &  {\cal O}_{\ft12+ip,\ft12}^{\varsigma_1} {\cal O}_{\ft12-ip,\ft12}^{\varsigma_2} \,  \frac{2\pi}{\G(ip)\G(-ip)}\nn\\ 
& \times & \int_0^\infty\frac{d\t_1}{\t_1}\int_0^\infty\frac{d\t_2}{\t_2}\left(\frac{\t_2}{\t_1}\right)^{ip} \d^2 (\hat y-i\theta^{\prime L}) e^{-\bar \theta^{\prime L}\yb}\ , 
\eea
where $\theta^{\prime L}_\a$ 
\bea
\theta^{\prime L}_\a & := & \left.\D \theta^L_\a\right|_{\D= 0} \ \equiv \ \left[(F(X) \t_1+G(X) \t_2)b^+_\a +(L(X)\t_1+K(X)\t_2)b^-_\a \right]_{\D= 0} \nn\\
& = & [l(X)\t_1+k(X)\t_2]b^-_\a  
\eea
loses completely one component, compared to the previous case, with
\bea  \displaystyle  l & := & \left.L\right|_{\D=0} \ = \  \left(\frac{X_1-iX_2}{X_1+iX_2}\right)^{1/4}\frac{X_3-X_0}{2}\left(X_{0'}+X_1+iX_2\right) \ , \\
\displaystyle  k & := & \left.K\right|_{\D=0} \ = \  -i\left(\frac{X_1-iX_2}{X_1+iX_2}\right)^{1/4}\frac{X_3-X_0}{2}\left(X_{0'}-X_1-iX_2\right) \ .
  \eea
As a consequence, 
\bea   
\lim_{\D\to 0} f^L_{\ft12+ip,\ft12;\ft12-ip,\ft12}\star\k_y & = &   {\cal O}_{\ft12+ip,\ft12}^{\varsigma_1} {\cal O}_{\ft12-ip,\ft12}^{\varsigma_2} \, \frac{2\pi\d(\hat y^{-})}{\G(ip)\G(-ip)}\nn\\
&\times& \int_0^\infty d\t_1\, \t_1^{-ip-1}\int_0^\infty d\t_2\, \t_2^{ip-1}\d (\hat y^{+}-i\theta^{\prime L+}) e^{-\bar \theta^{\prime L+}\yb^{L-}}\ . \label{limitfull2}
\eea
Representing the delta function in Fourier transform,
\be \d (\hat y^+-iU'^+) \ = \  \int_{-\infty}^{+\infty}\frac{dw}{2\pi} \, e^{i(\hat y^+-i(l\t_1+k\t_2))w} \ ,
 \ee
one can rewrite
\bea   
\lim_{\D\to 0} f^L_{\ft12+ip,\ft12;\ft12-ip,\ft12}\star\k_y & = & {\cal O}_{\ft12+ip,\ft12}^{\varsigma_1} {\cal O}_{\ft12-ip,\ft12}^{\varsigma_2} \, \frac{2\pi\d(\hat y^-)}{\G(ip)\G(-ip)}\int_{-\infty}^{+\infty}\frac{dw}{2\pi}\,e^{i\hat y^+ w} \nn \\ 
&\times& \ \ \int_0^\infty d\t_1\,e^{(lw-l^\ast \bar y^-)\t_1} \t_1^{-ip-1}\int_0^\infty d\t_2\, \t_2^{ip-1} \,e^{(kw-k^\ast\yb^-)\t_2}\ . 
\eea
Again, under the condition that ${\rm Re}(lw-l^\ast \bar y^-)<0$, ${\rm Re}(kw-k^\ast\yb^-)<0$, one can perform the two $\tau_i$ integrals to get
\bea   
\lim_{\D\to 0} f^L_{\ft12+ip,\ft12;\ft12-ip,\ft12}\star\k_y \ = \  {\cal O}_{\ft12+ip,\ft12}^{\varsigma_1} {\cal O}_{\ft12-ip,\ft12}^{\varsigma_2} \,2\pi\,\d(\hat y^-)\int_{-\infty}^{+\infty}\frac{dw}{2\pi}\,e^{i\hat y^+ w} \left(\frac{lw-l^\ast \bar y^-}{kw-k^\ast\yb^-}\right)^{ip}\ . 
\eea
As before, one can show that the $w$ dependence drops out entirely from the rational function in the integrand, due to $lk^\ast-l^\ast k=0$, in such a way that 
\bea   
\lim_{\D\to 0} f^L_{\ft12+ip,\ft12;\ft12-ip,\ft12}\star\k_y & = &   {\cal O}_{\ft12+ip,\ft12}^{\varsigma_1} {\cal O}_{\ft12-ip,\ft12}^{\varsigma_2}\,2\pi\,\left(\frac{l}{k}\right)^{ip}\d^2(\hat y) \nn \\   & = &   {\cal O}_{\ft12+ip,\ft12} {\cal O}_{\ft12-ip,\ft12} \, 2\pi\,\left(i\frac{X_{0'}+X_1+iX_2}{X_{0'}-X_1-iX_2}\right)^{ip}\d^2(\hat y)  \ . 
\eea

\vspace{0.5cm}

Finally, let us comment on the apparent singularity at $\D^2=0$ for more general eigenvalues. The most general element in the twisted sector that satisfies the conditions \eq{ident}-\eq{rebos} is (see Eq. \eq{fcond})
\bea f^L_{\boldsymbol{\l}} \star \k_y
& = & \frac{1}{\Gamma \left(-q-ip\right) \Gamma \left(-q+ip\right)} \,{\cal O}_{\boldsymbol{\lambda }_{1}}^{\varsigma_1}{\cal O}_{\boldsymbol{\lambda }_{2}}^{\varsigma_2}\nn\\
&  \times& \ \ \int_{0}^{+\infty }d\tau _{1} \int_{0}^{+\infty }d\tau _{2} \frac{1}{(\tau _{1}\tau _{2})^{q+1}}\left(\frac{\t_2}{\t_1}\right)^{ip}
 \notag \\
&  \times& \ \ \frac{1}{\sqrt{(\check \vark^L)^2}}\, \exp\left[-\frac1{2}(\ty^L-i\theta^L)(\check \vark^L)^{-1}(\ty^L-i\theta^L)+\frac{1}2\yb\check \varkb^L\yb-\bar \theta^L\yb\right] \ , 
\label{Psikappagen}
\eea
with $q:=  {\rm Re}(\D\lambda)\in \mathbb{Z}$, and it may have a membrane-like curvature singularity at $\D^2=0$ for our choice of $\l_{iR}+\frac{1}{2}\in\mathbb{Z}^+$.  Higher $\l_{iR}$ will in general increase the order of the pole in $\D^2=0$, and we shall defer a full analysis of the general case to future work, focussing here on elements with the lowest right eigenvalue $\l_{1R}=\l_{2R}=\frac{1}{2} $ . In such case, again omitting evanescent terms $O(\varsigma_2-\varsigma_1)$ and studying the $\D^2 \to 0$ limit from above, one is reduced to the expression 
\bea   
\lim_{\D\to 0}f^L_{\boldsymbol{\l}}\star \k_y & = & \frac{1}{\G(-q-ip)\G(-q+ip)}{\cal O}_{\boldsymbol{\lambda }_{1}}^{\varsigma_1}{\cal O}_{\boldsymbol{\lambda }_{2}}^{\varsigma_2}\lim_{\D\to 0} \int_{0}^{+\infty }d\tau _{1} \int_{0}^{+\infty }d\tau _{2} \frac{1}{(\tau _{1}\tau _{2})^{q+1}}\left(\frac{\t_2}{\t_1}\right)^{ip}\nn \\ 
&& \ \  \frac{1}{\varsigma\D}\, e^{\frac{i}{2\varsigma\D}(\ty^L-i\theta^L)D(\ty^L-i\theta^L)-\frac{i\varsigma\D}2\yb\bar{D}\yb-\bar \theta^L\yb} \  ,
\eea
and again rescaling the integration variables of the Mellin transforms as $\t_i \to \t'_i:=\frac{\t_i}{\sqrt{\D}}$, and then omitting the primes on $\t_i$, we get 
\bea   
\lim_{\D\to 0}f^L\star \k_y & = &  \frac{1}{\G(-q-ip)\G(-q+ip)}{\cal O}_{\boldsymbol{\lambda }_{1}}^{\varsigma_1}{\cal O}_{\boldsymbol{\lambda }_{2}}^{\varsigma_2}\lim_{\D\to 0} \int_{0}^{+\infty }d\tau _{1} \int_{0}^{+\infty }d\tau _{2} \frac{1}{(\tau _{1}\tau _{2})^{q+1}}\left(\frac{\t_2}{\t_1}\right)^{ip}\nn \\ 
&& \ \  \frac{1}{\varsigma\D^{1+q}}\, e^{\frac{i}{2\varsigma\D}(\ty^L-i\theta^L)D(\ty^L-i\theta^L)-\frac{i\varsigma\D}2\yb\bar{D}\yb-\sqrt{\D}\bar \theta^L\yb} \  .
\eea
Clearly, in the case that $q<0$ we end up with a master fields than is more regular that the $q=0$ case studied above. On the other hand, the higher powers of $\D$ that appear at the denominator for $q>0$ can be interpreted as giving rise to derivatives of a delta function. The latter can however still be considered part of an associative algebra, in the sense that they admit a star-factorization in terms of delta functions, as $Y$-derivatives of $\d^2(\hat y)$ can be rewritten as (linear combinations of) star products of the type $y_{\a} \star \d^2(\hat y)$ and $\bar y_{\ad} \star \d^2(\hat y)$.


\section{Removing the ambiguity between regular and twisted sector}\label{App:trouble}


In this Appendix we briefly explain in what way the choice of regular presentation we have made resolves the ambiguity of the choice of regular and twisted sector in the expansion of the Weyl zero-form. As mentioned in Section \ref{Section:recallproj}, regular and twisted sector, for the family ${\cal M}(iB;iP)$ that we are working with, are completely equivalent. However, as we shall see, the small-contour integral presentation \eq{solnint} is only suitable for the twisted sector in the expansion of $\Phi'$ (i.e., regular sector in the expansion of $\Psi$), in the sense it cannot provide an unambiguous realization of the fluctuation fields  in the regular sector everywhere in spacetime, thereby violating criterion iii) of our regular presentation scheme given in Section \ref{Sec:HSBTZ}. This is why we discarded the regular sector (see Sections \ref{Section:recallproj} and \ref{Sec gravitational notation}) in $\Phi'$ from the analysis of the present paper. 

$L$-rotating a generic element $f_{\boldsymbol{\l}}$ in an expansion $\Phi'=\sum_{\boldsymbol{\l}}\mu_{\boldsymbol{\l}}f_{\boldsymbol{\l}}$ over the regular sector results in a Weyl zero-form master field $\Phi^{(L)}$ expanded over
\be f^{L}_{\boldsymbol{\l}} \ = \ L^{-1}\star f_{\boldsymbol{\l}}\star\pi(L) \ = \ L^{-1}\star f_{\boldsymbol{\l}}\star \k_y\star L\star\k_y \ = \ (f_{\boldsymbol{\l}}\star \k_y)^{L} \star \k_y \ .\ee
Let us for simplicity consider first the diagonal case, $\l_{iL}-\l_{iR} =0$ and in particular let us focus on the lowest-weight state $\l_{1L}=\l_{2L}=\frac{1}{2}=\l_{1R}=\l_{2R} $. Then,
\be f^L_{\ft12,\ft12;\ft12,\ft12} \ = \ {\cal O}_{1;1}^{\varsigma}\,e^{\frac{\varsigma}{2}(y\vark y+\yb \varkb \yb)} \ , \ee
where ${\cal O}_{1;1}^{\varsigma}$ was defined in \eq{O11}, and 
\be f^L_{\ft12,\ft12;\ft12,\ft12}  \star \k_y \ = \ {\cal O}_{1;1}^{\varsigma}\,\frac{1}{\varsigma\sqrt{\vark^2}}\,e^{-\frac{1}{2\varsigma}y\vark^{-1} y+\frac{\varsigma}{2}\yb \varkb \yb} \ .  \ee
As $\vark^2=1$ and $\vark^{-1}=-\vark$, one can then write the exponent as $\frac12 Y \breve{K} Y$, with 
\bea \breve{K}  =  \left( \begin{array}{cc} \breve{\vark} & 0 \\ 
0 & \breve{\varkb} \end{array} \right) := 
\left( \begin{array}{cc} \frac{\vark}{\varsigma} & 0  \\  0 & \varsigma\varkb\end{array}\right)  \ ,\eea 
and repeat the steps above to obtain 
\be \Phi^{(L)} \ = \  (f^L_{\ft12,\ft12;\ft12,\ft12}\star \k_y)^{L} \star \k_y  \ \propto \ \oint_{C(1)}\frac{d\varsigma}{2\pi i \varsigma}\,\frac{\varsigma+1}{\varsigma-1} \,\frac{1}{\sqrt{(\breve \vark^L)^2}}\,e^{\frac{\varsigma}{2}\yb \varkb \yb-\frac{1}{2}\ty^L(\breve \vark^L)^{-1}\ty^L}    \ ,\label{PhiLpt}\ee
where this time $\ty^L:= y-i\breve v^L\yb$. The different treatment of the holomorphic and anti-holomorphic dependence induced by the star-multiplication with $\k_y$ give rise to all the difference with respect to the twisted sector, resulting in particular in a more complicated form of the Killing two-form and Killing vector:
\bea \breve \vark^L_{\a\b} & = & -\frac{i}{2\varsigma(1-X_{0'})}\left[(1-X_{0'})^2-\varsigma^2X^aX_a\right](\s_{03})_{\a\b}+\frac{2i\varsigma}{1-X_{0'}}X_{[0}X^a\s_{3]a}\\
\breve v^L_{\a\bd} & = & (\varsigma-\varsigma^{-1})X_{[1}(\s_{2]})_{\a\bd}+i(\varsigma+\varsigma^{-1})X_{[0}(\s_{3]})_{\a\bd} \ .\eea
Equivalently, in terms of a rigid spin-frame,
\bea \breve \vark^L_{\a\b} & = & -\frac{i}{2\varsigma(1-X_{0'})}\left[(1-X_{0'})^2-\varsigma^2X^aX_a\right](u^+_\a u^-_\b+ u^-_\a u^+_\b)\nn \\
&& \ \ +\frac{i\varsigma}{1-X_{0'}}\left[(X_0+X_3)(X_1-iX_2)u^+_\a u^+_\b+(X_0-X_3)(X_1+iX_2)u^-_\a u^-_\b\right]\ ,\\
\breve v^L_{\a\bd} & = & \frac{i}2(\varsigma^{-1}-\varsigma)\left[(X_1-iX_2)u^+_\a \bar u^-_{\bd}-(X_1+iX_2)u^-_\a \bar u^+_{\bd}\right]\nn \\
&& \ \ +\frac{i}2(\varsigma+\varsigma^{-1}) \left[(X_3-X_0)u^+_\a \bar u^+_{\bd}+(X_3+X_0)u^+_\a \bar u^+_{\bd}\right]\ .\eea
In particular, note that, while for $\varsigma=1$ the above expressions reduce to \eq{varkiB}-\eq{viB} and \eq{varkiB2}-\eq{viB2}, respectively, in this case the $\varsigma$-dependence cannot be factored out of each of the $2\times 2$ blocks of $\breve K^L$ as it happened for the twisted sector. This means that, as we shall see, the integrand of the contour integrals will differ from those so far examined, and will in fact be incompatible with a small-contour integral presentation of type \eq{fgen} that we consider here. In particular, the study of the limit $\D^2\to 0$ elucidates the problem.

In fact, recalling that $(\breve \vark^L)^{-1}_{\a\b}=-\frac{\breve \vark^L_{\a\b}}{(\breve \vark^L)^2}$ it is immediate to see that the $\varsigma$ dependence is now nested with the spacetime dependence in the integrand, via 
\bea (\breve \vark^L)^2 \ = \ \frac{1-\varsigma^2}{4\varsigma^2}[1-\varsigma-X_{0'}(1+\varsigma)] [1+\varsigma-X_{0'}(1-\varsigma)]+\D^2 \ .\label{varkpt}\eea
It is clear that, evaluating the contour integrals first, the Weyl zero-form \eq{PhiLpt} reduces to the corresponding one in the twisted sector. This is expected, since the element $4e^{-4iB}$, which corresponds to $f^L_{\ft12,\ft12;\ft12,\ft12}$ after the contour integral is evaluated, is an eigenstate of $\k_y$, so there is no distinction between the $x$-independent elements $f^L_{\ft12,\ft12;\ft12,\ft12}$ and $f^L_{\ft12,\ft12;\ft12,\ft12}\star\k_y$ on which \eq{PhiLpt} and \eq{PhiL} are based. As a consequence, as long as the non-integral presentation of such elements is concerned, the regular and the twisted sector are equivalent. However, as mentioned above, this is not the case at the level of the integral presentation. Indeed, the integrand in \eq{PhiLpt}, coming from the regular sector, develops a branch cut due to \eq{varkpt}, and for $\D^2 = 0$ the latter inevitably crosses over the integration contour, making the small-contour integral presentation ill-defined for the expansion of the Weyl zero-form over the regular sector. This conclusion still holds when one considers non-diagonal element and  gives an imaginary part to the left eigenvalues, as the extra dependence on $\t_i$ (contained in $\theta^L$) coming from the Mellin transform does not modify the poles in $\varsigma_1$ and $\varsigma_2$ of the contour integrals. This is the reason that we discarded the regular sector in the expansion of the Weyl zero-form in this paper. We defer the analysis of alternative, more general integral presentations to a future publication \cite{paper1prime}.

\section{An integral formula using parabolic cylinder functions}\label{App:integralformula}

In this Appendix we shall prove the formulae \eq{Mellinscalarsin} and \eq{Mellinscalar}, which are crucial to extract the scalar field fluctuation \eq{C1C2}. One way to do it is the following.  First, one can compute one of the two $\t_i$-integral, say the one in $\t_2$, by regularizing it via multiplication by a factor $\lim_{\e\to 0^+}\tau_2^\e e^{-\e\t_2^2}$. We will find in the end that the result can be analytically continued to $\e\to 0^+$, so it will be possible to remove the regulator from the final expression. We can then recognize that the $\t_2$-integral corresponds to the integral realization of a parabolic cylinder function 
\be D_{-\n}\left(\frac{\gamma}{\sqrt{2\b}}\right) \ = \  \frac{(2\b)^{\n/2}}{\Gamma(\n)}e^{-\gamma^2/8\b}\int_0^\infty dx\,x^{\n-1}\,e^{-\b x^2-\g x} \ee 
(for ${\rm Re}(\b)>0$, ${\rm Re}(\n)>0$),
\bea &\displaystyle  \int_{0}^{+\infty }d\tau _{2}\,\tau _{2}^{ip-1} \,e^{-i\left( c_{2}\tau _{1}\tau
_{2}+c_{3}\tau _{2}{}^{2}\right) }    \ = \ \lim_{\e\to 0^+}\int_{0}^{+\infty } d\tau _{2}\, \tau _{2}^{\e+ip-1}\,e^{-ic_{2}\tau _{1}\tau
_{2}-(\e+i c_{3})\tau _{2}{}^{2} }& \nn\\
& \displaystyle  \ = \ \lim_{\e\to 0^+}\frac{\G(\e+ip)}{[2(\e+ic_3)]^{\frac{\e+ip}{2}}}\,\exp\left[-\frac{c_2^2\t_1^2}{8(\e+ic_3)}\right]D_{-\e-ip}\left(\frac{ic_2\t_1}{\sqrt{2(\e+ic_3)}}\right)& \ . \eea
Inserting in \eq{Mellinscalar} one is left with the integral
\be \lim_{\e\to 0^+}\frac{\G(\e+ip)}{[2(\e+ic_3)]^{\frac{\e+ip}{2}}}\,  \int_{0}^{+\infty }d\tau _{1}\,\tau _{1}^{-ip-1}\,\,\exp\left[-\left(ic_1+\frac{c_2^2}{8(\e+ic_3)}\right)\t_1^2\right]\,D_{-\e-ip}\left(\frac{ic_2\t_1}{\sqrt{2(\e+ic_3)}}\right) \ . \label{leftover}\ee
We can now change variable to $t=\t^2_1$ and use the formula \cite{GradshteynRyzhik}
\be \int_0^\infty dt\,t^{\frac{\b}2-1}\,e^{-zt}D_{-\n}(2\sqrt{kt}) \ = \ \frac{\sqrt{\pi}\,2^{1-\b-\frac{\n}2}}{\G(\frac{\n+\b+1}{2})}\G(\b)(z+k)^{-\b/2}{}_2F_1\left(\frac{\n}2,\frac{\b}2;\frac{\n+\b+1}{2};\frac{z-k}{z+k}\right) \ , \ee
(${\rm Re}(z+k)>0$, ${\rm Re}(\frac{z}{k})>0$) with $\b=-ip$, $\n=\e+ip$, $z=ic_1+\frac{c_2^2}{8(\e+ic_3)}+\e'$ ($\e'>0$), $k=-\frac{c^2_2}{8(\e+ic_3)}$ to compute the integral in \eq{leftover} as
\bea  & \displaystyle \lim_{\e'\to 0^+} \frac12 \int_{0}^{+\infty }dt\,t^{-\frac{ip}2-1}\,\,\exp\left[-\left(ic_1+\frac{c_2^2}{8(\e+ic_3)}+\e'\right)t\right]\,D_{-\e-ip}\left(\frac{ic_2}{\sqrt{2(\e+ic_3)}}\sqrt{t}\right)&\nn\\
& \displaystyle  \ = \  \frac{\sqrt{\pi} \,2^{\frac{ip-\e}2}}{\Gamma\left(\frac{\e+1}2\right)}\, \G(-ip)\,(ic_1)^{\frac{ip}2}{}_2F_1\left(\frac{\e+ip}2,-\frac{ip}2;\frac{\e+1}{2};1-\frac{c_2^2}{4c_1c_3}\right)\ . &\eea
One can now collect all terms and take the $\e\to 0$ limit to get the final result
\bea &\displaystyle \int_{0}^{+\infty }d\tau _{1}\frac{\tau _{1}^{-ip-1}}{\Gamma \left(
-ip\right) }\ \int_{0}^{+\infty }d\tau _{2}\frac{\tau _{2}^{ip-1}}{\Gamma
\left( ip\right) }e^{-i\left( c_{1}\tau _{1}{}^{2}+c_{2}\tau _{1}\tau
_{2}+c_{3}\tau _{2}{}^{2}\right) } & \nn\\
&\displaystyle \ = \ \frac{1}{\Gamma \left(
ip\right)}\lim_{\e\to 0^+}\frac{\G(\e+ip)}{[2(\e+ic_3)]^{\frac{\e+ip}{2}}}\, \frac{\sqrt{\pi} \,2^{\frac{ip-\e}2}}{\Gamma\left(\frac{\e+1}2\right)}\,(ic_1)^{\frac{ip}2}{}_2F_1\left(\frac{\e+ip}2,-\frac{ip}2;\frac{\e+1}{2};1-\frac{c_2^2}{4c_1c_3}\right)  &\nn\\
&\displaystyle \ = \ \left(\frac{c_1}{c_3}\right)^{\frac{ip}{2}}\cosh\left[p\arcsin\left(\sqrt{1-\frac{c_2^2}{4c_1 c_3}}\right)\right]\ ,\eea
and, for $\frac{c_2}{\sqrt{c_1 c_3}}>0$, 
\be  \int_{0}^{+\infty }d\tau _{1}\frac{\tau _{1}^{-ip-1}}{\Gamma \left(
-ip\right) }\ \int_{0}^{+\infty }d\tau _{2}\frac{\tau _{2}^{ip-1}}{\Gamma
\left( ip\right) }e^{-i\left( c_{1}\tau _{1}{}^{2}+c_{2}\tau _{1}\tau
_{2}+c_{3}\tau _{2}{}^{2}\right) }  \ = \ \left(\frac{c_1}{c_3}\right)^{\frac{ip}{2}}\cosh\left[p\arccos\left(\frac{c_2}{2\sqrt{c_1 c_3}}\right)\right] \ ,\ee
which are the results that we wanted to prove.

\section{Apparent singularity at $X^{0}+X^{3}=0$}\label{App:03divergence}

In this appendix, we discuss a subtlety arising in the generalized Weyl tensor
computation for the particular choice of eigenvalues (\ref{eigenchoice}).

We first look at (\ref{Mellinscalar}) and (\ref{Mellinparam}) for the scalar
field computation. One can see that the integral may not be well-defined at $%
X^{0}+X^{3}=0$, in which case the exponent vanishes. In this situation, we
should first compute the integrals for $X^{0}+X^{3}\neq 0$ and then
analytically continue the result to $X^{0}+X^{3}=0$. 

For the scalar field it is easy to see that $X^{0}+X^{3}=0$ is not a real
problem, because the factor $X^{0}+X^{3}$ does not appear at all in the
result (\ref{C1C2}). Another way to see this is that in (\ref{Mellinscalar})
the factor $X^{0}+X^{3}$ can be simply absorbed by redefining the
integration variables 
\begin{equation}
\tau _{1,2}=\frac{1}{\sqrt{X^{0}+X^{3}}}\tau _{1,2}^{\prime }\text{ ,}
\label{tauredef}
\end{equation}%
without creating any extra factors in the integrand.

However, for spin $s>0$ we need a more careful discussion. In this case, due
to the derivatives of the Weyl zero-form master field with respect to the $y^{\alpha }$%
-coordinates, the integrand has an extra
factor in comparison to the spin-0 case, a polynomial in the $\tau_i $:
\begin{equation}
\int_{0}^{+\infty }d\tau _{1}\frac{\tau _{1}^{-ip-1}}{\Gamma \left(
p_{1}\right) }\ \int_{0}^{+\infty }d\tau _{2}\frac{\tau _{2}^{ip-1}}{\Gamma
\left( p_{2}\right) }\ \text{Polynomial}(\tau _{1},\tau _{2})\ e^{-i\left(
c_{1}\tau _{1}{}^{2}+c_{2}\tau _{1}\tau _{2}+c_{3}\tau _{2}{}^{2}\right) }%
\text{ ,}
\end{equation}
and thus the redefinition (\ref{tauredef}) may or may not give rise to
vanishing denominators in the limit $X^{0}+X^{3}\rightarrow 0$, depending on the
coefficients of the polynomial.

We have checked the integrands of the spin-1 Faraday tensor $
C_{\alpha \beta }$ and of the spin-2 Weyl tensor $C_{\alpha \beta
\gamma \delta }$, the \textquotedblleft Polynomial$(\tau _{1},\tau _{2})$%
\textquotedblright\ factors of which respectively correspond to 
\bea
(\vark^L)^{-1}_{\a\b}-W_\a W_\b \ ,
\eea
and
\bea
3(\vark^L)^{-1}_{(\a\b}(\vark^L)^{-1}_{\g\d)}-6(\vark^L)^{-1}_{(\a\b}W_{\g}W_{\d)}+W_\a W_\b W_\g W_\d \ ,
\eea
where
\begin{equation}
W_{\alpha } \ := \ \left[(\vark^L)^{-1}\theta^L\right]_\a \text{ \ .}
\end{equation}
After the redefinition (\ref{tauredef}) these polynomial
factors do contain components that blow up at $X^{0}+X^{3}=0$, but the Lorentz-invariants $C_{\alpha \beta
}C^{\alpha \beta }$, $C_{\alpha \beta \gamma \delta }C^{\alpha \beta \gamma
\delta }$, $C_{\alpha \beta \gamma \delta }C^{\gamma \delta \varepsilon
\zeta }C_{\varepsilon \zeta }{}^{\alpha \beta }$ and $C_{\alpha \beta \gamma
\delta }C^{\alpha \beta \gamma \delta }$ approach finite constant values in the limit $%
X^{0}+X^{3}\rightarrow 0$.

This suggests that the generalized Weyl tensors can be made finite by a
frame rotation%
\begin{equation}
\check{y}^{\alpha }=y^{\alpha ^{\prime }}\left( \Lambda ^{-1}\right) _{\alpha
^{\prime }}{}^{\alpha }\text{ ,}
\end{equation}%
i.e.\ the tensors are finite after the transformation $\Lambda _{\alpha
}{}^{\alpha ^{\prime }}\Lambda _{\beta }{}^{\beta ^{\prime }}C_{\alpha
^{\prime }\beta ^{\prime }}$, $\Lambda _{\alpha }{}^{\alpha ^{\prime
}}\Lambda _{\beta }{}^{\beta ^{\prime }}\Lambda _{\gamma }{}^{\gamma
^{\prime }}\Lambda _{\delta }{}^{\delta ^{\prime }}C_{\alpha ^{\prime }\beta
^{\prime }\gamma ^{\prime }\delta ^{\prime }}$, etc. For example, we checked
that by the rotation 
\begin{equation}
\Lambda _{\alpha }{}^{\alpha ^{\prime }}=\left( 
\begin{array}{cc}
\sqrt{X^{0}+X^{3}} & 0 \\ 
\sqrt{X^{0}+X^{3}} & \frac{1}{\sqrt{X^{0}+X^{3}}}%
\end{array}%
\right) \text{ \ , \ \ }\left( \Lambda ^{-1}\right) _{\alpha ^{\prime
}}{}^{\alpha }=\left( 
\begin{array}{cc}
\frac{1}{\sqrt{X^{0}+X^{3}}} & 0 \\ 
-\sqrt{X^{0}+X^{3}} & \sqrt{X^{0}+X^{3}}%
\end{array}%
\right) \text{ ,}
\end{equation}%
all components in the polynomial factors for spin-1 and spin-2 remain finite
after the redefinition (\ref{tauredef}) in the limit\footnote{Of course, the background connection $\Omega$ transforms, too, under the Lorentz rotation $\epsilon_L:=\frac{1}{4i}\Lambda^{\a\b}y_{\a}y_{\b}-{\rm h.c.}$, $ \Omega'  =  \e_L^{-1}\star\Omega\star \e_L + \e_L^{-1}\star d\e_L $. The first term on the r.h.s. is a bilinear in the oscillators $\check{y}^{\alpha }$ with bounded coefficients, while the second contains, as only potentially dangerous factor, the logarithmic derivative $d\rho/\rho$, where $\rho:=\sqrt{X^{0}+X^{3}}$, which can however be reabsorbed in a coordinate transformation, leaving a well-defined background connection.} $X^{0}+X^{3}\rightarrow
0$.


\bibliography{biblio}
\bibliographystyle{utphys}

\end{document}